\documentclass{pasa}%

\usepackage{graphicx}
\usepackage[T1]{fontenc}
\usepackage{aecompl}

\title[Spectroscopic and photometric analysis of 21 chromospherically active variables]{Spectroscopic and photometric analysis of 21 chromospherically active variables: activity cycles and differential rotation}

\author[\"Ozdarcan O.]{\"Ozdarcan O.$^{1,2}$\thanks{E-mail: orkun.ozdarcan@ege.edu.tr}
\affil{$^{1}$Ege University, Science Faculty, Department of Astronomy and Space Sciences, 
35100 Bornova, \.{I}zmir, Turkey\\}%
\affil{$^{2}$T\"UB\.ITAK National Observatory, Akdeniz University Campus, 07058 Konyaalt\i, Antalya, Turkey}
}%

\jid{PASA}
\doi{10.1017/pas.\the\year.xxx}
\jyear{\the\year}

\usepackage{aas_macros}
\usepackage{hyperref} 
\hypersetup{colorlinks,citecolor=blue,linkcolor=blue,urlcolor=blue}

\hypersetup{draft}

\newcommand{\kms}{\mbox{km s$^{-1}$}}
\newcommand{\fm}{\hbox{$.\!\!^{\rm m}$}}

\begin{document}

\begin{frontmatter}
\maketitle

\begin{abstract}
We investigate magnetic activity properties of 21 stars via medium resolution optical 
spectra and long-term photometry. Applying synthetic spectrum fitting method, we find that all 
targets are cool giant or sub-giant stars possessing overall [M/H] abundances between $0$ and 
$-0.5$. We find that six of these targets exhibit only linear trend in mean brightness while 
eight of them clearly shows cyclic mean brightness variation. Remaining seven target appear 
to exhibit cyclic mean brightness variation but this can not be confirmed due to the long time 
scales of the predicted cycle compared to the current time range of the photometric data.
We further determine seasonal photometric periods and compute average 
photometric period of each target. Analysed sample in this study provides a quantitative 
representation of a positive linear correlation between the inverse of the rotation period and 
the cycle period normalized to the rotation period, on the log-log scale. We also observe no 
correlation between the activity cycle length and the relative surface shear, indicating 
that the activity cycle must be driven by a parameter rather than the differential rotation. 
Our analyses show that the relative surface shear is positively correlated with the rotation
period and there is a noticeable separation between main sequence stars and our sample. Compared 
to our sample, the relative surface shear of a main sequence star is larger for a given 
rotation period. However, dependence of the relative surface shear on the rotation period appears 
stronger for our sample. Analysis of the current photometric data indicates that the photometric 
properties of the observed activity cycles in 8 targets seem dissimilar to the sunspot cycle.

\end{abstract}

\end{frontmatter}

\section{Introduction}\label{sec_intro}

\begin{table*}
	\centering
	\caption{Identifiers, J2000 equatorial coordinates, $V$ magnitudes and the rotation periods (P)
	of the the target stars. In the last column, the first reference is for $V$ magnitude and the 
	second reference is for the period.}
	\label{table_basic_data}
	\begin{tabular}{lccccc} 
		\hline
Identifier &	RA	(J2000)	&	Dec (J2000) 			&  $V$   &  P    &	Ref.	\\
				  &	(h m s)		&	($^{\circ}$~$'$~$''$)	&  (mag) & (day) &			\\
\hline\noalign{\smallskip}
V660\,Vir       &  14 22 50   &  $+$06 41 12   &  11.69   & 70.8    &  \citet{V_ref_UCAC4_2013AJ....145...44Z, gsc324_tyc1683_tyc723_ref_2009OEJV..105....1S} \\
DG\,Ari         &  02 55 21   &  $+$15 39 51   &  11.15   & 33.998  &  \citet{Tycho_2_cat_Hog_et_al2000AA, gsc1224_tyc1083_tyc1656_tyc2237_ref_2011OEJV..136....1L} \\
V1263\,Tau      &  03 24 06   &  $+$07 29 27   &  10.61   & 20.504  &  \citet{V_mag_ref_2009AJ....137.4598S, tyc1756_tyc63_tyc4912_ref_2008OEJV...86....1B} \\
FK\,CMi         &  07 36 42   &  $+$03 54 20   &  11.22   & 19.28   &  \citet{V_mag_ref_2009AJ....137.4598S, tyc186_tyc450_tyc4667_tyc424_ref_2011PZP....11...15B} \\
V383\,Vir       &  12 16 53   &  $+$05 41 26   &  10.20   & 14.26   &  \citet{V_mag_ref_2009AJ....137.4598S, tyc287_ref_2008PZ.....28....9B}        \\
BD+04\,3503     &  17 46 25   &  $+$03 58 49   &  9.54    &  8.447  &  \citet{V_ref_ASAS_Kiraga_2012AcA....62...67K, tyc186_tyc450_tyc4667_tyc424_ref_2011PZP....11...15B}  \\
BD+02\,3610     &  18 32 19   &  $+$02 14 54   &  11.55   & 12.84   &  \citet{V_ref_ASAS_Kiraga_2012AcA....62...67K, tyc186_tyc450_tyc4667_tyc424_ref_2011PZP....11...15B} \\
GH\,Psc         &  00 57 08   &  $+$10 25 57   &  10.05   & 34.81   &  \citet{V_mag_ref_2009AJ....137.4598S, tyc608_ref_2008OEJV...78....1B} \\
TYC\,723-863-1  &  05 48 23   &  $+$12 18 26   &  10.40   & 44.85   &  \citet{Tycho_2_cat_Hog_et_al2000AA, gsc324_tyc1683_tyc723_ref_2009OEJV..105....1S} \\
HD\,354410      &  19 57 53   &  $+$14 20 18   &  11.20   & 27.573  &  \citet{V_ref_ASAS_Kiraga_2012AcA....62...67K, gsc1224_tyc1083_tyc1656_tyc2237_ref_2011OEJV..136....1L} \\
TYC\,1094-792-1 &  20 55 51   &  $+$10 23 41   &  11.46   & 10.15   &  \citet{Tycho_2_cat_Hog_et_al2000AA, tyc1094_tyc1104_tyc1191_ref_2008OEJV...92....1B} \\
UY\,Equ         &  21 10 54   &  $+$08 58 16   &  11.58   & 14.06   &  \citet{V_mag_ref_2009AJ....137.4598S, tyc1094_tyc1104_tyc1191_ref_2008OEJV...92....1B} \\
FP\,Psc         &  00 43 49   &  $+$18 46 53   &  10.91   & 13.74   &  \citet{Tycho_2_cat_Hog_et_al2000AA, tyc1094_tyc1104_tyc1191_ref_2008OEJV...92....1B} \\
TYC\,1541-191-1 &  17 24 05   &  $+$18 29 37   &  11.23   & 11.563  &  \citet{Tycho_2_cat_Hog_et_al2000AA, tyc1541_ref_2008OEJV...82....1B} \\
V343\,Del       &  21 03 23   &  $+$19 30 56   &  10.50   & 10.377  &  \citet{V_mag_ref_2009AJ....137.4598S, gsc1224_tyc1083_tyc1656_tyc2237_ref_2011OEJV..136....1L} \\
V439\,Peg       &  21 30 41   &  $+$22 01 43   &  10.55   & 24.135  &  \citet{Tycho_2_cat_Hog_et_al2000AA, tyc1676_ref_2008OEJV...89....1B} \\
TYC\,1683-144-1 &  21 59 45   &  $+$16 57 38   &  11.27   & 44.22   &  \citet{Tycho_2_cat_Hog_et_al2000AA, gsc324_tyc1683_tyc723_ref_2009OEJV..105....1S} \\
BE\,Ari         &  01 47 10   &  $+$23 45 32   &  10.08   & 21.203  &  \citet{Tycho_2_cat_Hog_et_al2000AA, tyc1756_tyc63_tyc4912_ref_2008OEJV...86....1B} \\
V592\,Peg       &  23 21 53   &  $+$23 16 56   &  10.79   & 19.09   &  \citet{V_ref_UCAC4_2013AJ....145...44Z, gsc1224_tyc1083_tyc1656_tyc2237_ref_2011OEJV..136....1L} \\
TYC\,4667-90-1  &  00 15 08   &  $-$03 20 00   &  11.24   &  8.84   &  \citet{Tycho_2_cat_Hog_et_al2000AA, tyc186_tyc450_tyc4667_tyc424_ref_2011PZP....11...15B} \\
BC\,Sex         &  10 30 03   &  $-$00 47 32   &  11.79   & 15.37 &  \citet{V_mag_ref_2009AJ....137.4598S, tyc287_ref_2008PZ.....28....9B}  \\
		\hline
	\end{tabular}
\end{table*}

Solar irradiance measurements performed by Nimbus-7 satellite revealed that the solar 
constant coherently varied with daily sunspot number \citep{nimbus7_1988SSRv...48..321H}.
This result triggered the idea that the target stars observed in Mount Wilson Observatory 
$HK$ Project \citep{wilson_1978ApJ...226..379W, vaughan_preston_wilson_1978PASP...90..267V, 
duncan_1991ApJS...76..383D, baliunas_1995ApJ...438..269B} could exhibit variability in 
long-term mean brightness, like observed in their $S$ indexes. Long-term Str\"omgren photometry 
in $b$ and $y$ bands were carried out for a sample of target stars of the $HK$ project and
a correlation was found between chromospheric and photometric variations, but with a clear 
diversity between younger and older stars \citep{lockwood_1997ApJ...485..789L, 
radick_1998ApJS..118..239R}. Since the main reference is the Sun itself, almost all target
star were chosen as main sequence stars from different spectral types. 
\citet{Baliunas_pcycp_prot_1996ApJ...460..848B} used a sample from $HK$ project targets 
and discovered a correlation between the cycle length normalized to the rotation period 
($P_{cyc}/P_{rot}$) and the dynamo number, $D\sim1/P_{rot}$, which was expected to provide 
clues about the type of the dynamo operating in these stars. Further studies on the $HK$ 
project targets also revealed a linear positive correlation between $P_{rot}$ and the 
rotational period variation amplitude $\Delta\,P$ \citep{donahue_saar_baliunas_1996ApJ...466..384D}.
Since $\Delta\,P$ is commonly accepted as the proxy of the surface differential rotation, 
the correlation indicates that slow rotators possess stronger surface differential rotation 
compared to fast rotators.

In addition to the Str\"omgren photometry, some other projects on the long-term photometric 
observations of chromospherically active stars from various luminosity classes have started 
by different groups \citep{APTs_Henry_1995ASPC, APTs_Strassmeier_1997PASP, 
APTs_Rodono_2001AN....322..333R} with automated photoelectric telescopes (APTs). 
These efforts enabled to trace long-term photometric behaviour of chromospherically active 
giant and sub-giant stars, in addition to the $HK$ project stars.

Efficient and precise broad-band $V$ observations obtained from APTs allowed studying 
photometric activity cycles for main sequence and giant stars 
\citep{Olah_strassmeier_cycles_2002AN, Messina_Guinan_2003A&A...409.1017M, Olah_2013AN....334..625O}.
Furthermore, these observations enabled studies on photometric period variations and 
long-term tracing of the light curve properties of giant and sub-giant stars 
\citep{Jetsu_Henry_long_term_data_2017ApJ}. These studies enable to make a comparison 
between photometric properties of the magnetic activity on these stars and 
the sunspot cycle becomes possible \citep[see, e.g.][]{V1149Ori_Fekel_et_al_2005AJ, 
Ozdarcan_V2075Cyg_2010AN}. For instance, \citet{FG_IS_Fekel_et_al_2002AJ} analysed long-term 
photometric data of two SB1 systems, HD\,89546 and HD\,113816 and did not find any 
correlation between the photometric period and the mean brightness, which is a typical 
properties of the sunspot cycle. However, ten years later, \citet{Ozdarcan_FGUMa_2012AN} 
re-analysed the long-term photometry of HD\,89546, which was doubled as the time coverage 
of the data compared to the former study, and showed that the distribution of the mean 
brightness with respect to the photometric period is similar to one observed in the sunspot 
cycle (see Fig. 6 in their study). Beside these APT observations, photometric sky surveys, 
such as The All Sky Automated Survey \citep[ASAS3,][]{ASAS_Pojmanski_1997, ASAS3_Pojmanski_2005AcA, 
ASAS_Pojmanski_2002AcA} and All-Sky Automated Survey for Supernovae Sky Patrol 
\citep[ASAS-SN,][]{ASAS_SN_2014ApJ, ASAS_SN_2017PASP..129j4502K} also provided long-term 
photometric data for a given position on the sky plane, which is also useful for long-term 
photometric analyses of chromospherically active stars 
\citep[see, e.g.,][]{Ozdarcan_Dal_2018AN....339..277O}. All these studies led to an increase 
in the number of well studied giant and sub-giant stars, thus reliable statistical evaluation of 
magnetic activity properties of these stars became possible. Owing to these studies, it came out 
that both main sequence and giant stars obey the relation between ($P_{cyc}/P_{rot}$) and ($1/P_{rot}$), 
without a significant diversity between the luminosity classes \citep{Olah_strassmeier_cycles_2002AN, 
Olah_cycles_2009A&A}. However, giant and sub-giant stars still need more attention. We still do not 
know precisely if the photometric properties of the magnetic activity in these stars are similar 
or dissimilar to the photometric properties of the solar activity. On the other hand, test bench 
of the theoretical studies, in the scope of the mean field dynamo models, has not been fulfilled 
sufficiently for cool giant and sub-giant region in the Hertzsprung-Russell diagram.

In this study we present analyses of spectroscopic and long-term photometric observations of 
21 stars, which are listed in Table~\ref{table_basic_data} along with their basic properties. 
These stars were suggested as candidate RS\,CVn variables in the studies listed in the table. 
We give technical details of the spectroscopic observations and data reductions in the next section. 
Then we focus on spectral types, spectral features in terms of chromospheric activity and global 
atmospheric properties of the target stars. In Section~\ref{sec_photometry}, we describe sources 
of the photometric data and analyse long-term brightness variation, together with seasonal light 
curves. We investigate possible relations between the activity cycle length, the rotation period 
and the relative surface shear (i.e. the differential rotation) in Section~\ref{sec_cycle_rot_dif_rot}. 
We summarize and discuss our findings in the last section.

\section{Spectroscopy}\label{sec_spectroscopy}

\subsection{Data}\label{sec_data}

We obtained intermediate resolution optical \'echelle spectra of the target stars at T\"UB\.ITAK 
National Observatory (TNO) with 1.5 m Russian -- Turkish telescope and Turkish Faint 
Object Spectrograph Camera 
(TFOSC\footnote{\url{https://tug.tubitak.gov.tr/en/teleskoplar/rtt150-telescope-0}}).
Until 8th August 2017, optical spectra were recorded by a back illuminated CCD camera 
with 15 $\times$ 15 $\mu m^{2}$ pixel size and 2048 $\times$ 2048 pixels. After that 
date, CCD camera was replaced by a new Andor DW436-BV 2048 $\times$ 2048 pixels CCD 
camera with a pixel size of 13.5 $\times$ 13.5 $\mu m^{2}$. This instrumental set-up 
enabled us to record optical spectrum between 3900 -- 9100 \AA\@ in 11 \'echelle orders 
with actual spectral resolution (R = $\lambda/\Delta\lambda$) of 2700 $\pm$ 500 around 
$\lambda=5500$ \AA\@ for both CCD cameras. 

Raw object spectra were reduced by following conventional \'echelle reduction steps,
starting with bias correction and followed by flat-field division by bias corrected 
and normalized flat-field image, scattered light correction, cosmic rays removal and 
extraction of the spectra from the \'echelle orders. Wavelength calibration of the object spectra
were done by Fe-Ar calibration lamp spectra. In the final step, extracted and wavelength calibrated 
object spectra were normalized to the unity order by order, by using 4th or 5th order cubic spline 
functions. All these steps were carried out under IRAF environment\footnote{The Image Reduction 
and Analysis Facility is hosted by the National Optical Astronomy Observatories in Tucson, Arizona 
at URL iraf.noao.edu}. Since normalization of spectra is based on fitting a mathematical function to 
a stellar spectrum, we use these normalized spectra as the initial input for atmosphere analysis. During the 
atmosphere analysis, these spectra are re-normalized with respect to a proper template synthetic spectrum, 
which is more reliable for spectrum normalization.

\subsection{Spectral types and features}\label{sec_sp_type_features}

\begin{figure*}
	\includegraphics[width=\textwidth, height=9.5cm]{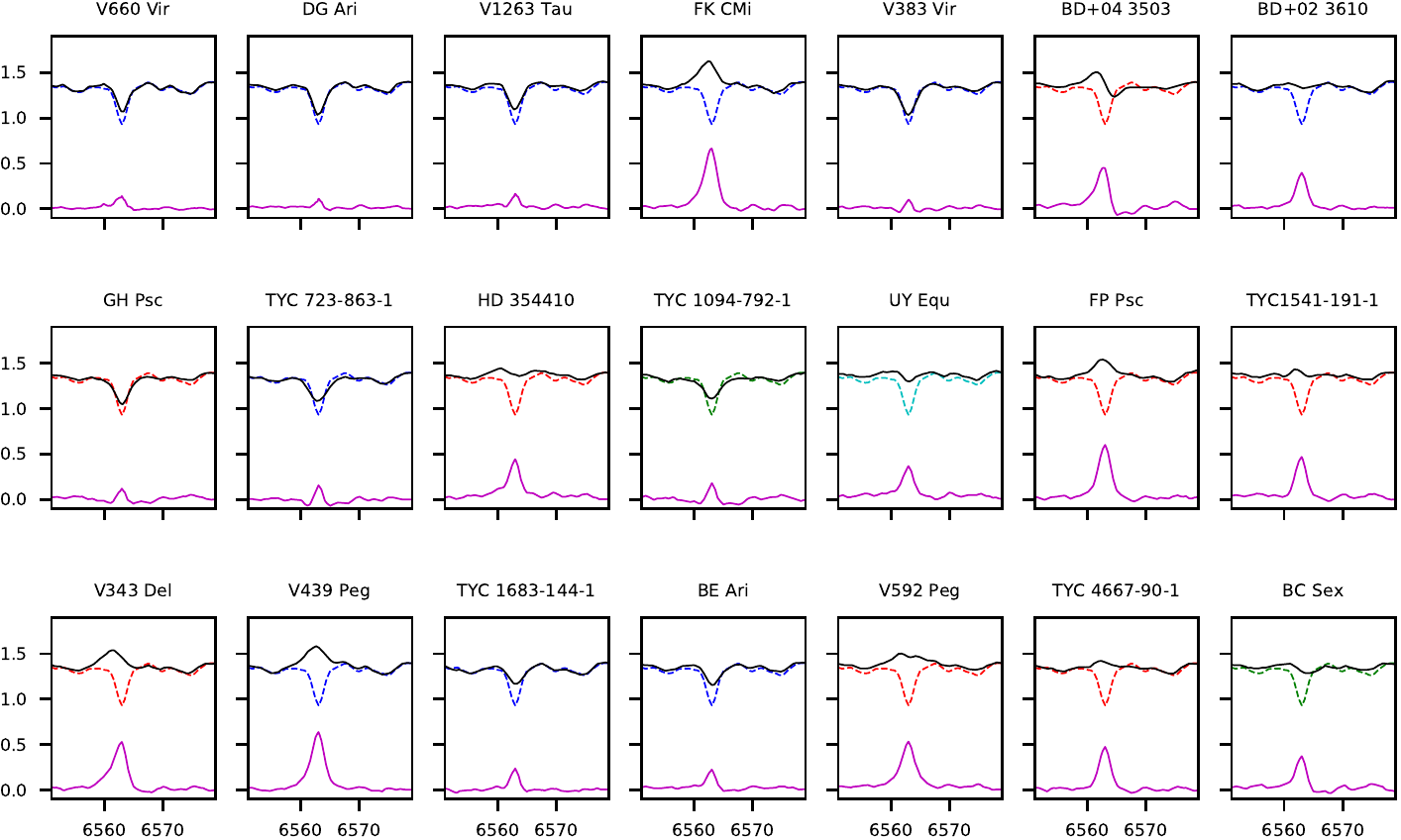}
	\caption{H$_{\alpha}$ line profiles of the target stars (black lines), comparison stars 
	(dashed lines) and the difference spectrum in the sense of target-minus-comparison 
	(magenta lines). Horizontal and vertical axes show wavelength (\AA) and normalized flux,
	respectively. We shift line profiles of the target and comparison stars upwards by 0.4 
	in order to see difference spectrum separately. The spectra of comparison stars $\kappa$\,Oph, 
	35\,Peg, $o$\,Psc and $\epsilon$\,Psc are in blue, red, green and cyan colours, respectively.}
    \label{fig_halpha_profiles}
\end{figure*}

\begin{figure*}
	\includegraphics[width=\textwidth, height=9.5cm]{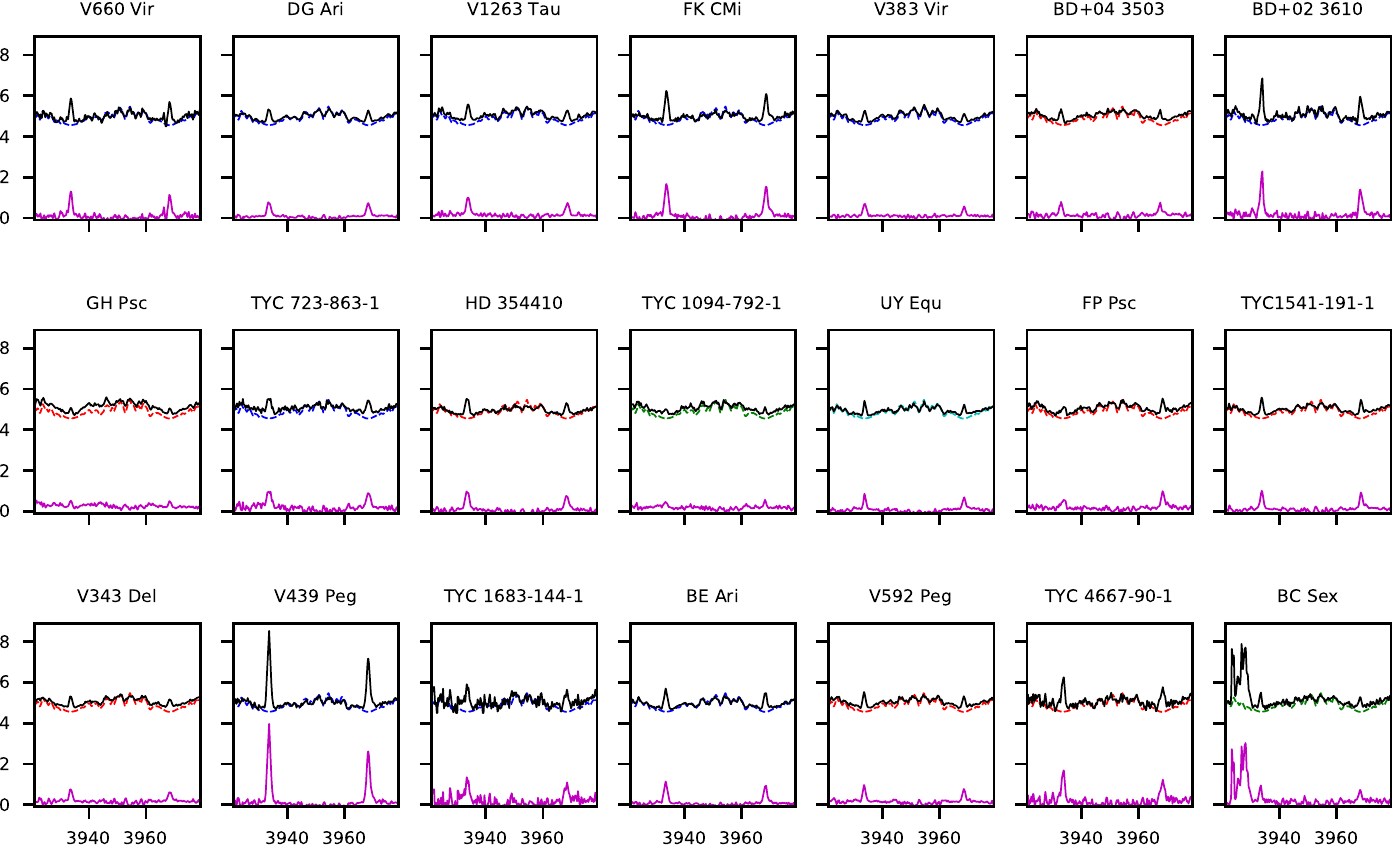}
	\caption{Same as Fig.~\ref{fig_halpha_profiles} but for Ca\,{\sc ii} H\& K line profiles.
	Due to the very strong emission in V439\,Peg spectrum, we shift all observed spectra
	upwards by 4.5 so that emission strength can be seen separately. Note that 
	emission-like feature around 3925 \AA\, seen in the spectrum of BC\,Sex is a very strong 
	cosmic spike, which could not be removed properly. However, emission feature of 
	Ca\,{\sc ii} K line profile is still visible next to the spike.}
    \label{fig_CaIIHK_profiles}
\end{figure*}

Beside target stars, we obtained optical \'echelle spectra of four more stars, 
$\kappa$\,Oph \citep[K2 III,][]{kappa_oph_ref_2003AJ....126.2048G}, 
35\,Peg \citep[K0 III,][]{35Peg_ref_2009A&A...508.1313F}, 
$o$\,Psc \citep[G8 III,][]{omicron_psc_ref_2015A&A...574A..50J} and $\epsilon$\,Psc
\citep[G9 III,][]{epsilon_psc_ref_2008A&A...480...91S} with the same instrumental 
set-up, which are to be used as observational comparison templates. Comparing
the template spectra with each of target star spectrum individually, we see that each 
of the target star spectrum nicely match one of the comparison template spectra, meaning that 
our sample include only giant and sub-giant stars. In Fig.~\ref{fig_halpha_profiles} and
Fig.~\ref{fig_CaIIHK_profiles}, we show H$_{\alpha}$ and Ca\,{\sc ii} H\& K
line profiles of each target star and the best-matched observational comparison spectrum.
These spectral lines are very sensitive to the chromospheric activity observed in cool stars.
Some of our target stars exhibit direct emission in H$_{\alpha}$ line profile
while emission features become visible for other stars in the difference spectrum, 
in the sense of target-minus-comparison. Emission features are more pronounced in 
Ca\,{\sc ii} H\& K line profiles. These line profiles strongly indicate chromospheric 
activity on these targets. Among our sample, V439\,Peg exhibits the strongest 
emission features both in H$_{\alpha}$ and Ca\,{\sc ii} H\& K lines and appears as 
the most active star in the sample. Another noticeable feature can be seen in H$_{\alpha}$ line 
profile of BD+04\,3503, which resembles P\,Cyg line profile. Since the spectral features 
strongly indicate chromospheric activity, we may speculate that it might be a
spectroscopic binary candidate, that one component might be very active and show 
emission in H$_{\alpha}$, while secondary component is a less active or inactive
star, thus contributing to composite H$_{\alpha}$ line profile in form of absorption. 
Since we only have a single spectrum per target and resolution of TFOSC spectra is 
limited, it is not possible to detect any shift in spectral lines as an evidence of 
orbital motion or any spectral line of a possible secondary component.

\begin{table*}
	\centering
	\caption{Atmospheric analysis results of target stars. In the last two columns, resolution 
	($R=\lambda/\Delta\,\lambda$) and estimated spectral types of the target stars are given. 
	Spectral type estimation is done by comparing final atmospheric parameters with calibration
	given in \citet{Gray_2005oasp.book.....G}.}
	\label{table_atmos_analysis}
	\resizebox{\textwidth}{!}{\begin{tabular}{lcccccccccc} 
		\hline
Target  &  Comparison &HJD  & Exp. (s) & SNR & $T_{\rm eff}$ (K) &  log$g$ (cgs) &  $[$M/H$]$  &   $v_{\rm mic}$ (\kms) & R & Sp  \\
\hline\noalign{\smallskip}
V660\,Vir         & $\kappa$\,Oph   & 2458227.4379   &   3600    &    97      &   4468$\pm$136   &   2.59$\pm$0.30   &$~~$0.02$\pm$0.15   &   1.78$\pm$0.32   &   3411$\pm$227 & K3 III \\
DG\,Ari           & $\kappa$\,Oph   & 2457671.5264   &   3600    &    210     &   4444$\pm$157   &   2.37$\pm$0.44   & $-$0.02$\pm$0.17   &   1.72$\pm$0.34   &   2757$\pm$200 & K3 III \\
V1263\,Tau        & $\kappa$\,Oph   & 2457997.5884   &   2700    &    147     &   4426$\pm$217   &   2.14$\pm$0.42   & $-$0.53$\pm$0.23   &   1.56$\pm$0.41   &   2768$\pm$241 & K3 III \\
FK\,CMi           & $\kappa$\,Oph   & 2458194.2858   &   3600    &    85      &   4497$\pm$168   &   2.67$\pm$0.51   & $-$0.49$\pm$0.19   &   1.86$\pm$0.46   &   2740$\pm$231 & K2 III \\
V383\,Vir         & $\kappa$\,Oph   & 2458227.3942   &   3600    &    225     &   4655$\pm$206   &   2.62$\pm$0.45   &$~~$0.05$\pm$0.19   &   1.56$\pm$0.37   &   2611$\pm$207 & K1 III \\
BD+04\,3503       & 35\,Peg         & 2458394.2246   &   2400    &    205     &   4858$\pm$304   &   3.11$\pm$0.59   & $-$0.05$\pm$0.25   &   1.26$\pm$0.55   &   1648$\pm$183 & K0 III-IV \\
BD+02\,3610       & $\kappa$\,Oph   & 2458394.3054   &   3600    &    131     &   4489$\pm$128   &   3.02$\pm$0.30   &$~~$0.00$\pm$0.14   &   1.93$\pm$0.41   &   2474$\pm$183 & K2 III-IV \\
GH\,Psc           & 35\,Peg         & 2458345.4909   &   2400    &    178     &   4866$\pm$278   &   2.78$\pm$0.53   & $-$0.05$\pm$0.24   &   1.34$\pm$0.41   &   2828$\pm$251 & K0 III \\
TYC\,723-863-1    & $\kappa$\,Oph   & 2458194.2365   &   2700    &    123     &   4414$\pm$166   &   2.11$\pm$0.56   & $-$0.10$\pm$0.19   &   1.84$\pm$0.35   &   2323$\pm$180 & K3 III \\
HD\,354410        & 35\,Peg         & 2458227.5840   &   3600    &    119     &   4729$\pm$273   &   2.76$\pm$0.57   & $-$0.47$\pm$0.25   &   2.05$\pm$0.53   &   2235$\pm$216 & K1 III \\
TYC\,1094-792-1   & $o$\,Psc        & 2458265.5389   &   3000    &    89      &   5240$\pm$399   &   3.59$\pm$0.53   & $-$0.49$\pm$0.29   &   1.74$\pm$0.70   &   2999$\pm$329 & G4 III-IV \\
UY\,Equ           & $\epsilon$\,Psc & 2458346.3991   &   3600    &    123     &   4706$\pm$216   &   2.98$\pm$0.54   & $-$0.54$\pm$0.20   &   1.71$\pm$0.48   &   3367$\pm$295 & K1 III \\
FP\,Psc           & 35\,Peg         & 2458395.5739   &   3600    &    147     &   4662$\pm$253   &   2.65$\pm$0.48   & $-$0.50$\pm$0.24   &   1.69$\pm$0.47   &   2841$\pm$259 & K1 III \\
TYC\,1541-191-1   & 35\,Peg         & 2458226.5153   &   3600    &    123     &   4705$\pm$225   &   2.94$\pm$0.75   & $-$0.53$\pm$0.22   &   1.55$\pm$0.53   &   2783$\pm$266 & K1 III \\
V343\,Del         & 35\,Peg         & 2458265.5785   &   2400    &    92      &   4629$\pm$214   &   2.86$\pm$0.59   & $-$0.36$\pm$0.21   &   1.77$\pm$0.47   &   2599$\pm$226 & K1 III \\
V439\,Peg         & $\kappa$\,Oph   & 2458395.3852   &   3000    &    135     &   4363$\pm$145   &   2.32$\pm$0.42   & $-$0.10$\pm$0.16   &   1.90$\pm$0.35   &   2614$\pm$189 & K3 III \\
TYC\,1683-144-1   & $\kappa$\,Oph   & 2458395.4778   &   3600    &    100     &   4456$\pm$141   &   2.49$\pm$0.39   & $-$0.03$\pm$0.15   &   1.92$\pm$0.35   &   2582$\pm$188 & K3 III \\
BE\,Ari           & $\kappa$\,Oph   & 2458345.5277   &   2400    &    170     &   4433$\pm$182   &   2.23$\pm$0.37   & $-$0.02$\pm$0.19   &   1.58$\pm$0.34   &   2625$\pm$197 & K3 III \\
V592\,Peg         & 35\,Peg         & 2458345.4500   &   3600    &    134     &   4741$\pm$243   &   2.60$\pm$0.56   & $-$0.19$\pm$0.22   &   1.49$\pm$0.40   &   2902$\pm$250 & K1 III \\
TYC\,4667-90-1    & 35\,Peg         & 2458395.5278   &   3600    &    113     &   4614$\pm$178   &   2.70$\pm$0.52   &$~~$0.02$\pm$0.17   &   1.60$\pm$0.40   &   2259$\pm$188 & K1 III \\
BC\,Sex           & $o$\,Psc        & 2458227.3453   &   3600    &    101     &   4947$\pm$331   &   3.32$\pm$0.85   & $-$0.54$\pm$0.28   &   1.56$\pm$0.64   &   2575$\pm$274 & G9 III-IV \\
		\hline
	\end{tabular}}
\end{table*}

\subsection{Atmosphere analysis}\label{sec_atmos_analysis}

For further detailed atmospheric analysis of the target stars, we adopt synthetic 
spectrum fitting method. Main idea of the method is to minimize the difference between 
the observed and the synthesized spectra for defined wavelength ranges. We apply the 
method by the newest version of \textsc{iSpec} software \citep{iSpec_Cuaresma_2014A&A}, 
which includes various radiative transfer codes, line lists and model atmospheres in a 
python framework with a user friendly graphical user interface. Since each target star 
appears as a cool giant or sub-giant star, we use \textsc{Turbospectrum} 
\citep{turbospectrum_ref_1998A&A...330.1109A, turbospectrum_ref2_2012ascl.soft05004P} 
radiative transfer code in conjunction with MARCS model atmospheres \citep{MARCS_Gustafsson_2008}.
These models are one-dimensional hydrostatic plane-parallel and spherical models computed 
under local thermodynamic equilibrium (LTE) assumption. In synthetic spectrum fitting and 
computing, we adopt spherical models, which are more proper for giant stars due to their 
large radii. Line list compiled from Vienna Atomic Line Database \citep[VALD3,][]{VALD3_Ryabchikova_2015} 
provided by \textsc{iSpec} software is adopted for spectrum synthesizing.

For each target star, the first step is to adopt a set of atmospheric parameters 
estimated in spectrum comparison described in the previous section and synthesize a 
template spectrum. Then, we shift observed spectrum of the target star to the rest wavelength 
and re-normalize it with respect to the template. In the normalization, we adopt the wavelength 
region between 3900 \AA\@ and 6800 \AA\@ and compute median for every 5 \AA\@ bin within this range
for the template spectrum. Computed median values within this wavelength range form a function, which
we adopt as the continuum function. Then we divide object spectrum by the continuum function and 
obtained normalized spectrum of the object. Shifted and re-normalized target spectrum 
is our input observed data. We set the effective temperature ($T_{\rm eff}$), logarithm of 
the surface gravity (log$g$), metallicity ($[$M/H$]$), micro turbulence velocity 
($v_{\rm mic}$) and spectral resolution as adjustable parameters during synthetic spectrum 
fitting. Considering resolution of TFOSC spectra and reported rotation periods of target stars 
in Table~\ref{table_basic_data}, we neglect line broadening due to projected rotational 
velocity ($v\,sin\,i$) and macro turbulence velocity, i.e. fix them to zero. We adopt 
wavelength range between 4900 \AA\@ and 6800 \AA\@ for synthetic spectrum fitting, but 
avoid highly blended absorption lines, very strong and broad spectral lines, such as 
H$_{\alpha}$ and NaI D resonant lines. We did not extend wavelength range towards bluer
part of the spectra since flux of target stars steeply decreases at these wavelengths, leading
to considerably lower signal-to-noise ratio. This is natural indication of that we are dealing 
with cool and red stars. Beyond the 6800 \AA\@, telluric lines and bands are dominant structures 
thus we omit this part of the spectrum in atmosphere analysis. When fitting process converges to a 
solution, we synthesize final template spectrum by adopting solution parameters. Since resolution of 
TFOSC spectra is not high, it was not necessary to repeat this process iteratively, i.e., 
applying this procedure only for once is fairly enough to achieve an acceptable fit to the 
observed spectrum and reasonable atmospheric parameters. We tabulate synthetic spectrum 
fitting results in Table~\ref{table_atmos_analysis}.

\begin{figure}
	\includegraphics[scale=0.55]{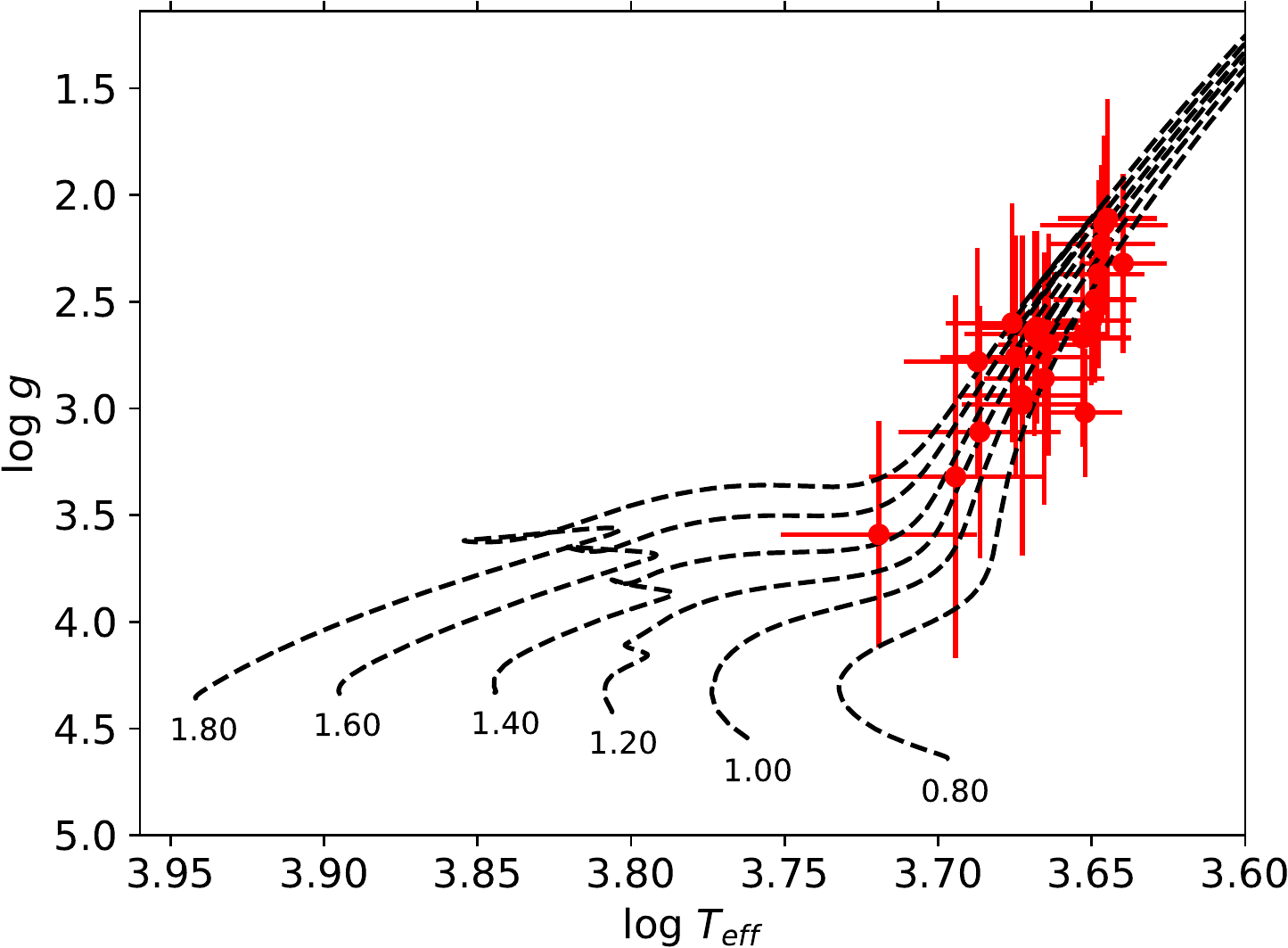}
	\caption{Positions of the target stars on  $log~T_{eff}-log~g$ plane. Evolutionary 
	tracks are from \citet{Bressan_et_al_2012MNRAS.427..127B} for $Z=0.014$ and $Y=0.273$. 
	We also show corresponding mass of each track in unit of the solar mass.}
	\label{figure_HRDiagram}
\end{figure}

We show observed spectrum of each target with its best-fitting synthetic spectrum in the 
appendix (Fig.~\ref{figure_spectrum_fit}). It can be noticed that the spectral lines of 
BD+04\,3503 are shallower and broader compared to the other target stars with similar atmospheric 
parameters. Furthermore, resulting spectral resolution for this star is considerably lower 
than the others. These results, together with spectral features of this star reported in 
Section~\ref{sec_sp_type_features}, support our speculation on spectroscopic binary 
possibility. For this star, we repeat synthetic spectrum fitting process with 
including $v\,sin\,i$ as additional adjustable parameters, or neglecting rotational 
broadening and fixing resolution to different values between 2200 and 3300. All these tests 
result in slightly different atmospheric parameters which basically agree within the uncertainties 
given in Table~\ref{table_atmos_analysis}. We show positions of the target stars on 
$log~T_{eff}-log~g$ plane in Fig.~\ref{figure_HRDiagram}, which clearly shows that our
targets are located in giant and sub-giant regions in the plane.

\section{Photometry}\label{sec_photometry}
Long-term $V$-band photometry of each target star were extracted from The All Sky Automated 
Survey \citep[ASAS,][]{ASAS_Pojmanski_1997, ASAS_Pojmanski_2002AcA, ASAS3_Pojmanski_2005AcA} 
and All-Sky Automated Survey for Supernovae Sky Patrol
\citep[ASAS-SN,][]{ASAS_SN_2014ApJ, ASAS_SN_2017PASP..129j4502K} databases. Compiled photometric
data from these databases provide observational time base of 16-18 years, depending on the 
target. There is a 3 or 4 years time gap between the last observation of ASAS and the first 
observation of ASAS-SN data. This gap was filled by ASAS-3N data \citep{ASAS-3N}, which enables
us to have continuous $V$-band photometry of each target without considerable data gap.
There are time ranges where data overlap occurs between ASAS and ASAS-3N data. Similar overlap
exists for ASAS-3N and ASAS-SN data. These overlaps provide an advantage of comparing different
data sets and check for any systematic shift in magnitude axis. ASAS and ASAS-3N data mostly agree, 
but considerable shifts (maximum 0\fm2) of ASAS-SN data with respect to the ASAS and ASAS-3N data 
were observed. In such cases, we adopt ASAS data as the reference data set and apply constant shift
in magnitudes with respect to the ASAS data. We do not expect any major effect in cycle search and
photometric period analysis results due to such shifts, since we only remove systematic offsets
by comparing overlapped data of different data sets.

Before starting analysis, we plot compiled data against time and check for highly deviated 
observations with respect to the general trend of the whole light curve. The data is mainly 
used to investigate long-term brightness variation and seasonal photometric behaviour. Therefore, 
any highly deviated data point, regardless if it shows an instantaneous brightness variation 
(e.g. flare) or not, were discarded from the data set. We further check outlier data points in 
by inspecting phase-folded seasonal light curves. For each seasonal light curve possessing 
sufficient data points and considerable light curve amplitude, we find a photometric period 
(see Section~\ref{subsec_seasonal_lc}) and compute phase-folded light curve with respect to 
that period. In order to detect outliers, we fit a cubic spline to the phase folded light curve, 
compute residuals from the best-fitting cubic spline and their mean absolute deviation ($\sigma$). 
If any data point is out of $\pm3\sigma$ band of the residuals, it is removed from the data.
Repeating this procedure iteratively until there is no outlier in the corresponding light curve, 
we obtain final light curves for photometric analysis.

\subsection{Mean brightness variations}\label{subsec_mean_bright}
In the first step of the photometric analysis, we inspect long-term mean brightness variation
of each target. For a given target, we first apply linear fit to all data in order to remove 
any sign of a possible very long-term variation with a time scale much longer than
the time base of current photometric data. Then we obtain residuals from the linear fit 
(here after, linear residuals), which scatters around zero level, and search for any long-term 
periodicities in the linear residuals via Lomb-Scargle periodogram\citep{Lomb_1976Ap&SS, Scargle_1982ApJ}. 
Inspecting compiled photometry of target stars given in Fig.~\ref{figure_long_term_photometry} 
we did not expect any cyclic variation time scale shorter than one year. In order to avoid rotational 
modulation signal and its harmonics, which could suppress cycle signal in the amplitude spectrum, we did 
not include periods shorter than 100 days in cycle search. Thus, we scan period range between 100 days 
and the time-span of the compiled photometric data ($\Delta\,t$) given in the last column of the 
Table~\ref{table_cycles}. We find that 8 of 21 target stars show
cyclic mean brightness variation with one significant period. We present compiled photometric data 
and of all target stars and detected mean brightness variations for 8 stars in 
Fig.~\ref{figure_long_term_photometry} and tabulate detected cyclic variation periods in 
Table~\ref{table_cycles}. Among remaining 13 targets, six of them, DG\,Ari, V1263\,Tau, HD\,354410, 
TYC\,1541-191-1, V343\,Del and V592\,Peg only exhibit linear increase or decrease in mean brightness.
For the each of remaining 7 targets, GH\,Psc, TYC\,723-863-1, TYC\,1094-792-1, BC\,Sex, BE\,Ari, V439\,Peg 
and TYC\,1683-144-1, we still find statistically significant cycles signal above 3$\sigma$ threshold, but 
resulting cycles do not repeat itself more than 1.5 times in the current data set. Further observations 
are needed to check the reliability of the detected signals, therefore we did not consider these stars 
as cyclic variables and did not include them in Table~\ref{table_cycles}.

We also provide amplitude spectra of the Lomb-Scargle periodogram in Fig.~\ref{figure_amplitude_spectra}, 
where one may see two peaks above 3$\sigma$ threshold for some the target stars. However, these secondary 
peaks hardly exceed 3$\sigma$ threshold or vanish below that level after prewhitening of the primary period,
therefore did not considered as significant signal.

\begin{figure*}
	\includegraphics[width=\textwidth,height=21cm]{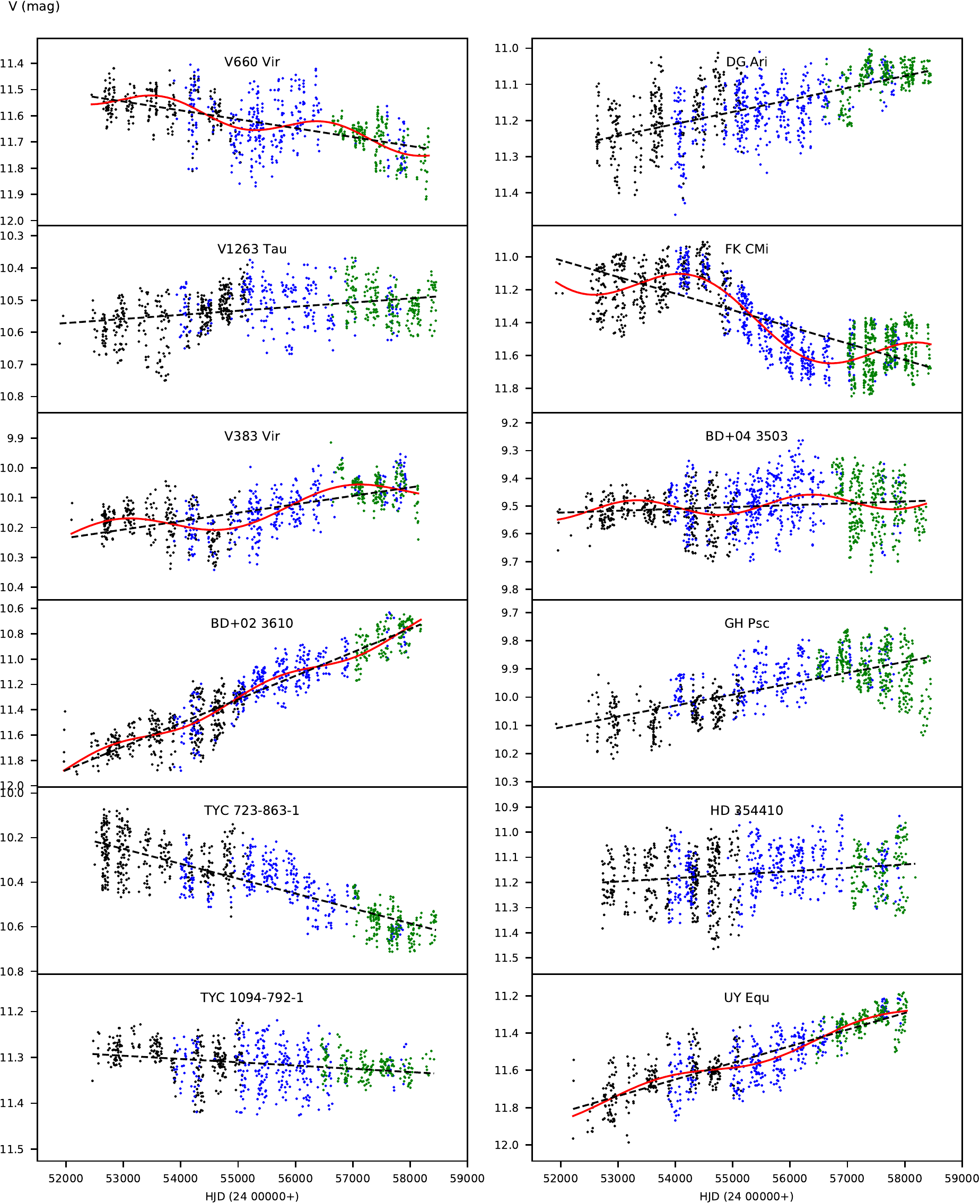}
	\caption{Compiled long-term $V$-band photometric data of target stars. Black, blue and green points denote
	ASAS, ASAS-3N and ASAS-SN data respectively. For each star, dashed line shows linear fit to the data. If the 
	target possesses a significant periodic signal, then the signal is shown with continuous (red) curve, which is 
	the combination of periodic signal(s) and linear fit.}
	\label{figure_long_term_photometry}
\end{figure*}

\addtocounter{figure}{-1}
\begin{figure*}
	\includegraphics[width=\textwidth,height=17cm]{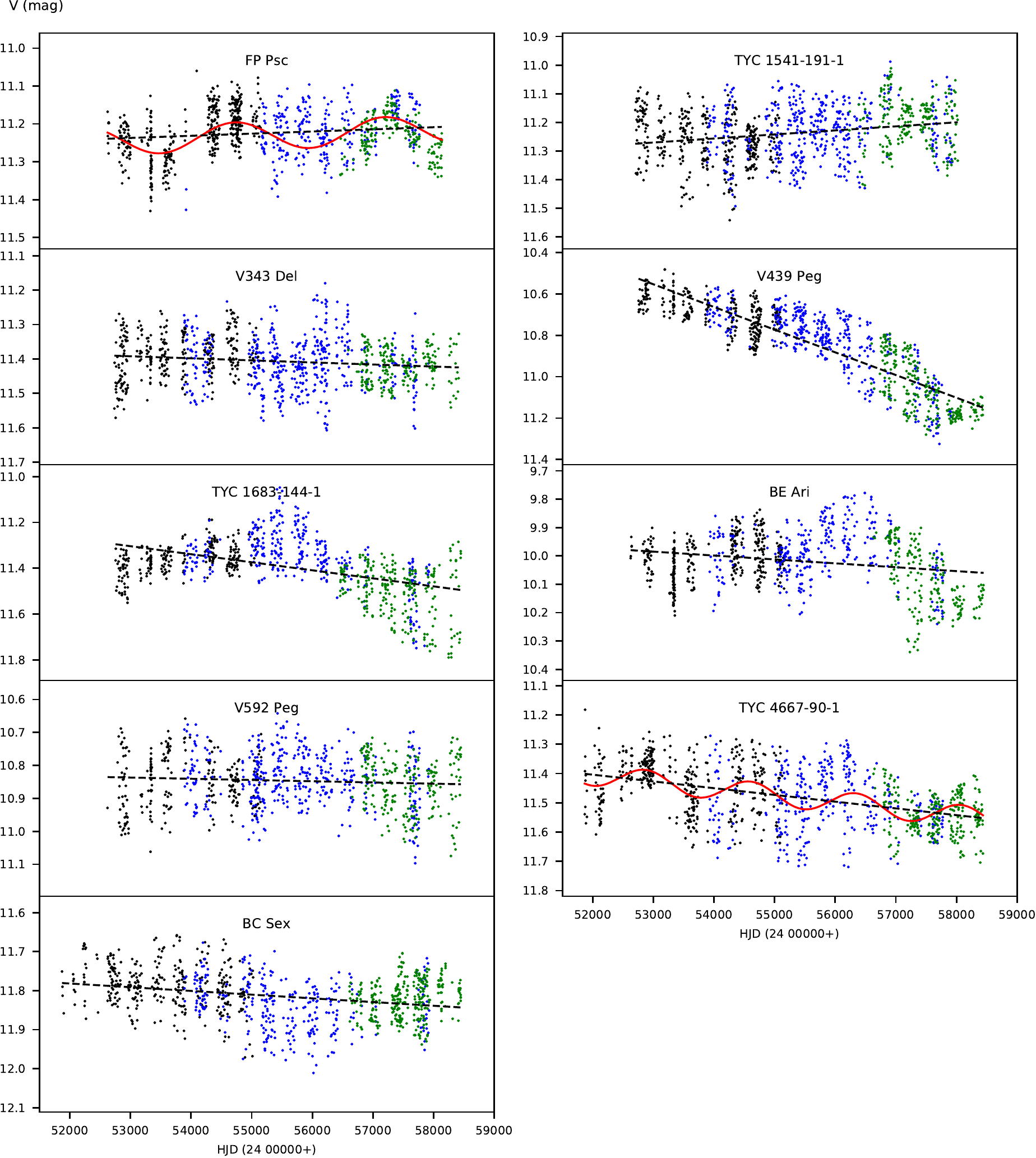}
	\caption{Continued.}
\end{figure*}

\begin{table}
	\centering
	\caption{Cycle period $P$ and amplitude $A$	found in lomb-scargle periodogram. 
	In the last column, $\Delta\,t$ is the time-span of the compiled photometric data.}
	\setlength{\tabcolsep}{0.1cm}
	\label{table_cycles}
	\begin{tabular}{lccc}
		\hline
Target	&	$P$			&	$A$    &	   $\Delta\,t$\\
        &    (year)			&	(mmag)     &	      (year)  \\
\hline\noalign{\smallskip}
V660\,Vir      &   7.9$\pm$0.2 &  38$\pm$4  &  16   \\
FK\,CMi        &  11.3$\pm$0.1 &  152$\pm$5 &  18   \\
V383\,Vir      &  11.0$\pm$0.3 &  43$\pm$3  &  16.5 \\
BD+04\,3503    &   8.3$\pm$0.3 &  30$\pm$3  &  17.6 \\
BD+02\,3610    &   8.0$\pm$0.3 &  48$\pm$6  &  17   \\
UY\,Equ        &  10.0$\pm$0.4 &  34$\pm$4  &  16   \\
FP\,Psc        &   6.8$\pm$0.1 &  36$\pm$2  &  15   \\
TYC\,4667-90-1 &   4.8$\pm$0.8 &  37$\pm$4  &  18   \\		
\hline
	\end{tabular}
\end{table}

\subsection{Analysis of seasonal light curves}\label{subsec_seasonal_lc}
After analysing global brightness variation, we focus on analysis of seasonal light 
curves of each target. For each seasonal light curve, we try to find photometric 
period and peak-to-peak light curve amplitude. Inspecting seasonal light curves, we see 
that light curve shapes are asymmetric or double-humped, which is a typical light curve 
feature of a chromospherically active star. 

Such light curves can be fitted satisfactorily by combining a number of signals with 
different periods, amplitudes and phases. One of these signals is the main signal and the 
remaining signals have periods which are harmonics of the period of the main signal. 
In this case, one may find many periods, which are statistically significant, for a given 
single light curve. Our aim is neither make a perfect trigonometric fit to a light curve 
nor to study harmonics of the main period, but to find a unique period, which represents 
the light curve as good as possible. Therefore, we adopt ANOVA method \citep{ANOVA_Czerny_1996ApJ} 
which combines strengths of Fourier method and ANOVA statistics. ANOVA method is capable 
to detect a unique period for an asymmetric or double-humped light curve by evaluating 
scatter of the phase-folded observational data with respect to a set of trial periods. 
This method is also very efficient in peak detection and damping alias periods.

\begin{figure*}
	\includegraphics[scale=0.92]{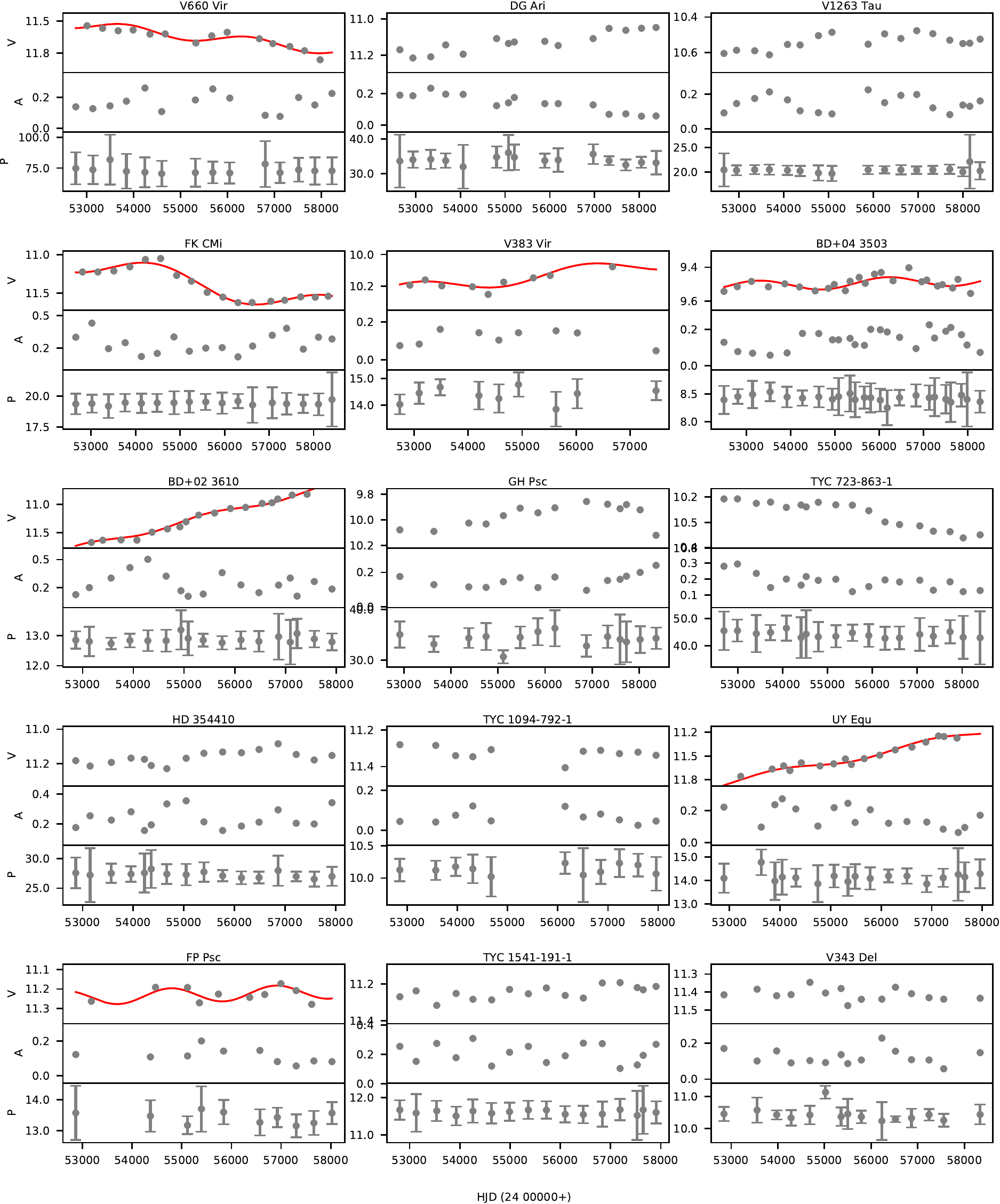}
	\caption{Seasonal mean brightness (V, mag), peak-to-peak light curve amplitude (A, mag) and 
	photometric period (P, day) of each target star are plotted versus time. Detected 
	long-period cycles are overplotted with red continuous line.}
	\label{seasonal_plots}
\end{figure*}

\addtocounter{figure}{-1}
\begin{figure*}
	\includegraphics[scale=0.92]{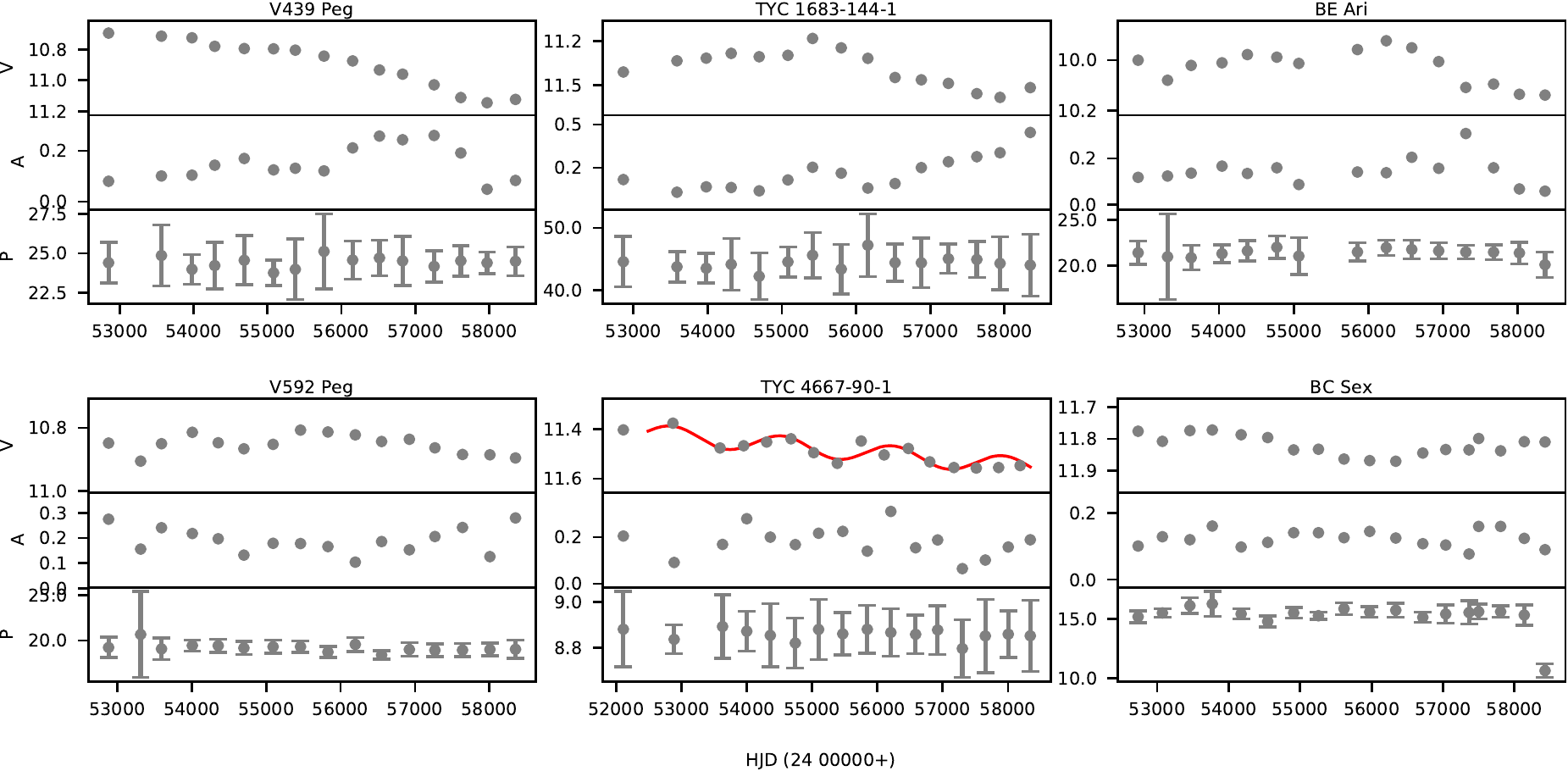}
	\caption{Continued.}
	\label{seasonal_plots2}
\end{figure*}

We carry out analysis of each target by using the residuals obtained from the linear fit 
in Section~\ref{subsec_mean_bright}. The purpose of using the residuals is to reduce 
unwanted effects of any global (linear) brightening or dimming trend on period and 
peak-to-peak amplitude detection from seasonal light curves. For each season of each 
target, we determine start, end and mean HJD values together with photometric period and 
its statistical uncertainty, maximum, minimum and mean brightnesses, peak-to-peak light 
curve amplitude and number of data points in the season. We tabulate numerical values of 
these parameters in Table~\ref{table_seasonal_phot_analysis}. We adopt the method proposed
by \citet{ANOVA_error_czerny1991MNRAS.253..198S} to estimate uncertainty of the seasonal
photometric periods. We also compute rotation period of each target by averaging 
corresponding seasonal photometric periods tabulated in Table~\ref{table_seasonal_phot_analysis}.
Standard deviation of seasonal photometric periods of each target is adopted as the
uncertainty of computed rotation period. Computed rotation periods and uncertainties are
listed at the end of Table~\ref{table_seasonal_phot_analysis}. Comparing computed rotation
periods with the previously reported periods given in Table~\ref{table_basic_data}, we see
that the computed periods in this study agrees with the previously reported periods in 1$\sigma$
error bar, except FK\,CMi and FP\,Psc. However, for these two stars, difference between previously 
reported periods and computed periods in this study slightly exceed 1$\sigma$ error bar.

In the context of solar-stellar connection, one may expect to observe relations between 
mean brightness, peak-to-peak light curve amplitude and photometric period. In the Sun, 
when a new sunspot cycle begins, small sunspots appear at higher latitudes which rotate 
slower than the solar equator, hence these sunspots have longer rotation periods and 
causes low amplitude light loss due to their small size. At the middle of the cycle, 
sunspots mostly emerge at mid-latitudes with a larger areas and shorter rotation periods 
compared to the beginning of the cycle. At the end of the cycle, sunspots emerge at lower 
latitudes with smaller areas and much shorter rotation periods. In summary, photometric 
period continuously decreases through the cycle while light loss amplitude and mean 
brightness gradually increase in the first half of the cycle, and then gradually decrease 
towards the end of the cycle\footnote{Note that the Sun appears brighter at the middle of 
a sunspot cycle due to the more dominant facular contribution to the brightness, compared 
to the spot contribution.}. At that point, we emphasize that the scenario outlined above 
does not strictly work through the cycle since sunspots could also be emerged at high or 
low latitudes in the middle of a sunspot cycle. However, these cases do not change the 
global trends observed in photometric properties.

In order to evaluate each star in the context described above, we present seasonal 
behaviours of the mean $V$ brightness, peak-to-peak light curve amplitude ($A$) and 
photometric period ($P$) of each target star in Fig.~\ref{seasonal_plots}. Although 
photometric data spans over 14-18 years, one may notice some missing seasons in the figure. 
These missing seasons are due to the sparse data or very low light curve amplitude, which 
prevent us from finding reliable photometric period and light curve amplitude. Still, we 
inspect Fig.~\ref{seasonal_plots}, especially for stars exhibiting photometric cycles. For a 
given target with a cyclic mean brightness variation, general distribution of seasonal photometric 
periods does not appear coherent with the corresponding cycle and exhibits scattered pattern 
rather than a systematic decrease trend that repeats itself in each cycle as observed in the 
sunspot cycle. Regarding light curve amplitude and mean brightness, it is noticeable that 
the mean brightness tends to increase as the peak-to-peak light curve amplitude decreases.

\section{Activity cycle length, rotation period and relative surface shear}\label{sec_cycle_rot_dif_rot}
\begin{figure}
	\includegraphics[scale=0.55]{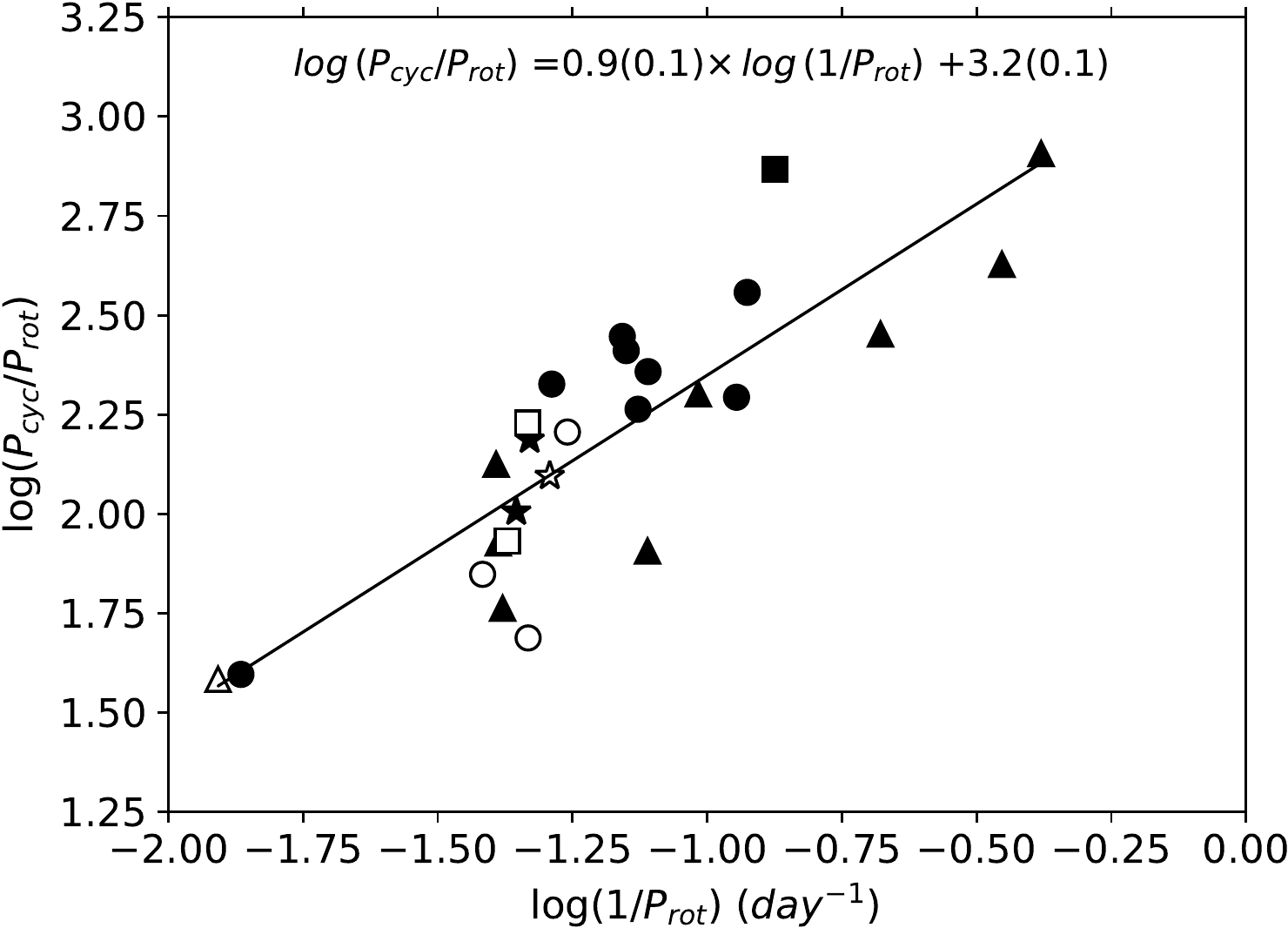}
	\caption{Relation between $log(1/P_{rot})$ and $log(P_{cyc}/P_{rot})$ for giant and sub-giant 
	stars. Filled circles are stars listed in Table~\ref{table_cycles}, open circles show BD+13\,5000, 
	TYC\,5163-1764-1 and BD+11\,3024 \citep{Ozdarcan_Dal_2018AN....339..277O}, filled stars denote 
	HD\,208472 \citep{Ozdarcan_V2075Cyg_2010AN} and HD\,89546 \citep{Ozdarcan_FGUMa_2012AN}, filled 
	triangles are for IM\,Peg, HK\,Lac, XX\,Tri, IL\,Hya, HU\,Vir, UZ\,Lib, V711\,Tau and FK\,Com 
	\citep{Olah_cycles_2009A&A}, open squares show IS\,Vir and V2253\,Oph \citep{Olah_2013AN....334..625O}
	open triangle is BM\,Cam \citep{Zboril_BMCam_2009AN....330..377Z}, filled square is DM\,UMa 
	\citep{Tas_Evren_DM_UMa_2012BaltA..21..435T} and open star is $\sigma$\,Gem 
	\citep{Kajatkari_sigma_Gem_2014A&A...562A.107K}. Equation of the linear fit is indicated in 
	the insert where numbers in parentheses show statistical errors for the last two digits.}
	\label{figure_pcyc_porb}
\end{figure}

We plot target stars listed in Table~\ref{table_cycles} on $log(1/P_{rot})-log(P_{cyc}/P_{rot})$ 
plane in Fig.~\ref{figure_pcyc_porb}, where $P_{rot}$ is the rotation period in day. We also overplot 
same values of previously published giant or sub-giant stars in the figure. We observe a clear linear 
trend as reported in previous studies \citep{Baliunas_pcycp_prot_1996ApJ...460..848B, Olah_strassmeier_cycles_2002AN,
Olah_cycles_2009A&A,mascareno_et_al_2016A&A...595A..12S}, however, we provide quantitative 
representation of the trend by only using giant and sub-giant stars. Linear representation of 
the data leads to the relation inserted in the figure with a correlation coefficient of 0.85
and probability of no correlation $p=3.5\times10^{-8}$.


\begin{figure}
	\includegraphics[scale=0.55]{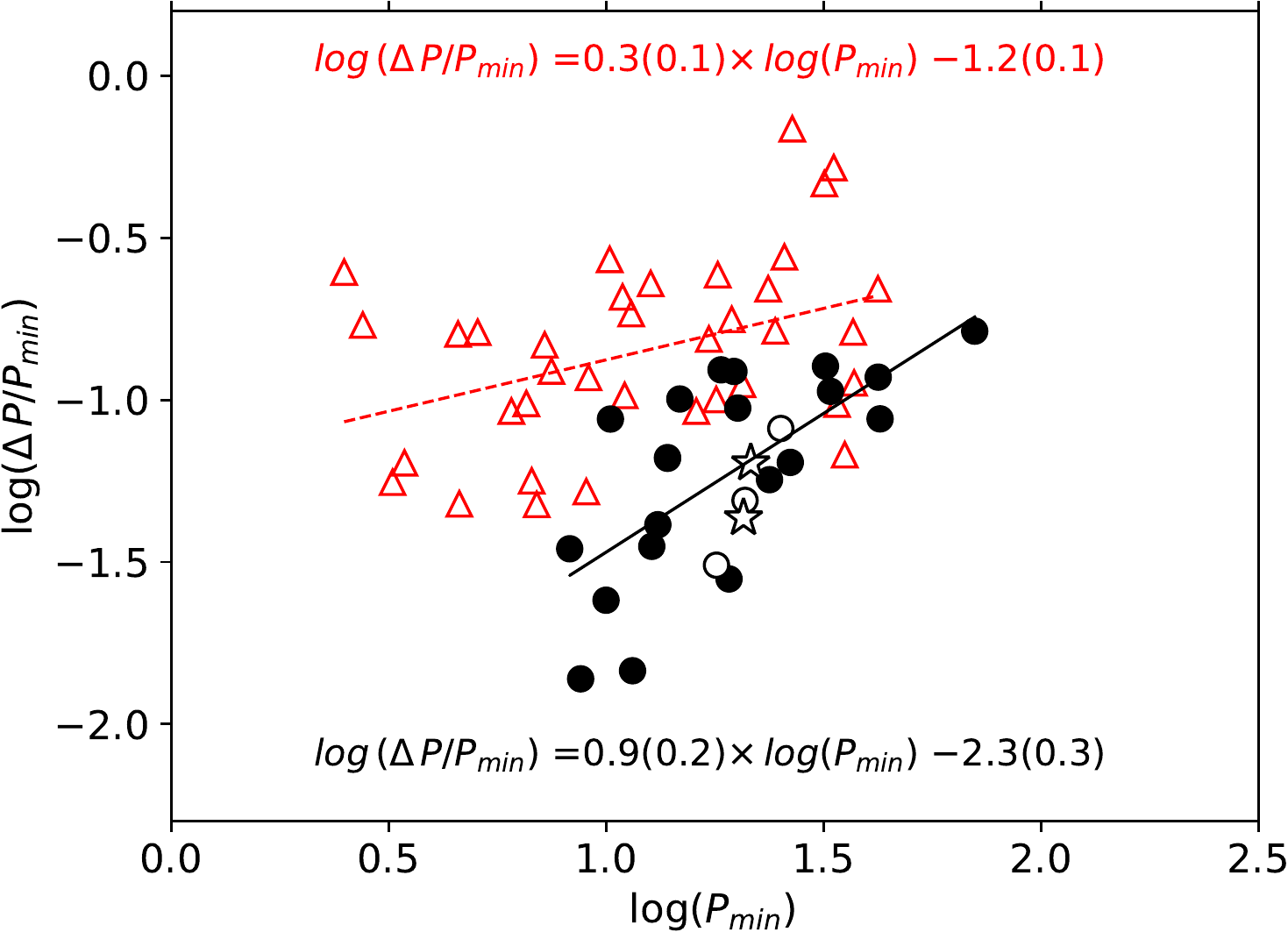}
	\caption{Relation between the minimum rotation period $P_{min}$ and the relative surface 
	shear $\Delta\,P/P_{min}$ on the logarithmic scale. Filled circles are stars listed in 
	Table~\ref{table_cycles}, open circles show BD+13\,5000, TYC\,5163-1764-1 and BD+11\,3024 
	\citep{Ozdarcan_Dal_2018AN....339..277O}, open stars denote HD\,208472 
	\citep{Ozdarcan_V2075Cyg_2010AN} and HD\,89546 \citep{Ozdarcan_FGUMa_2012AN}. Equation 
	of the linear fit (black continuous line) is indicated in the insert (bottom). Open red 
	triangles show stars taken from \citet{donahue_saar_baliunas_1996ApJ...466..384D} and the 
	linear representation of their distribution is shown by red dashed line and given in the insert (top).}
	\label{figure_dif_rot}
\end{figure}

Using maximum and minimum photometric periods of each target tabulated in 
Table~\ref{table_seasonal_phot_analysis}, we compute relative surface shear of each target in 
terms of $\Delta\,P/P_{min}$. Here $\Delta\,P=P_{max}-P_{min}$, where $P_{max}$ and $P_{min}$ 
denote maximum and minimum period, respectively. We plot logarithm of the relative surface shear 
$log(\Delta\,P/P_{min})$ versus $log(P_{min})$ in Fig.~\ref{figure_dif_rot} and we find another 
positive linear correlation with a correlation coefficient of 0.67 and probability of no 
correlation $p=1.9\times10^{-4}$. We also overplot the main 
sequence stars studied by \citet{donahue_saar_baliunas_1996ApJ...466..384D}, who obtained 
results with a method similar to one described in our study. Here, we note that we consider 
minimum periods tabulated in Table 1 of \citet{donahue_saar_baliunas_1996ApJ...466..384D} study 
and compute $\Delta\,P/P_{min}$ from these data, in order to evaluate both data sets properly. 
Linear regression to the main sequence star data in Fig.~\ref{figure_dif_rot} give a slope of
0.3, a correlation coefficient of 0.4 and probability of no correlation $p=0.014$.
One may easily observe the clear separation of two samples. For a given period, magnitude of 
the relative surface shear is clearly larger in main sequence stars than giant and sub-giant 
stars. The correlation denotes that the relative surface shear increases with increasing 
rotational period. Although both samples show positive linear correlations, giant and 
sub-giant sample exhibit linear trend with steeper slope and larger correlation coefficient 
than the main sequence sample.

\begin{figure}
	\includegraphics[scale=0.55]{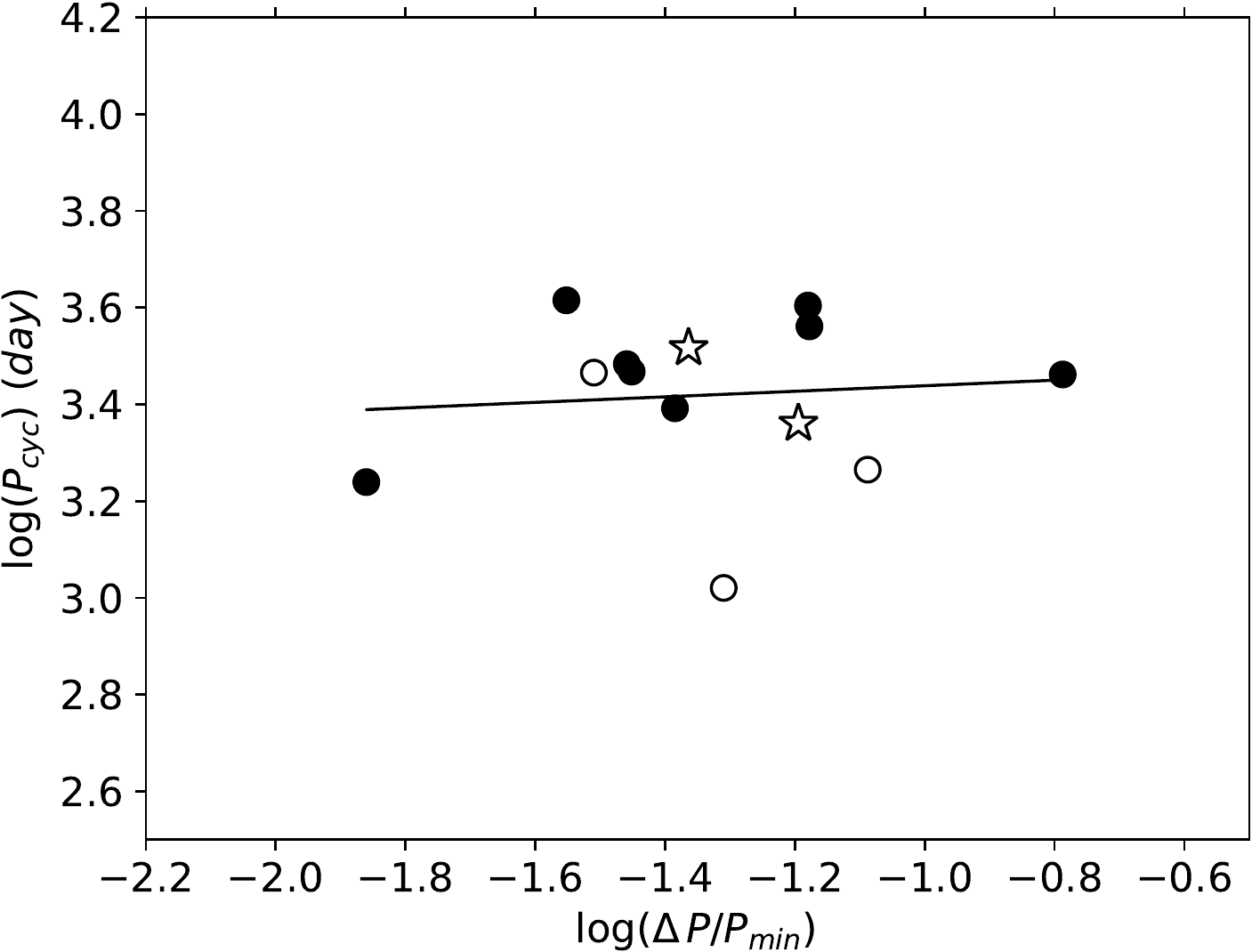}
	\caption{Relation between activity the cycle length and the relative surface shear on 
	the logarithmic scale. The symbols have the same meaning as in Fig.~\ref{figure_dif_rot}.}
	\label{figure_dif_rot_cyc}
\end{figure}

We also inspect possible relation between the activity cycle length and the relative surface 
shear in Fig.~\ref{figure_dif_rot_cyc}. Distribution of the data practically indicates no 
correlation between the cycle length and the relative surface shear.

\section{Summary and discussion}
We present analysis of medium resolution optical spectra and long-term V-band photometry of 21 
cool stars. All stars exhibit direct emission features in Ca\,{\sc ii} H\& K line profiles, while 
we observe both filled emission and direct emission in H$_{\alpha}$ lines. Emission features in 
these spectral lines are clear evidences of chromospheric activity on the target stars. LTE spectrum 
synthesis shows that the target stars are cool red giant or sub-giant stars with an overall 
metallicity, [M/H], between $0$ and $-0.5$.

Analysis of long-term $V$-band photometry of target stars reveals that DG\,Ari, V1263\,Tau, HD\,354410, 
TYC\,1541-191-1, V343\,Del and V592\,Peg undergo linear mean brightness variation, which may indicate 
activity cycles in very long time scales. We detect single significant period for 8 of the target stars, 
V660\,Vir, FK\,CMi, V383\,Vir, BD+04\,3503, BD+02\,3610, UY\,Equ, FP\,Psc and TYC\,4667-90-1. For each 
target, resulting cycle signal is superimposed on a linear brightening or dimming trend. Linear trend
might show a part of a possible second cycle with a much longer time scale compared to the current 
data set. In other words, these stars might possess multiple cycles. Remaining target stars appear to 
possess photometric cycles but time base of the current data is not sufficient to confirm these cycles.
Possible long term cycles can be checked only with further $V$-band observations. Here, we also stress that 
the cycle periods found in our analysis may only be average values. In case of the sunspot cycle, one 
can only give an average 11 years of period, which actually varies between 9 and 14 years. Therefore, 
in the context of solar-stellar connection, it is not reasonable to accept strict period values in case 
of stellar magnetic activity cycles. \citet{Olah_kollath_strassmeier2000A&A} and \citet{Olah_cycles_2009A&A} 
clearly showed that photometric cycle periods of cool active stars are variable.

Further analysis of long-term photometric data in terms of seasonal light curve analysis yields
photometric period, peak-to-peak light curve amplitude, minimum, maximum and mean brightnesses of
each observing season. For a given target, we compute average value of the seasonal photometric 
periods and adopt it as the photometric rotation period, $P_{rot}$, of the target. Then, we inspect 
the relation between $log(1/P_{rot})$ and $log(P_{cyc}/P_{rot})$ of giant and sub-giant stars. 
We observe clear linear trend on the logarithmic scale with a slope of $0.9\pm0.1$, which is 
consistent with that derived by the slope reported by \citet[][$0.81\pm0.05$]{Olah_cycles_2009A&A} 
within 1$\sigma$ level, but is considerably higher than the 
slope reported by \citet[][0.74]{Baliunas_pcycp_prot_1996ApJ...460..848B}, who originally tested the 
equivalency of the stellar dynamo number ($D\sim1/P_{rot}$) and $P_{cyc}/P_{rot}$ ratio. We note that 
our sample includes only giant and sub-giant stars while \citet{Olah_cycles_2009A&A} used a homogeneous 
sample, including main sequence and giant stars. If we do not consider the giant stars taken from their 
study, the slope of the linear fit in Fig.~\ref{figure_pcyc_porb} increases to $1.0\pm0.2$. In any case, 
these slopes are considerably smaller than the one found in theoretical computations of 
\citet[][1.47]{dube_charbonneau_2013ApJ...775...69D}, who mentioned that the observed linear relation 
is the robust property of kinematic mean-field $\alpha\Omega$ dynamo models with $\alpha$ quenching. 
Our results support this picture since the activity cycle length is weakly anti-correlated with the 
relative surface shear, indicating that the activity cycle should be mainly driven by turbulent 
diffusivity, rather than the differential rotation as anticipated in non-linearly saturated dynamo 
models. However, although activity cycle length does not appear sensitive to the rotation period 
for our sample (see Fig.~\ref{figure_pcyc_porb}), longer period stars still tends to have longer period 
activity cycles, which contradicts with the prediction given by \citet{dube_charbonneau_2013ApJ...775...69D}.

Assuming that the $P_{min}$ corresponds to the equatorial rotation period, we find that the relative 
surface shear, $\Delta\,P/P_{min}$, is positively correlated with the equatorial rotation period, 
which means differential rotation tends to be stronger for slow rotators. \citet{reinhold_2015A&A...583A..65R}
made photometric analysis of 12\,319 Kepler stars, which are homogeneously distributed in temperature 
and luminosity class, and found that the positive correlation between $P_{min}$ and the differential 
rotation exists for stars cooler than 6\,000 K. At that point, our study provides quantitative 
description of the relation for cool giants and sub-giant. Moreover, overplotting main sequence stars 
from the study of \citet{donahue_saar_baliunas_1996ApJ...466..384D} in Fig.~\ref{figure_dif_rot}, we 
observe clear separation between two samples and more steep slope for giant and sub-giant stars. 
The separation becomes more remarkable towards the shorter periods. Considering evolutionary 
properties of two samples, we may conclude that the relative surface shear is more sensitive to the 
rotation period in giant and sub-giant stars compared to the main sequence stars. However, magnitude of 
the relative surface shear for a given period is larger for main sequence stars than giant and sub-giant 
stars. This observational finding appear contradictory to the theoretical model predictions given in 
\citet{kitchatinov_rudiger_1999A&A...344..911K}. On the other hand, \citet{kovari_et_Al_2017AN....338..903K}
studied the rotation and the differential rotation relationship via a homogeneous sample in luminosity 
class whose differential rotation parameter were obtained from Doppler imaging studies. They find a 
correlation similar to one shown in Fig.~\ref{figure_dif_rot}, but also noticed a distinction between 
single and binary stars, where the relative surface shear in single stars shows stronger dependency on 
the rotation period compared to the close binaries. It could be speculated that the separation observed 
in Fig.~\ref{figure_dif_rot} might be partly caused by being single or binary star. Nevertheless, this 
needs to be checked by high resolution spectroscopy for all stars plotted in the figure.

Further analyses of individual targets indicate that temporal behaviour of the photometric period 
does not appear analogous to solar cycle and we do not observe any significant relation between 
photometric period, mean brightness and light curve amplitude. We can only say that the mean brightness 
generally tends to increase as the peak-to-peak light curve amplitude decreases. One may conclude that 
photometric cycle properties of target stars are not similar to the photometric properties of the sunspot 
cycle. However, as mentioned in \citet{FG_IS_Fekel_et_al_2002AJ}, the observed photometric period could be 
affected by the growth and decay of spots or spot groups at various latitudes and longitudes, which means 
the observed photometric may not be adopted as a true period of any stellar latitude. Therefore, observed 
temporal distributions of photometric periods in Fig.~\ref{seasonal_plots} might partly originate from 
spot growth and decay. With the current data, it is not possible to arrive at a conclusive result. Further
regular $V$ band photometric observations should be continued for successive cycles, so that global 
photometric behaviour could show itself more clearly in longer time scales. That was the case for HD\,89546, 
where 11 years of observations did not yield any significant correlation between the photometric period and 
the mean brightness \citep{FG_IS_Fekel_et_al_2002AJ}, but doubling the time base of the photometric data 
yielded the correlation \citep{Ozdarcan_FGUMa_2012AN}. Another possibility is that the distribution of the 
photometric period might be related to a possible cycle which has a time scale longer than the current time 
base of the photometric data. Observed linear trends in the mean brightnesses of our target stars might indicate 
such longer-term cycles. Consequently, we may observe only a part of the whole photometric period variation 
pattern, which appears as a scatter. 

\begin{acknowledgements}
I am indebted to Dr. Grzegorz Pojma{\'n}ski for reducing and providing unpublished ASAS-3N data, which
compensate long data gap between published ASAS3 and ASAS-SN data and increase the precision of the main results
of this study. I also express my thanks to Dr. Hasan Ali Dal for fruitful discussion on statistical analyses.
I acknowledge the anonymous referee for his/her helpful comments and valuable suggestions that have significantly 
contributed to improve the paper. I further thank T\"UB\.ITAK National Observatory for a partial support in using 
RTT150 (Russian-Turkish 1.5-m telescope in Antalya) with project numbers 14BRTT150-678 and 18BRTT150-1275. 
This research has made use of the SIMBAD database, operated at CDS, Strasbourg, France. This research received 
no specific grant from any funding agency, commercial, or not-for-profit sectors.
\end{acknowledgements}

\begin{appendix}

\section{Observed and the best-fitting synthetic spectra.}
We present plots of observed and the best-fitting synthetic 
spectrum for each target star separately in Fig.~\ref{figure_spectrum_fit}.

\begin{figure*}
	%
	%
	%
	%
	\includegraphics[scale=0.47]{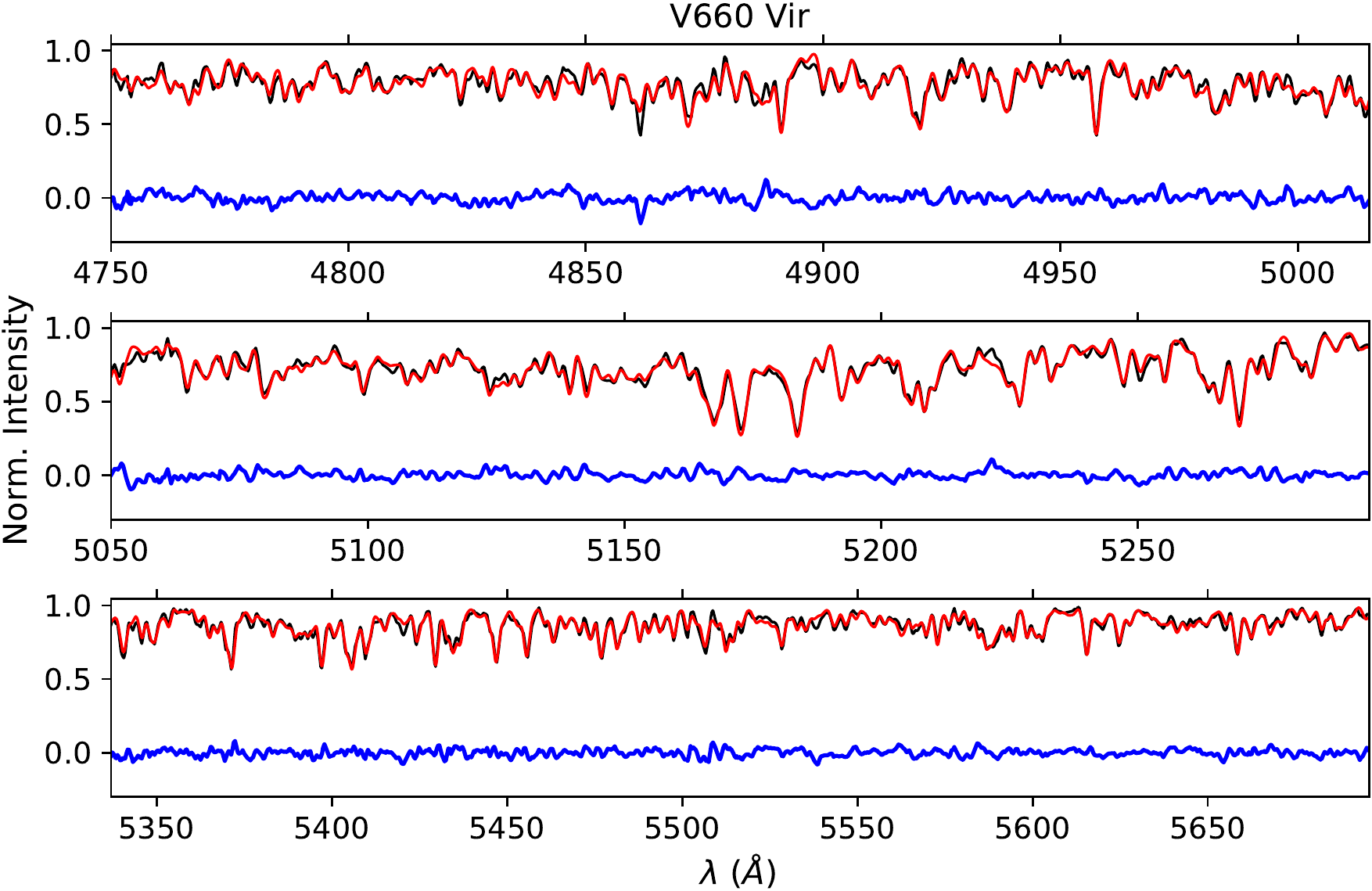}\hspace{0.3cm}
	\includegraphics[scale=0.47]{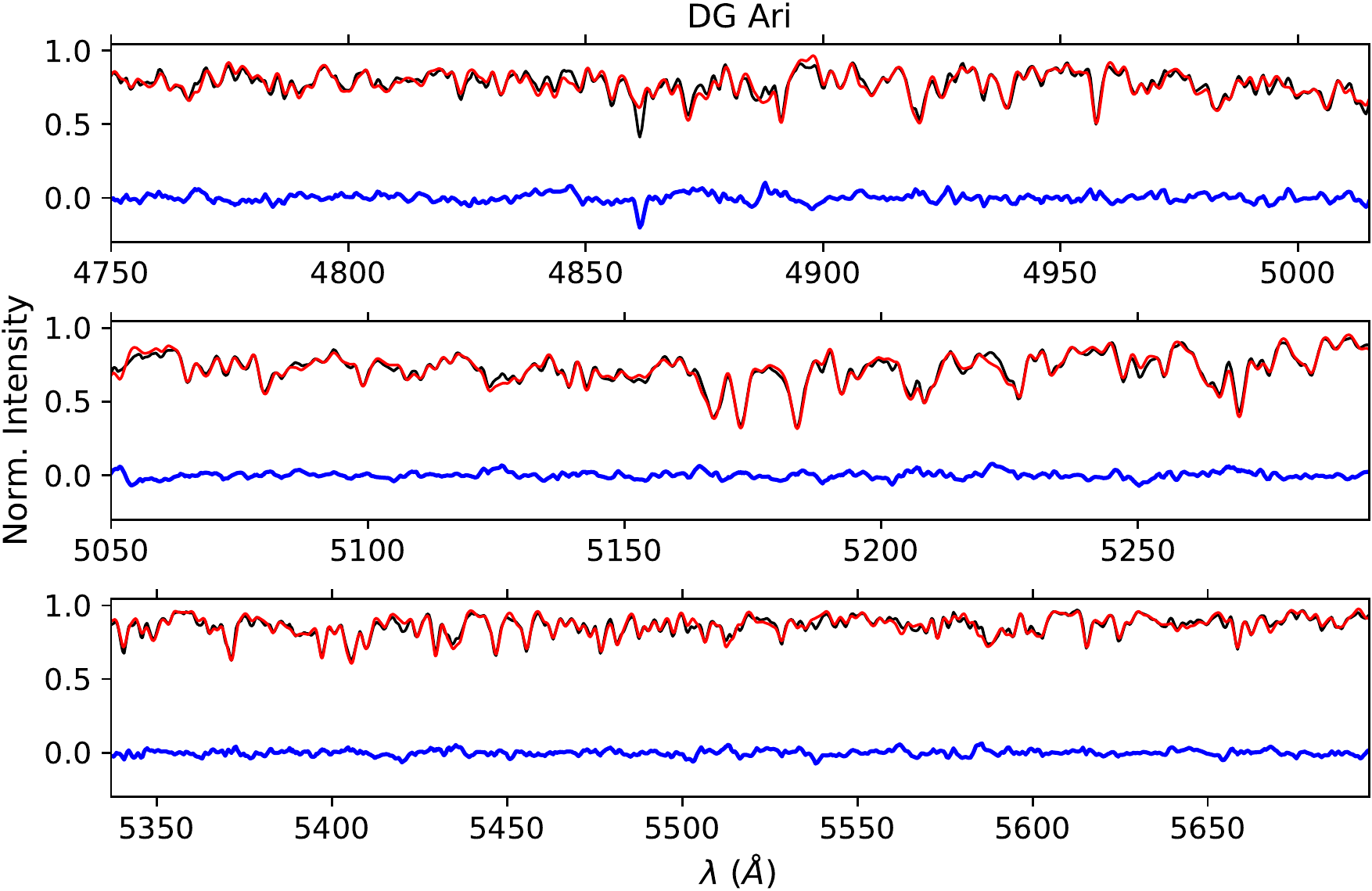}\vspace{0.3cm}
	\includegraphics[scale=0.47]{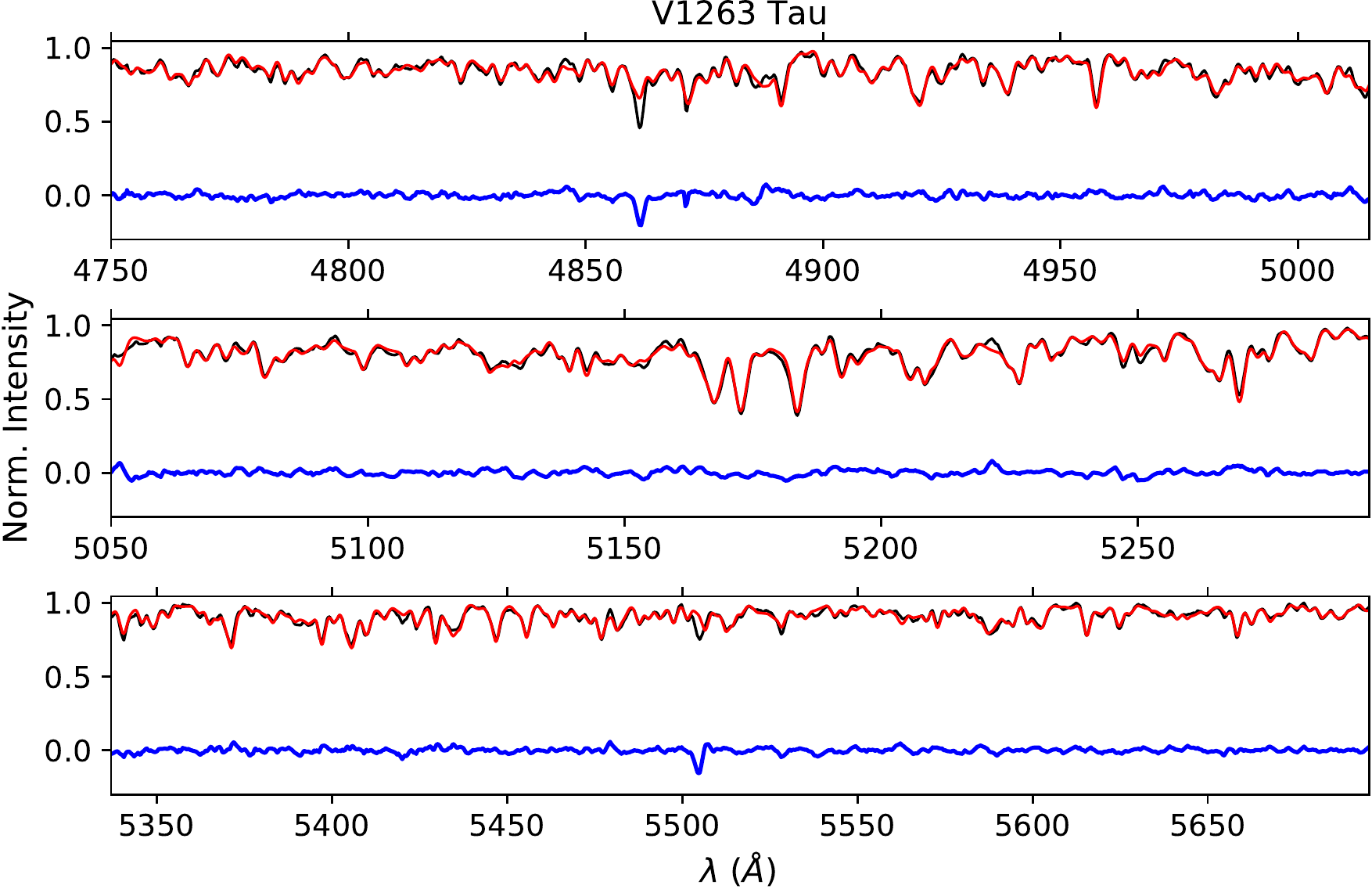}\hspace{0.3cm}
	\includegraphics[scale=0.47]{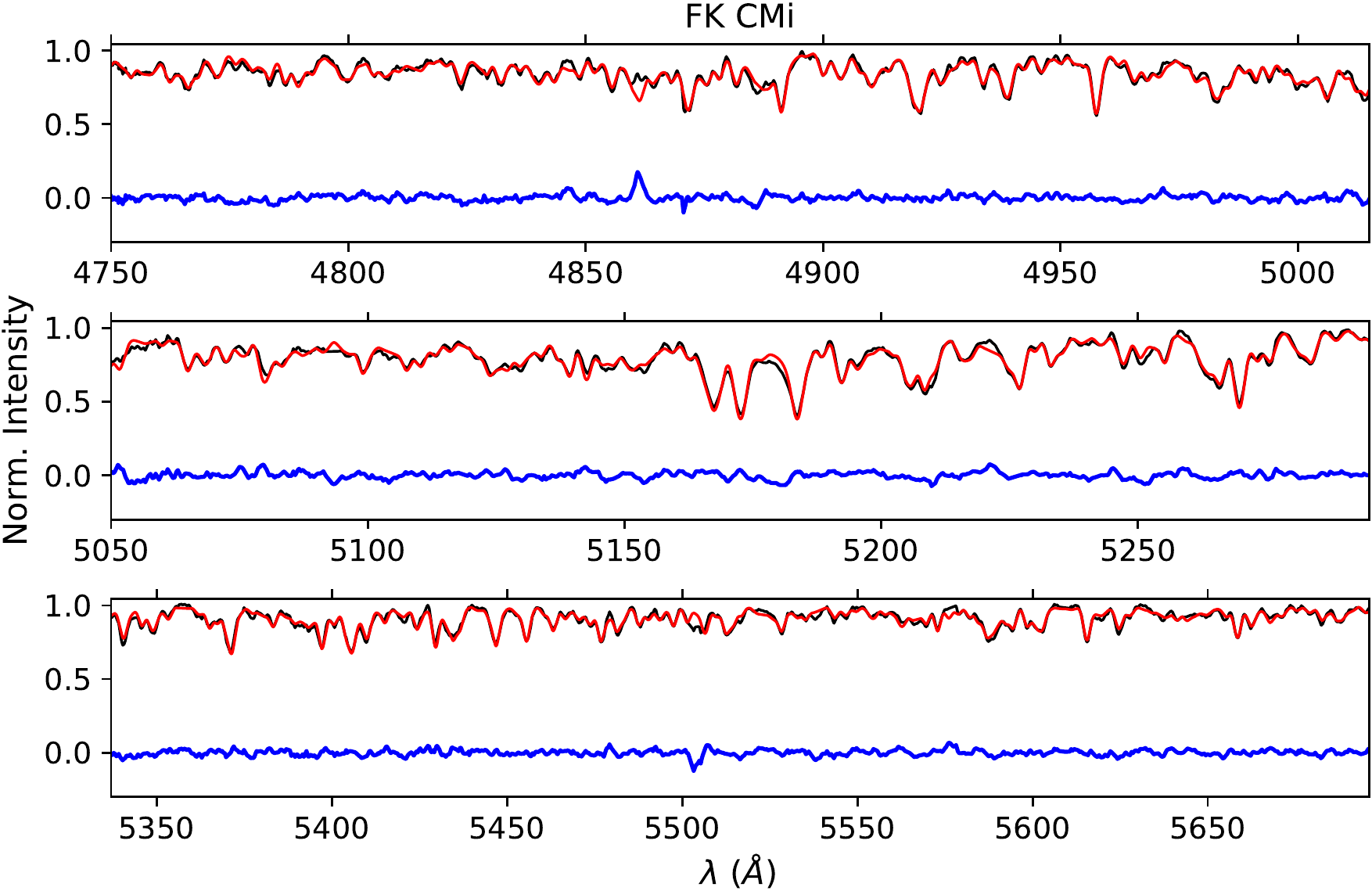}\vspace{0.3cm}
	\includegraphics[scale=0.47]{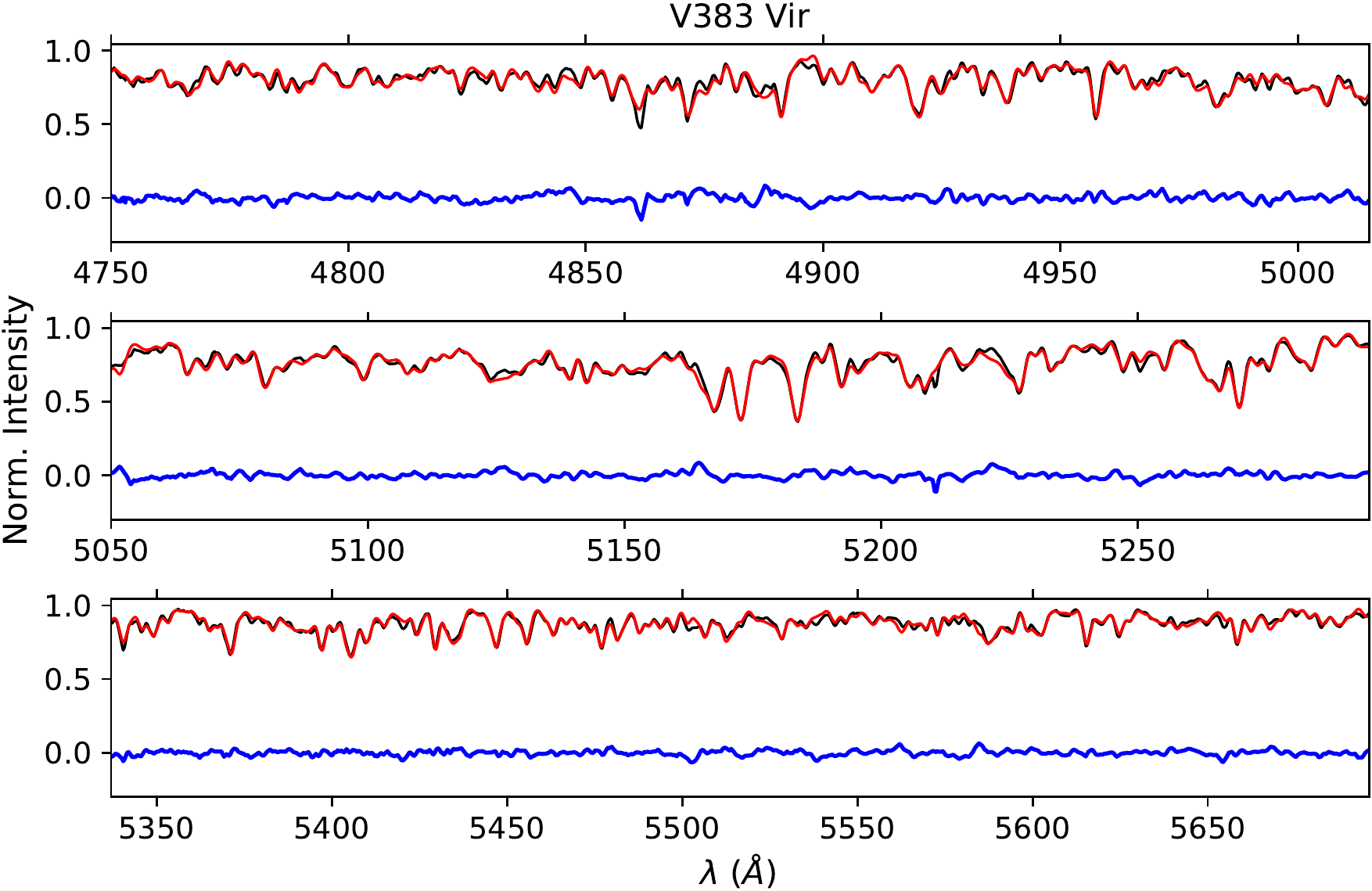}\hspace{0.3cm}
	\includegraphics[scale=0.47]{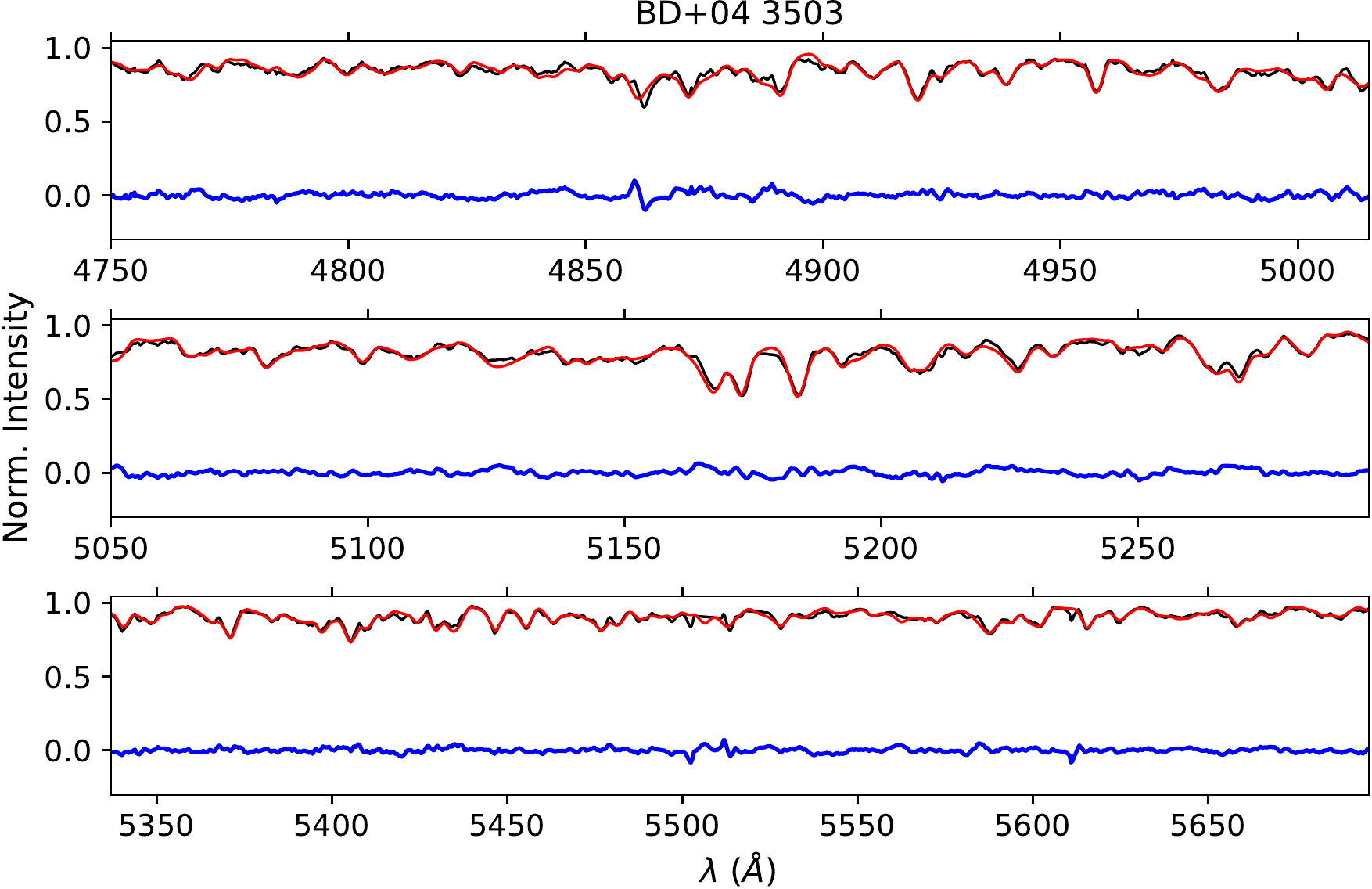}\vspace{0.3cm}
	\includegraphics[scale=0.47]{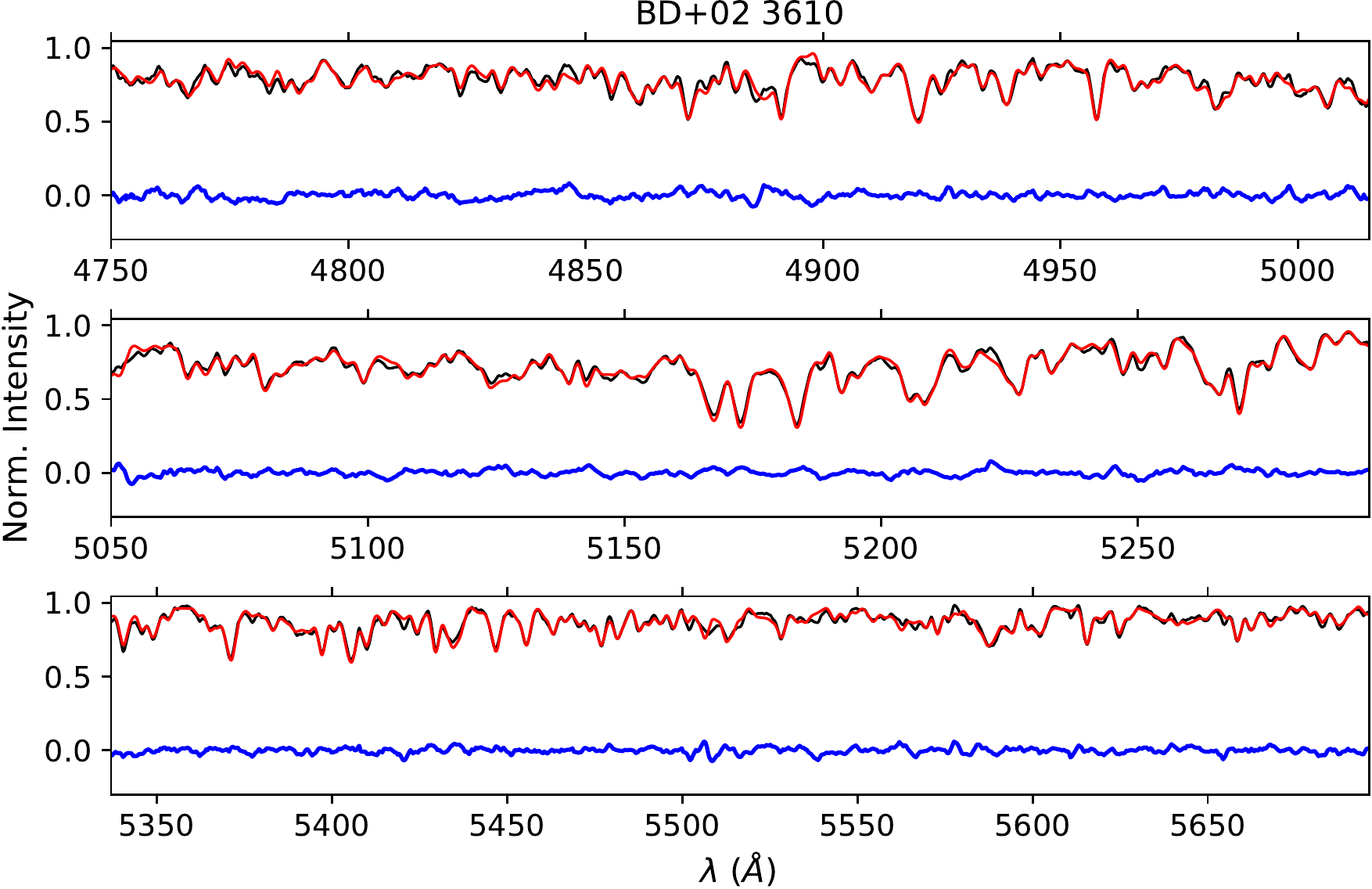}\hspace{0.3cm}
	\includegraphics[scale=0.47]{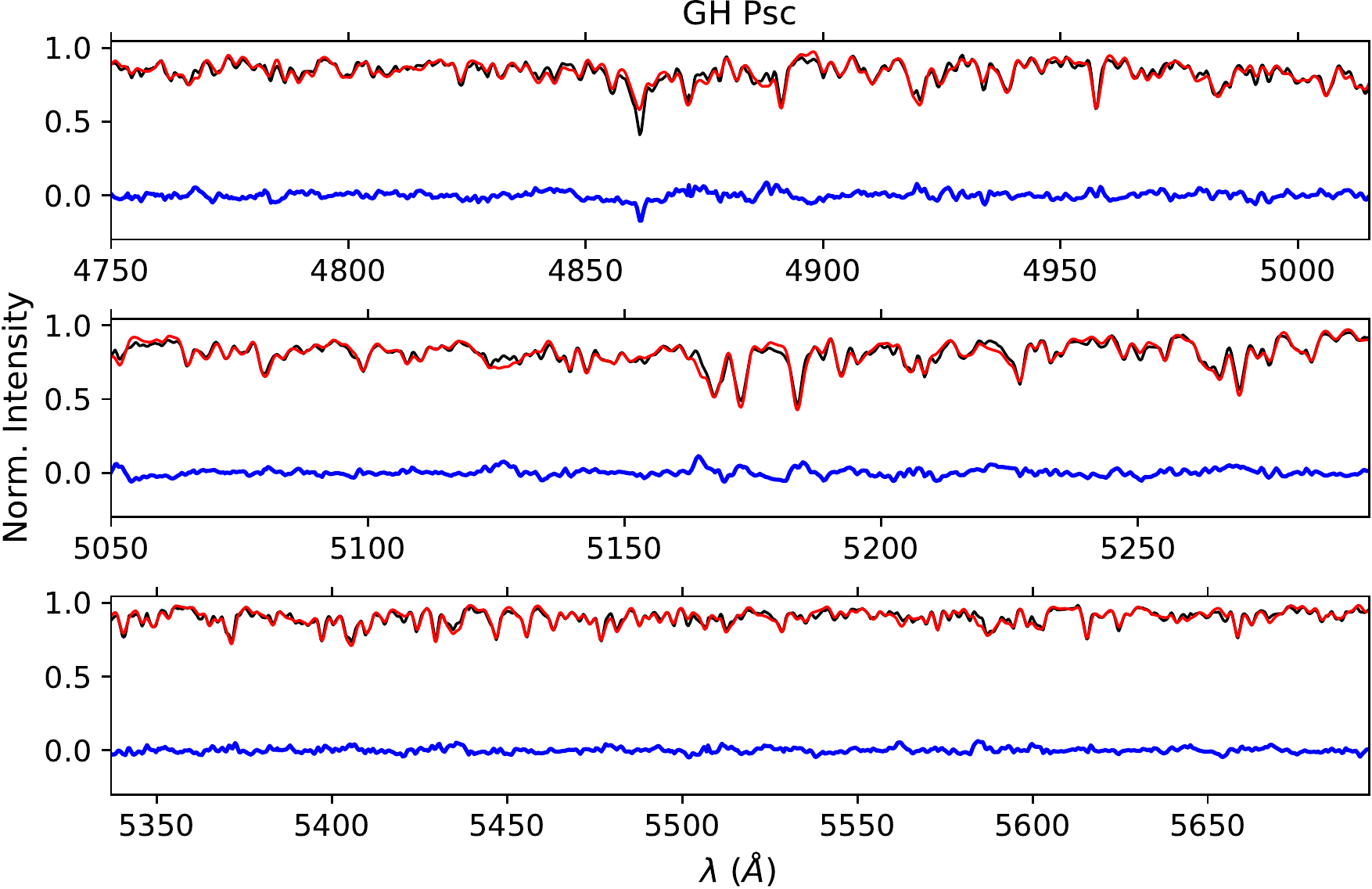}
	\caption{Representative plots of atmospheric analysis results for four stars. Observed (black) 
	and best fitting (red) synthetic spectra together with residuals (blue) are shown for three 
	regions in three windows one under the other.}
	\label{figure_spectrum_fit}
\end{figure*}

\addtocounter{figure}{-1}
\begin{figure*}
	%
	%
	%
	%
	\includegraphics[scale=0.47]{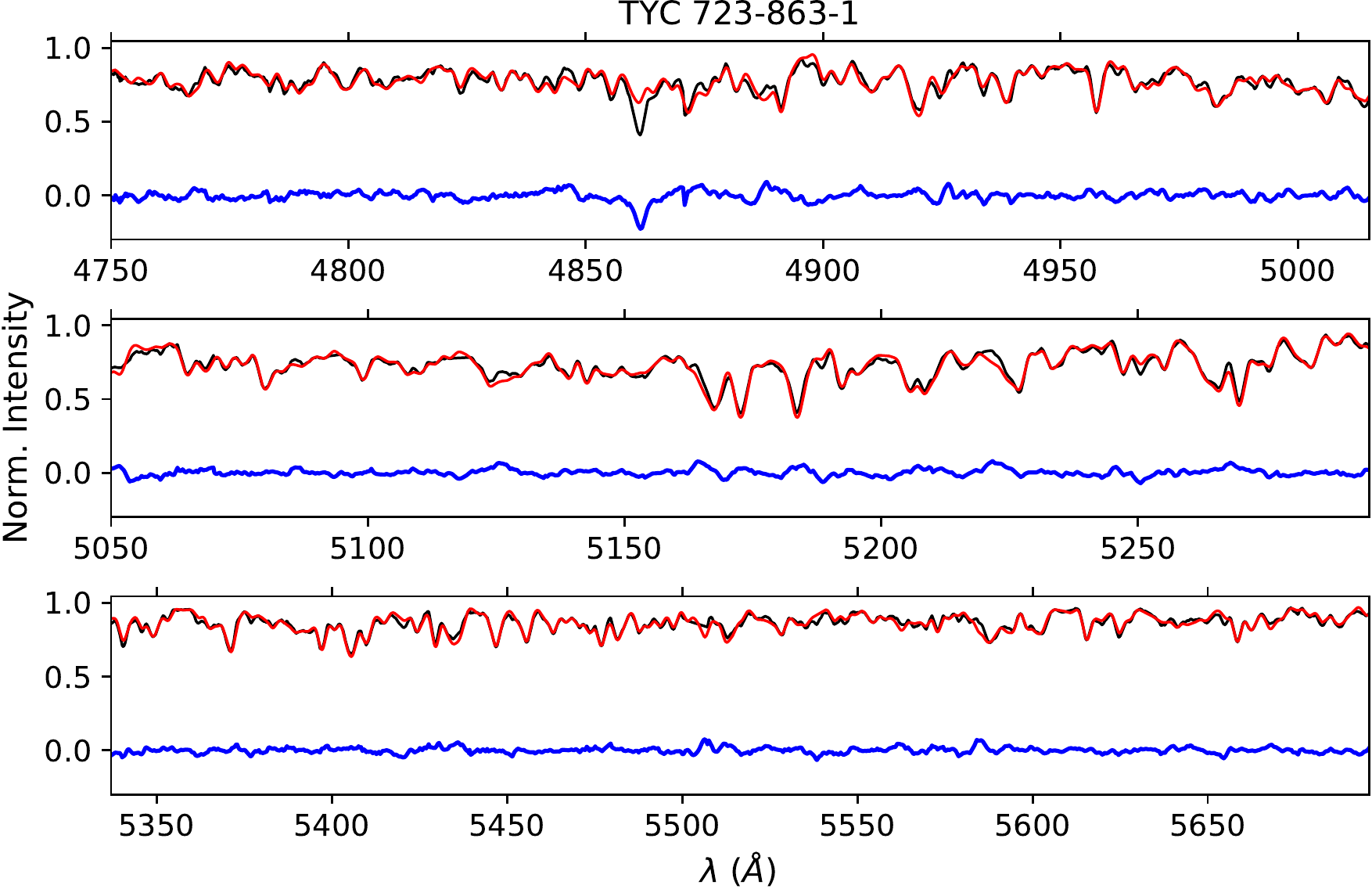}\hspace{0.3cm}
	\includegraphics[scale=0.47]{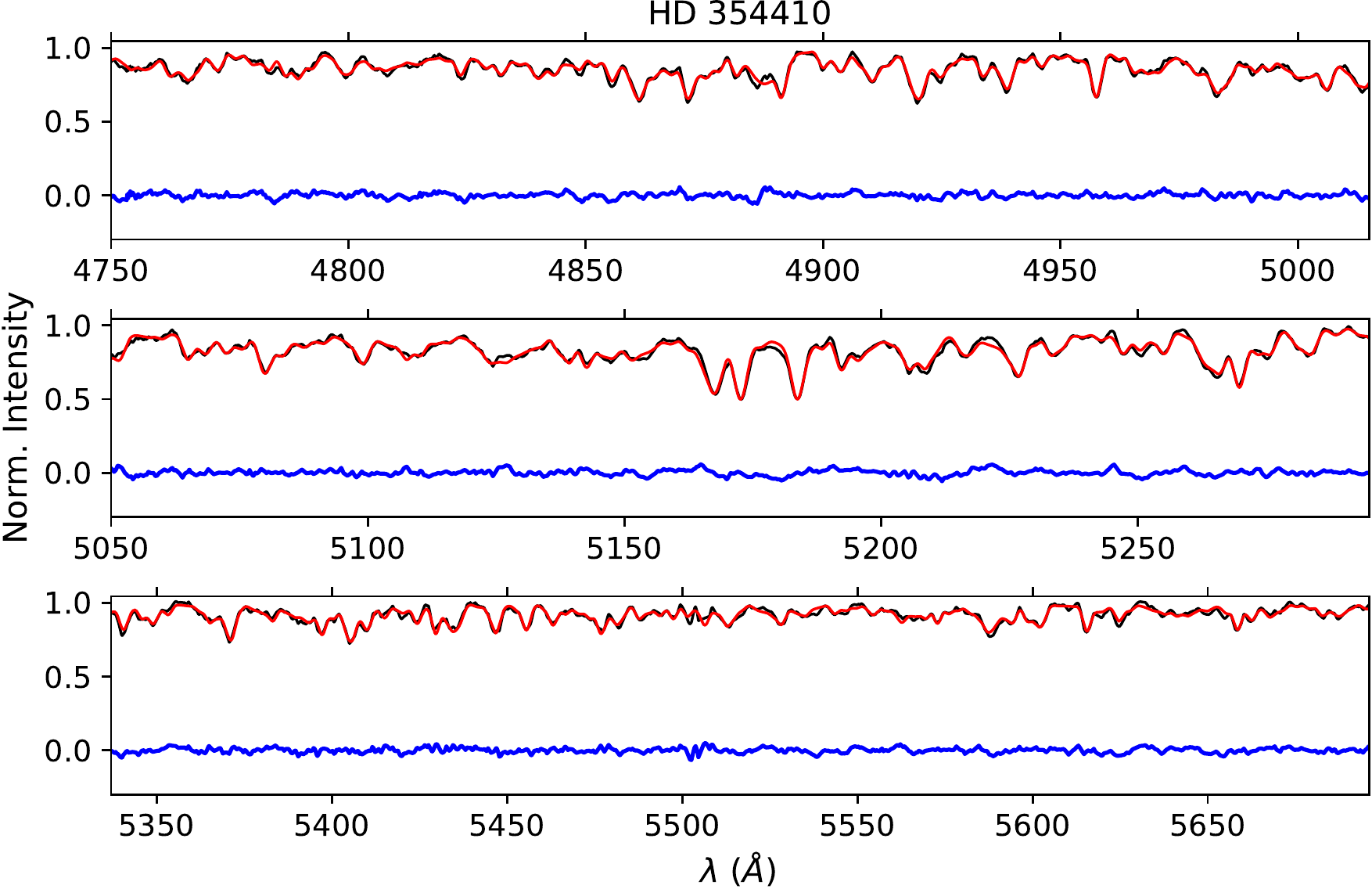}\vspace{0.3cm}
	\includegraphics[scale=0.47]{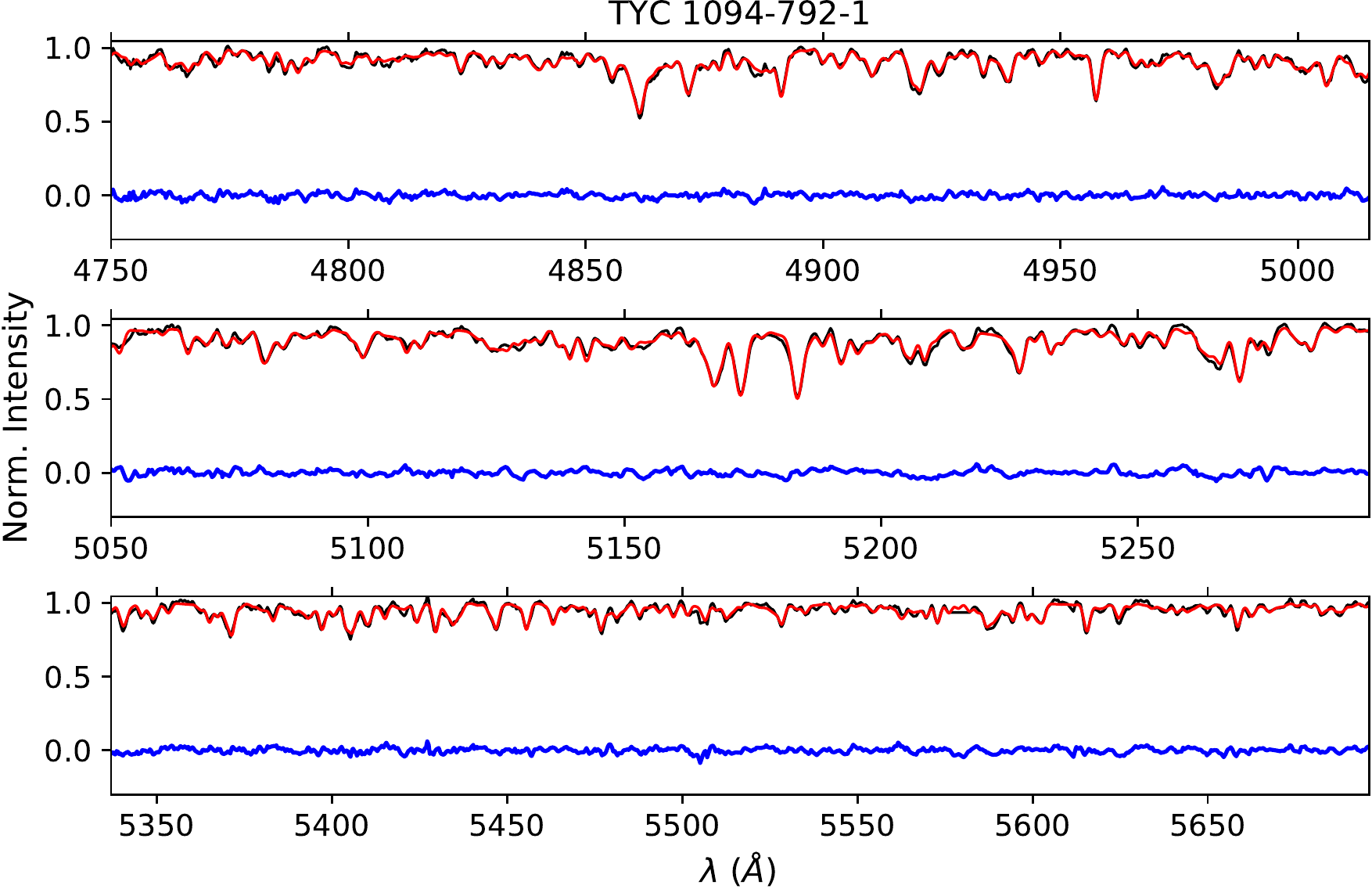}\hspace{0.3cm}
	\includegraphics[scale=0.47]{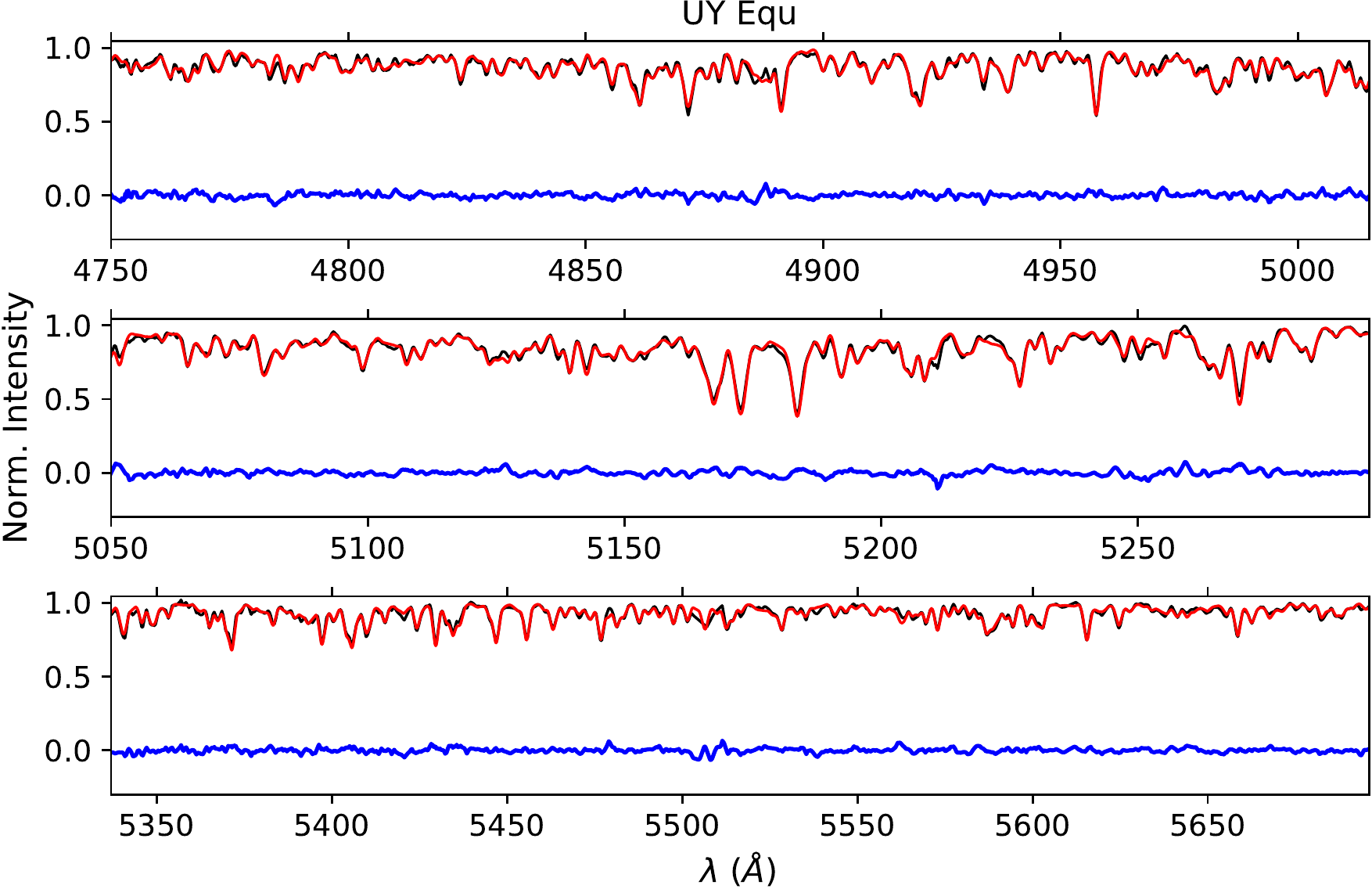}\vspace{0.3cm}
	\includegraphics[scale=0.47]{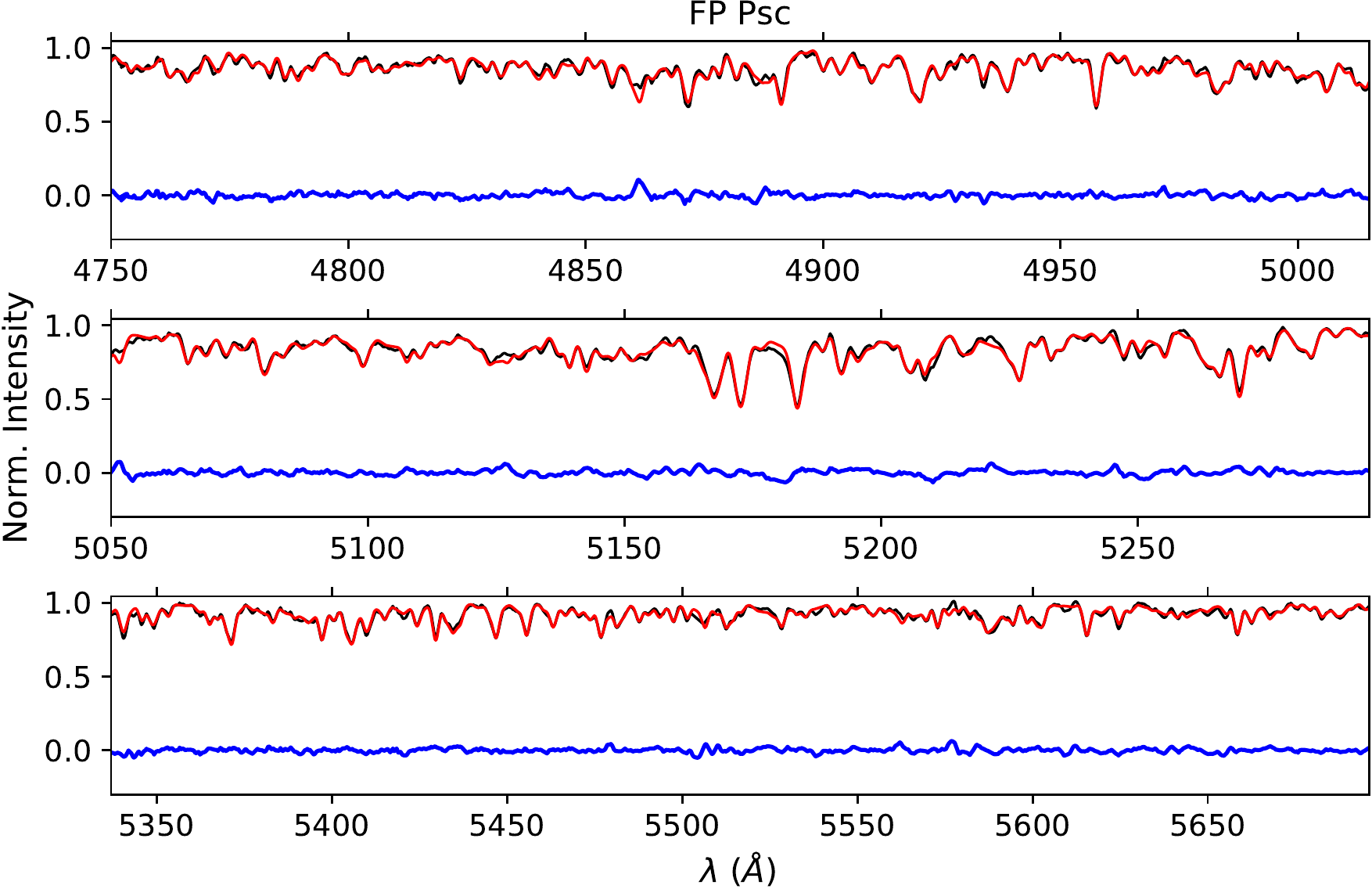}\hspace{0.3cm}
	\includegraphics[scale=0.47]{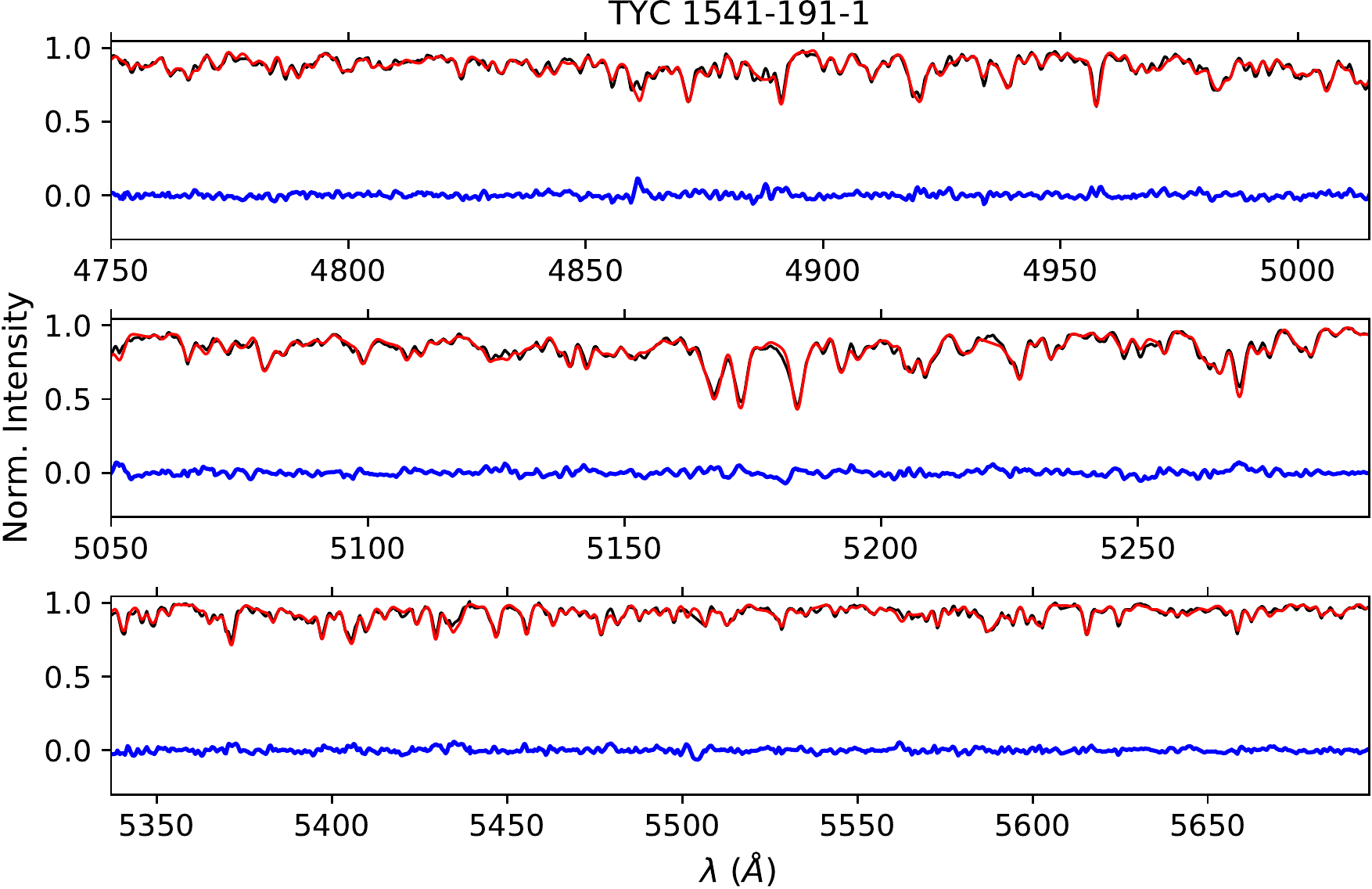}\vspace{0.3cm}
	\includegraphics[scale=0.47]{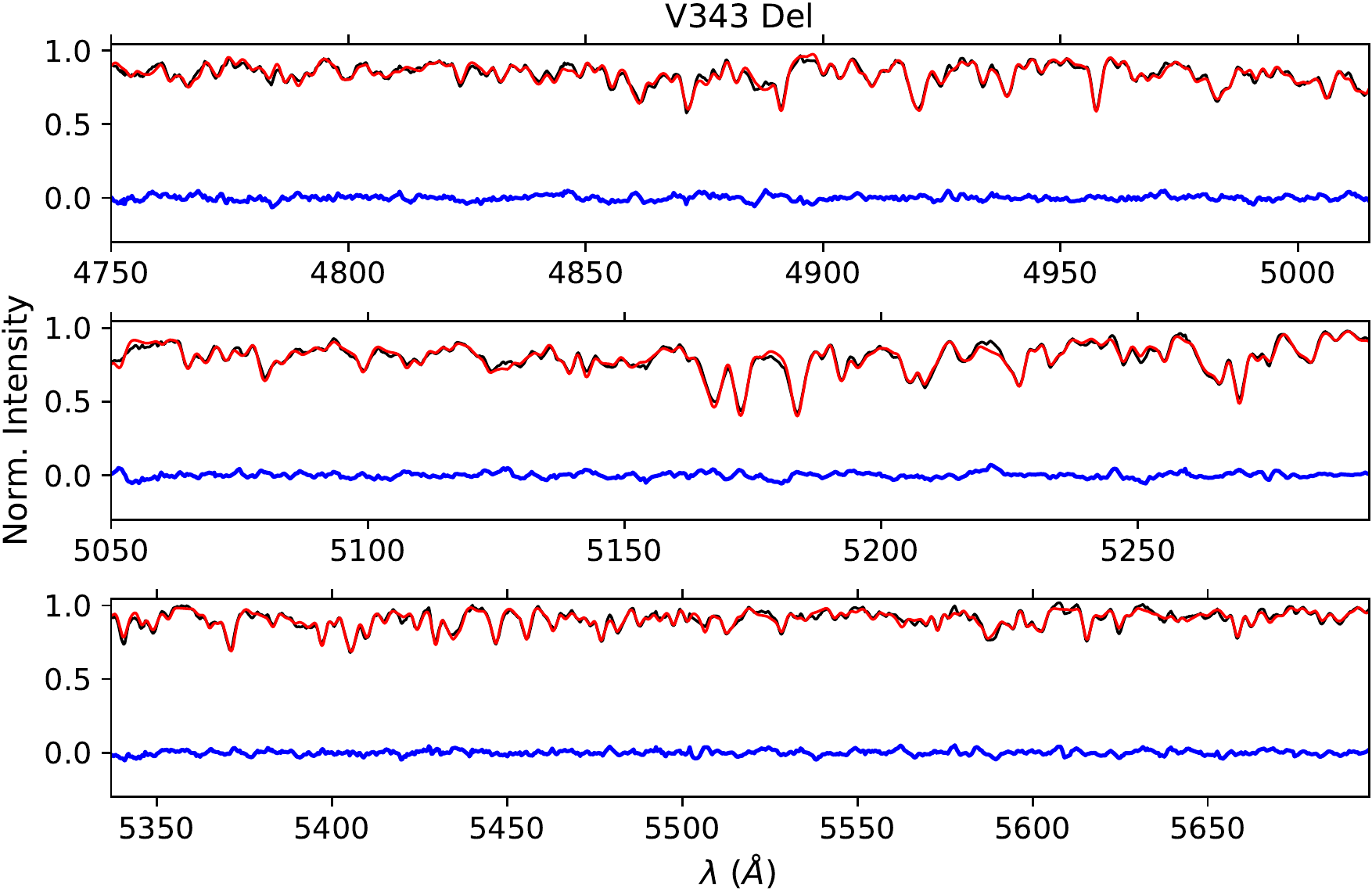}\hspace{0.3cm}
	\includegraphics[scale=0.47]{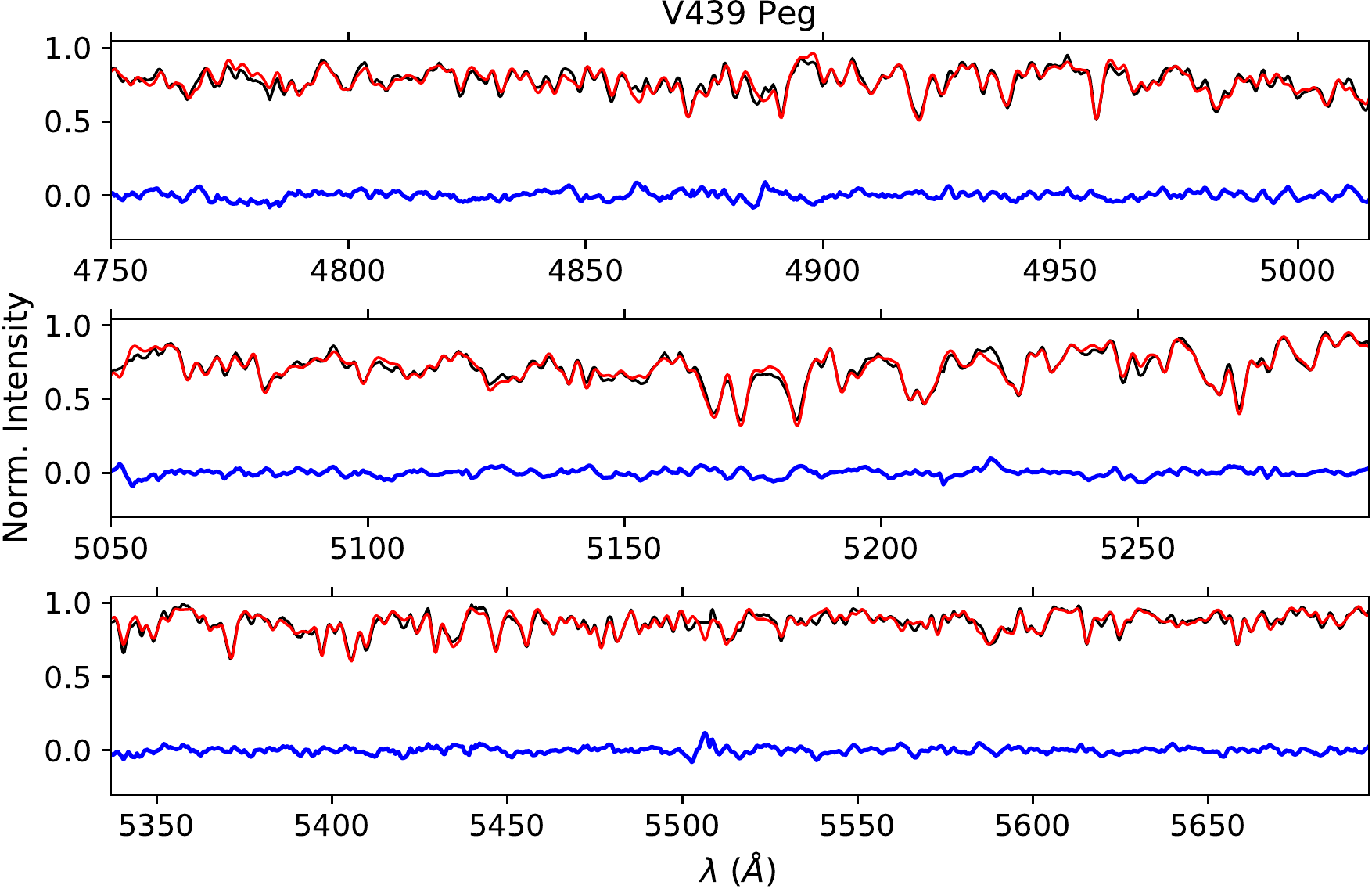}
	\caption{Continued.}
\end{figure*}

\addtocounter{figure}{-1}
\begin{figure*}
\begin{flushleft}
	%
	%
	%
	%
	\includegraphics[scale=0.47]{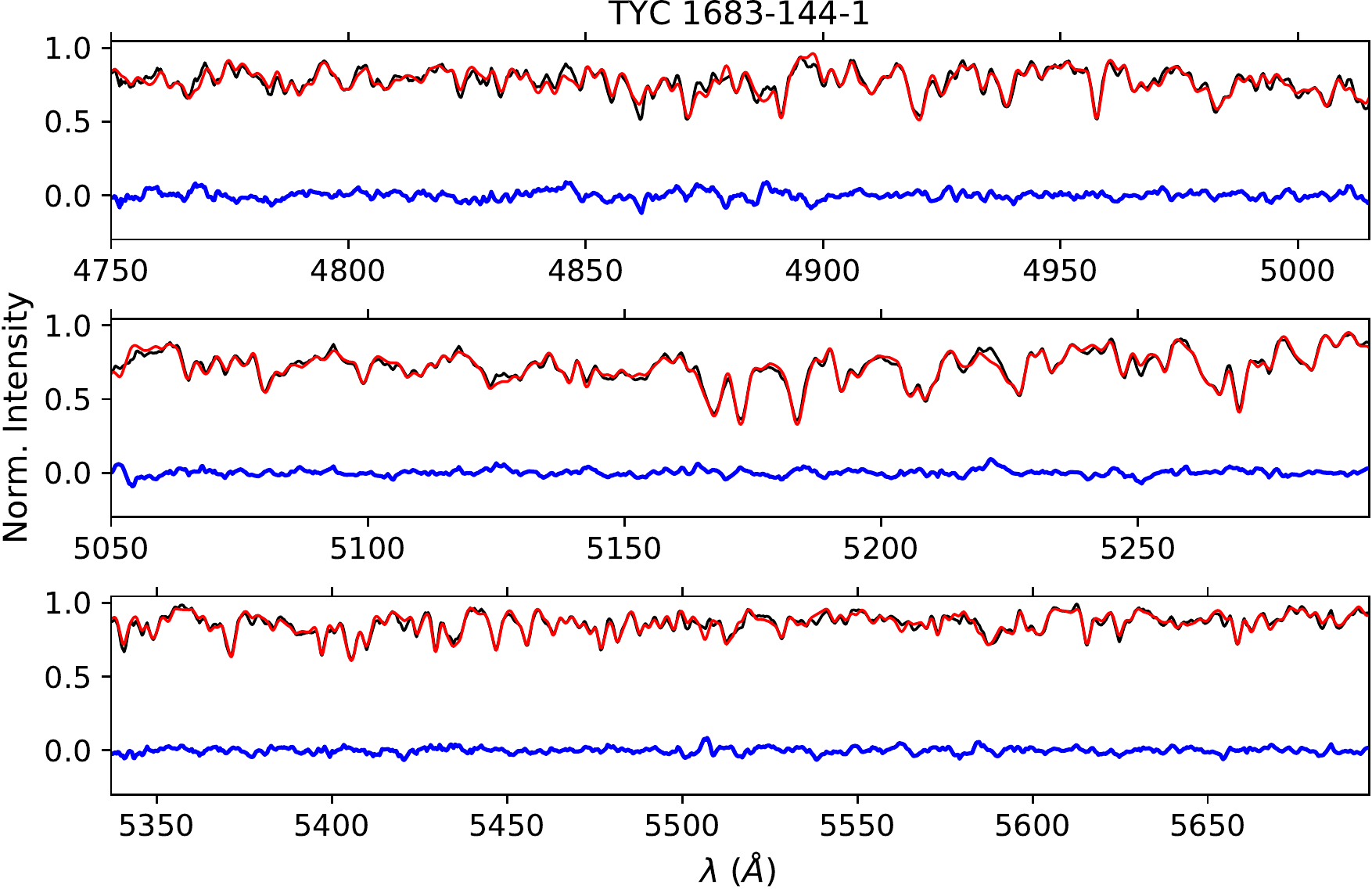}\hspace{0.3cm}
	\includegraphics[scale=0.47]{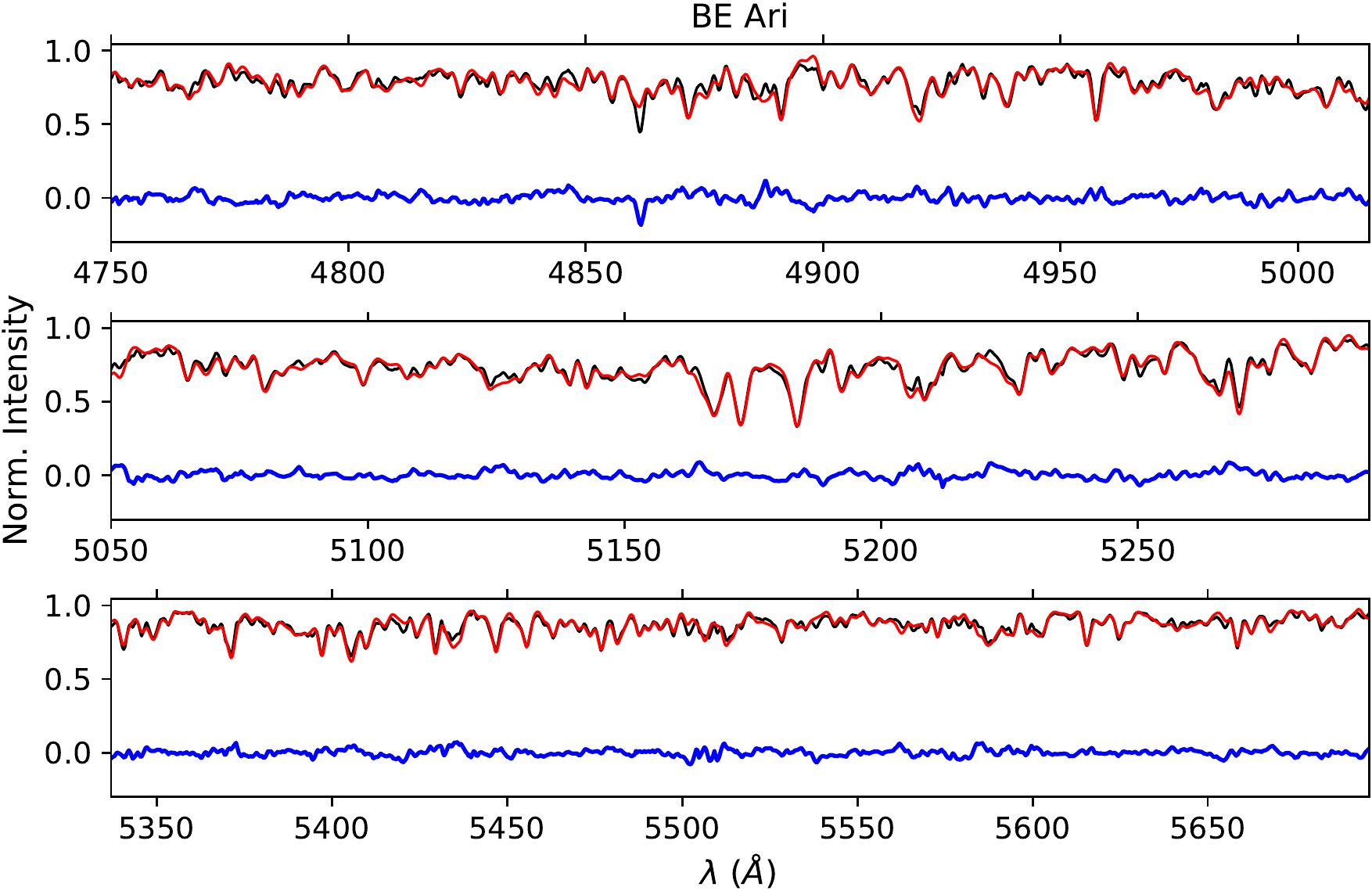}\vspace{0.3cm}
	\includegraphics[scale=0.47]{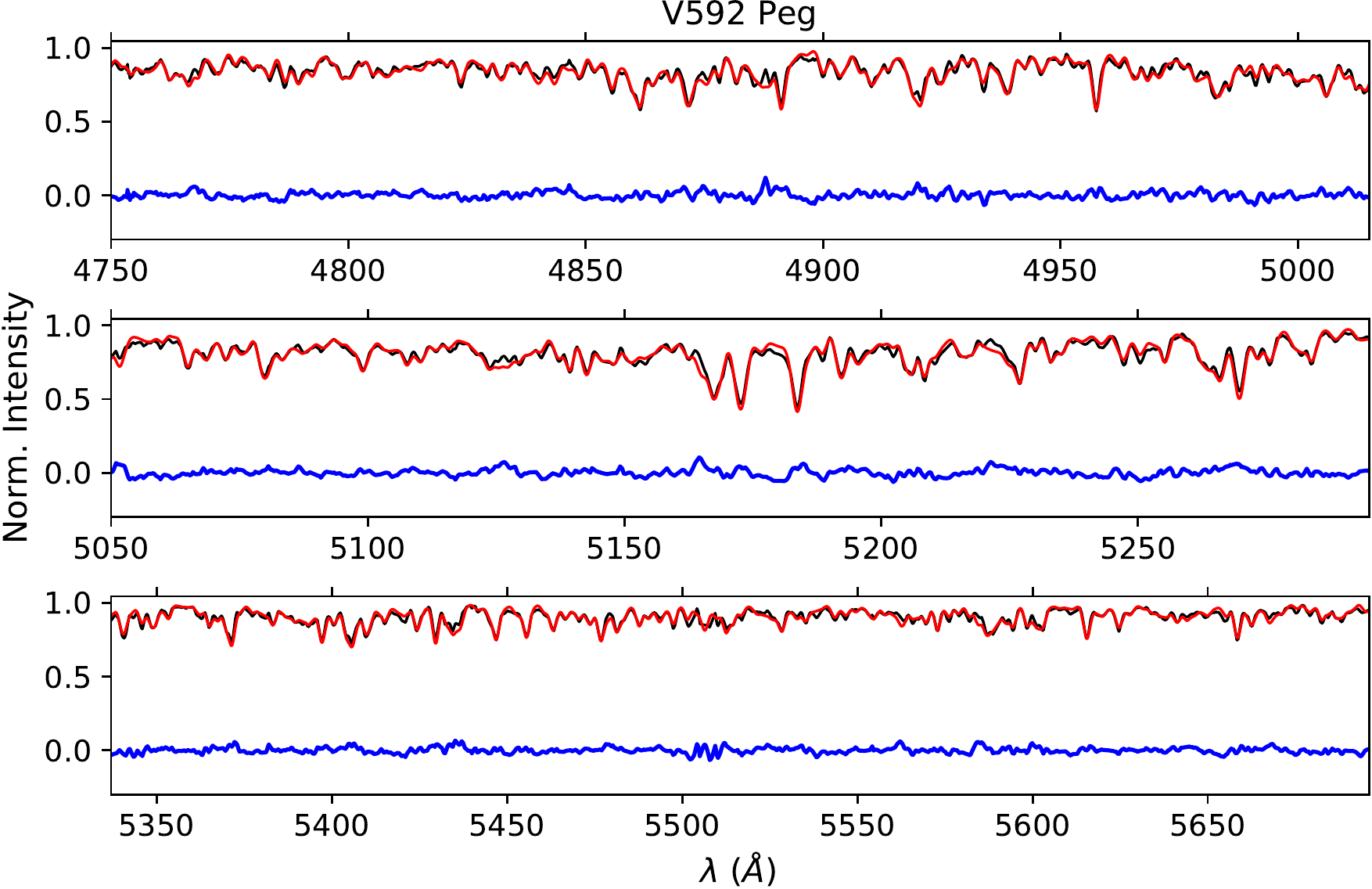}\hspace{0.3cm}
	\includegraphics[scale=0.47]{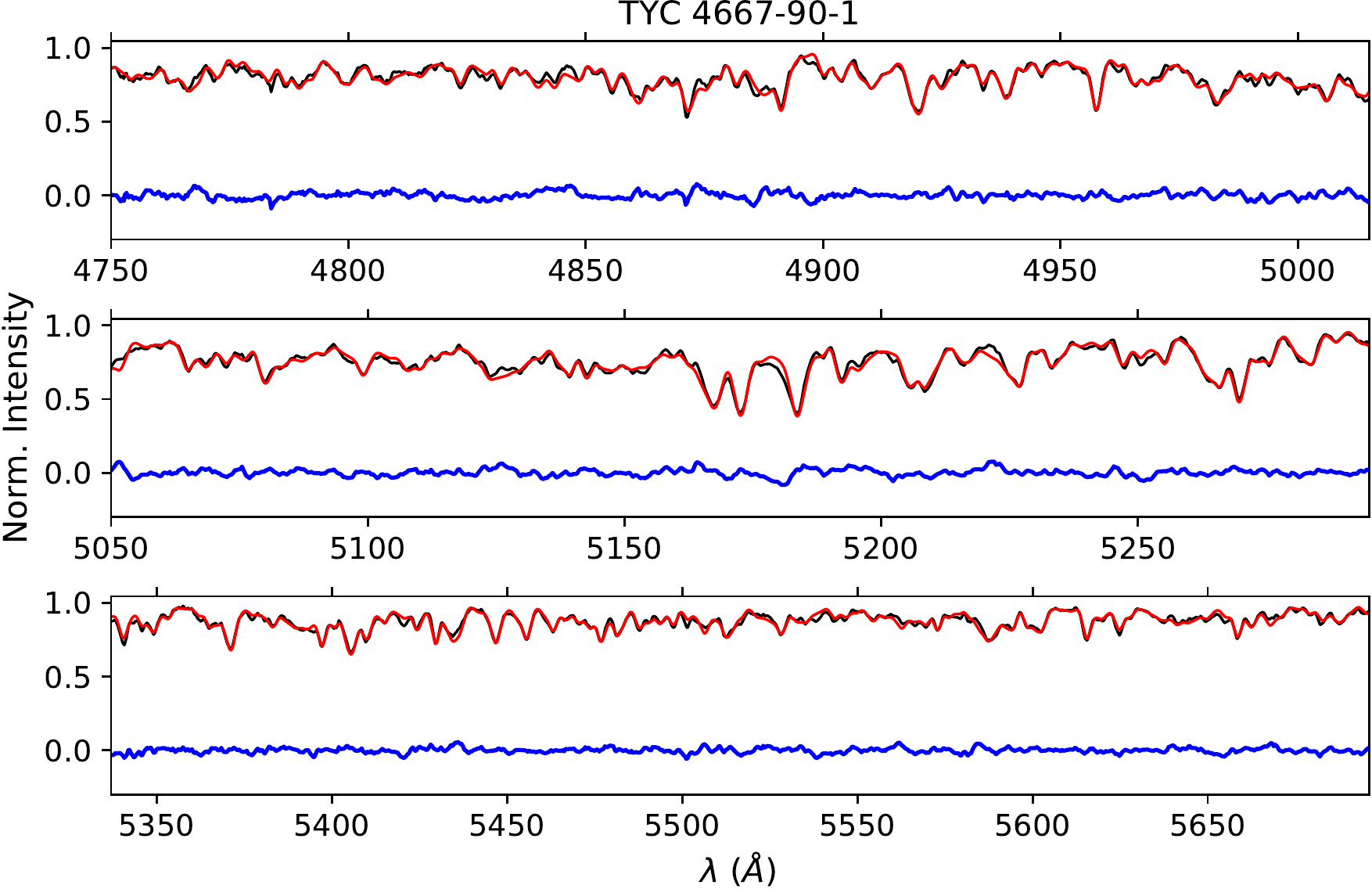}\vspace{0.3cm}
	\includegraphics[scale=0.47]{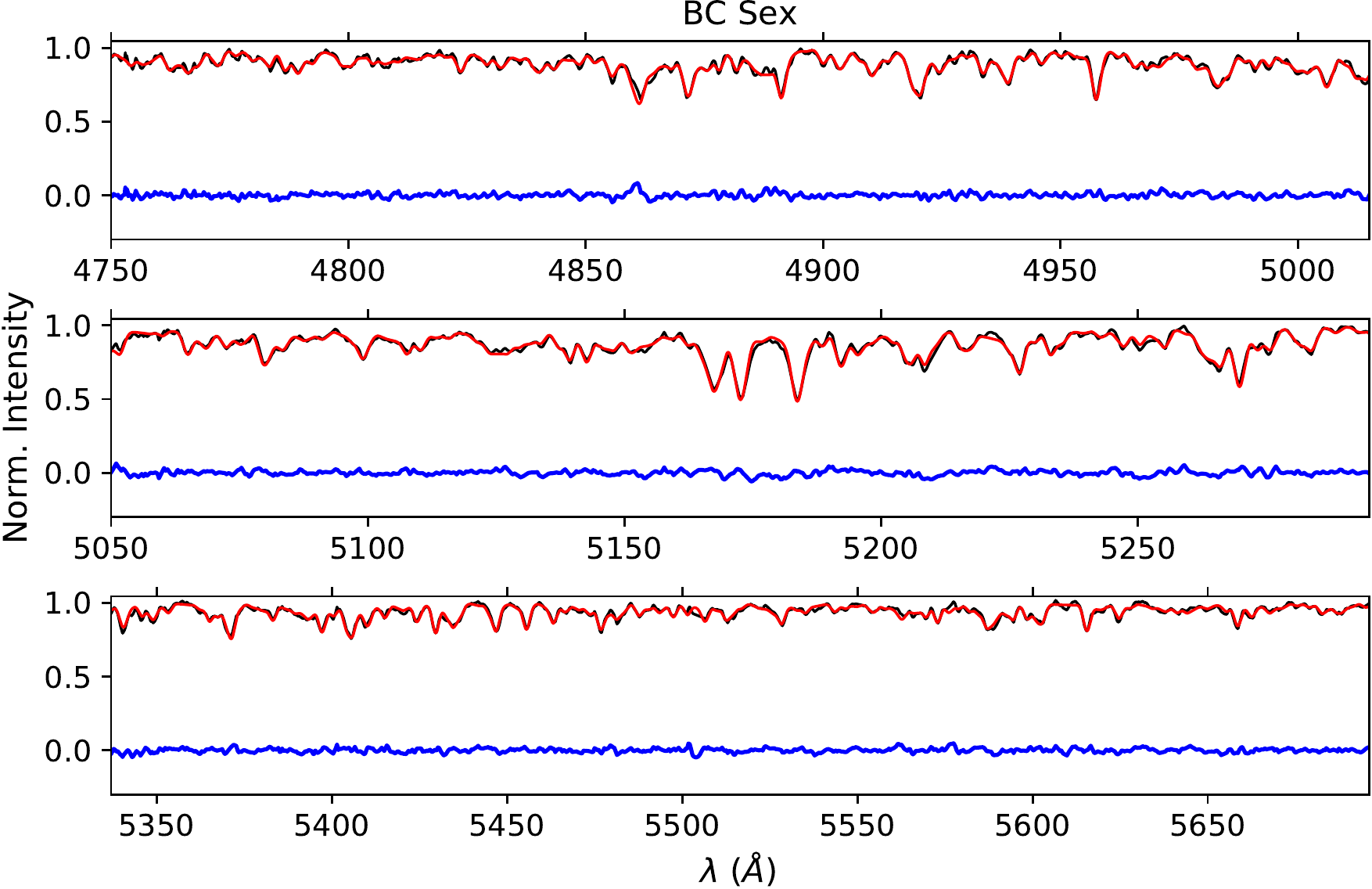}\vspace{5cm}
	\caption{Continued.}
\end{flushleft}
\end{figure*}

\section{Lomb-Scargle amplitude spectra}
We present Lomb-Scargle amplitude spectra of the stars with cyclic mean brightness 
variation shown in Fig.~\ref{figure_long_term_photometry}.

\begin{figure*}
	\includegraphics[width=\textwidth]{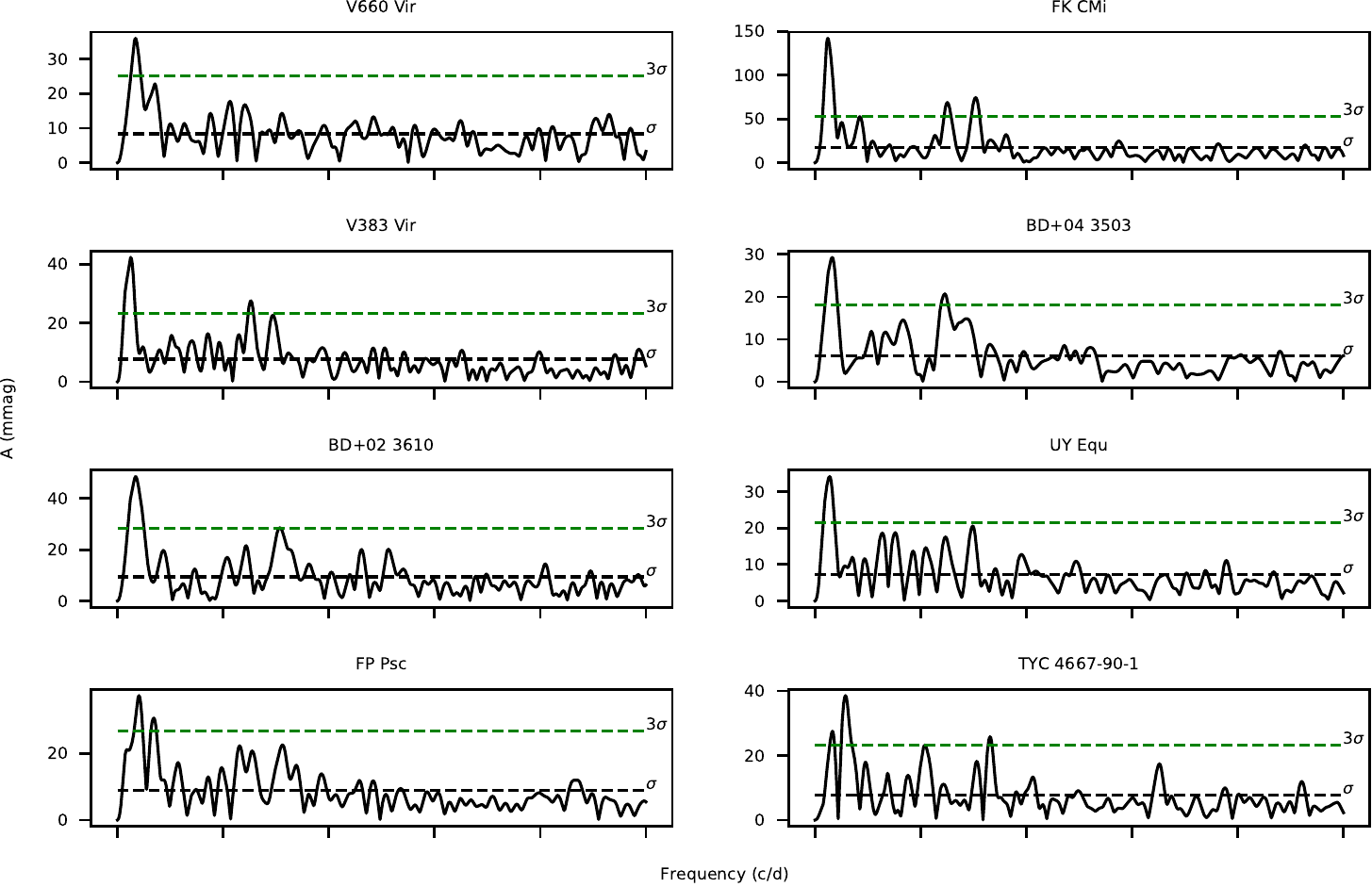}
	\caption{Lomb-Scargle amplitude spectra of the stars with detected cycle frequencies. 
	Horizontal dashed lines show 1$\sigma$ and 3$\sigma$ uncertainty levels. 1$\sigma$ level 
	is the mean of the amplitude spectrum, which is computed by excluding the dominant peak and
	its close vicinity.}
	\label{figure_amplitude_spectra}
\end{figure*}

\section{Seasonal photometric analysis of target stars.}
We tabulate seasonal photometric analysis results in Table~\ref{table_seasonal_phot_analysis}.

\begin{table*}
	\centering
	\caption{Photometric analysis results of seasonal light curves. Columns are start, end and mean HJD of the corresponding light curve,
	time span of the data set ($\Delta\,t$), computed photometric period (P) and its uncertainty, maximum, minimum and mean brightnesses, 
	peak-to-peak amplitude of the light curve (A) and number of data points in the light curve (N). We list computed average photometric period 
	($P_{rot}$) of each target as two columns list at the end of the table. Uncertainty of $P_{rot}$ is computed via standard deviation of seasonal
	photometric periods.}
	\label{table_seasonal_phot_analysis}
	\resizebox{17cm}{!}{%
	\begin{tabular}{cccccccccc}
		\hline
HJD 	&	HJD 	&	HJD 	& $\Delta\,t$	& P	&	max	 &	 min	&	mean	&	A	 &	N	\\				
start        &  end  	&	mean		    & (day) & (day) &  (mag) &  (mag)   &  (mag)	&  (mag) &     \\																
\hline\noalign{\smallskip}																					
V660\,Vir	&	&	&	&	&		&	&	&	&	\\										
52652.8631	&	52875.5046	&	52764.1839	&	222.64	&	75	$\pm$	13	&	11.468	&	11.607	&	11.537	&	0.139	&	58	\\
53036.8567	&	53219.5272	&	53128.1920	&	182.67	&	74	$\pm$	12	&	11.494	&	11.621	&	11.557	&	0.127	&	28	\\
53396.8736	&	53589.5255	&	53493.1996	&	192.65	&	82	$\pm$	20	&	11.504	&	11.649	&	11.576	&	0.145	&	46	\\
53754.8625	&	53931.8480	&	53843.3553	&	176.99	&	72	$\pm$	14	&	11.482	&	11.657	&	11.570	&	0.175	&	30	\\
54136.8491	&	54339.4810	&	54238.1651	&	202.63	&	72	$\pm$	12	&	11.473	&	11.733	&	11.603	&	0.260	&	70	\\
54480.1409	&	54708.4795	&	54594.3102	&	228.34	&	70	$\pm$	11	&	11.547	&	11.655	&	11.601	&	0.108	&	50	\\
55176.1530	&	55448.7186	&	55312.4358	&	272.57	&	71	$\pm$	11	&	11.580	&	11.763	&	11.672	&	0.183	&	80	\\
55567.1414	&	55801.7374	&	55684.4394	&	234.60	&	71	$\pm$	9	&	11.489	&	11.743	&	11.616	&	0.254	&	41	\\
55926.1492	&	56165.7471	&	56045.9482	&	239.60	&	71	$\pm$	9	&	11.491	&	11.686	&	11.588	&	0.195	&	49	\\
56740.8665	&	56857.8001	&	56799.3333	&	116.93	&	78	$\pm$	19	&	11.597	&	11.681	&	11.639	&	0.084	&	24	\\
56992.1584	&	57240.7435	&	57116.4510	&	248.59	&	71	$\pm$	8	&	11.639	&	11.716	&	11.677	&	0.077	&	66	\\
57386.1205	&	57639.7231	&	57512.9218	&	253.60	&	74	$\pm$	10	&	11.601	&	11.800	&	11.700	&	0.199	&	73	\\
57752.1058	&	57961.7584	&	57856.9321	&	209.65	&	73	$\pm$	11	&	11.658	&	11.809	&	11.733	&	0.151	&	56	\\
58120.1164	&	58325.7653	&	58222.9409	&	205.65	&	73	$\pm$	11	&	11.691	&	11.916	&	11.804	&	0.225	&	26	\\
	&	&	&	&	&	&	&	&	&		\\										
DG\,Ari	&	&	&	&	&	&	&	&	&		\\										
52621.5762	&	52674.5304	&	52648.0533	&	52.95	&	34	$\pm$	8	&	11.118	&	11.308	&	11.213	&	0.190	&	22	\\
52831.9138	&	53046.5255	&	52939.2197	&	214.61	&	34	$\pm$	2	&	11.177	&	11.363	&	11.270	&	0.186	&	55	\\
53266.8093	&	53410.5355	&	53338.6724	&	143.73	&	34	$\pm$	3	&	11.145	&	11.379	&	11.262	&	0.234	&	22	\\
53564.9302	&	53779.5349	&	53672.2326	&	214.60	&	34	$\pm$	2	&	11.082	&	11.278	&	11.180	&	0.196	&	88	\\
54012.8949	&	54111.6050	&	54062.2500	&	98.71	&	32	$\pm$	6	&	11.146	&	11.341	&	11.244	&	0.195	&	47	\\
54757.7669	&	54861.7684	&	54809.7677	&	104.00	&	35	$\pm$	3	&	11.075	&	11.198	&	11.136	&	0.123	&	28	\\
55014.1144	&	55135.7166	&	55074.9155	&	121.60	&	36	$\pm$	5	&	11.099	&	11.240	&	11.170	&	0.141	&	52	\\
55144.7181	&	55268.7446	&	55206.7314	&	124.03	&	35	$\pm$	4	&	11.072	&	11.247	&	11.160	&	0.175	&	48	\\
55775.0789	&	56000.7400	&	55887.9095	&	225.66	&	34	$\pm$	2	&	11.087	&	11.223	&	11.155	&	0.136	&	50	\\
56127.1154	&	56246.0212	&	56186.5683	&	118.91	&	34	$\pm$	3	&	11.116	&	11.252	&	11.184	&	0.136	&	39	\\
56836.1130	&	57098.7185	&	56967.4158	&	262.61	&	36	$\pm$	3	&	11.072	&	11.200	&	11.136	&	0.128	&	61	\\
57203.1194	&	57444.7839	&	57323.9517	&	241.66	&	34	$\pm$	1	&	11.032	&	11.100	&	11.066	&	0.068	&	74	\\
57564.1174	&	57831.7193	&	57697.9184	&	267.60	&	33	$\pm$	2	&	11.044	&	11.115	&	11.079	&	0.071	&	78	\\
57930.1178	&	58167.5161	&	58048.8170	&	237.40	&	33	$\pm$	2	&	11.037	&	11.093	&	11.065	&	0.056	&	78	\\
58297.1116	&	58450.9446	&	58374.0281	&	153.83	&	33	$\pm$	3	&	11.031	&	11.089	&	11.060	&	0.058	&	22	\\
	&	&	&	&	&	&	&	&	&		\\										
V1263\,Tau	&	&	&	&	&	&	&	&	&		\\										
52627.5901	&	52710.4978	&	52669.0440	&	82.91	&	20.4	$\pm$	3.4	&	10.561	&	10.652	&	10.606	&	0.091	&	27	\\
52831.9085	&	53068.5029	&	52950.2057	&	236.59	&	20.4	$\pm$	1.0	&	10.516	&	10.661	&	10.588	&	0.145	&	57	\\
53269.7895	&	53439.4971	&	53354.6433	&	169.71	&	20.5	$\pm$	0.9	&	10.504	&	10.678	&	10.591	&	0.174	&	28	\\
53564.9349	&	53801.5031	&	53683.2190	&	236.57	&	20.5	$\pm$	1.0	&	10.508	&	10.720	&	10.614	&	0.212	&	47	\\
53987.0826	&	54184.7266	&	54085.9046	&	197.64	&	20.4	$\pm$	0.9	&	10.472	&	10.638	&	10.555	&	0.166	&	45	\\
54293.9180	&	54439.6196	&	54366.7688	&	145.70	&	20.2	$\pm$	1.0	&	10.507	&	10.610	&	10.558	&	0.103	&	48	\\
54725.8015	&	54825.7000	&	54775.7508	&	99.90	&	19.8	$\pm$	1.6	&	10.459	&	10.551	&	10.505	&	0.092	&	45	\\
55021.9329	&	55132.7437	&	55077.3383	&	110.81	&	19.7	$\pm$	1.5	&	10.443	&	10.528	&	10.486	&	0.085	&	38	\\
55763.1018	&	56006.7237	&	55884.9128	&	243.62	&	20.4	$\pm$	0.9	&	10.442	&	10.666	&	10.554	&	0.224	&	40	\\
56144.1033	&	56359.7326	&	56251.9180	&	215.63	&	20.5	$\pm$	0.7	&	10.420	&	10.570	&	10.495	&	0.150	&	36	\\
56527.0755	&	56700.7722	&	56613.9239	&	173.70	&	20.4	$\pm$	1.0	&	10.425	&	10.617	&	10.521	&	0.192	&	24	\\
56865.1191	&	57070.7448	&	56967.9320	&	205.63	&	20.4	$\pm$	0.8	&	10.378	&	10.575	&	10.477	&	0.197	&	59	\\
57209.1200	&	57444.7155	&	57326.9178	&	235.60	&	20.4	$\pm$	0.8	&	10.434	&	10.554	&	10.494	&	0.120	&	53	\\
57581.1132	&	57829.7404	&	57705.4268	&	248.63	&	20.6	$\pm$	1.0	&	10.491	&	10.572	&	10.531	&	0.081	&	73	\\
57934.9328	&	58056.6148	&	57995.7738	&	121.68	&	20.0	$\pm$	0.9	&	10.482	&	10.618	&	10.550	&	0.136	&	22	\\
58121.8736	&	58179.5138	&	58150.6937	&	57.64	&	22.1	$\pm$	5.5	&	10.484	&	10.613	&	10.548	&	0.129	&	25	\\
58312.1149	&	58451.8172	&	58381.9661	&	139.70	&	20.2	$\pm$	1.8	&	10.445	&	10.605	&	10.525	&	0.160	&	26	\\
	&	&	&	&	&	&	&	&	&		\\										
FK\,CMi	&	&	&	&	&	&	&	&	&		\\										
52519.9188	&	52794.4482	&	52657.1835	&	274.53	&	19.4	$\pm$	0.8	&	11.059	&	11.392	&	11.225	&	0.333	&	67	\\
52894.9005	&	53142.4838	&	53018.6922	&	247.58	&	19.4	$\pm$	0.7	&	11.004	&	11.446	&	11.225	&	0.442	&	73	\\
53271.8805	&	53512.4605	&	53392.1705	&	240.58	&	19.2	$\pm$	1.0	&	11.089	&	11.333	&	11.211	&	0.244	&	50	\\
53650.8099	&	53883.4558	&	53767.1329	&	232.65	&	19.5	$\pm$	0.7	&	11.012	&	11.302	&	11.157	&	0.290	&	62	\\
54014.1014	&	54247.4538	&	54130.7776	&	233.35	&	19.4	$\pm$	0.8	&	10.971	&	11.154	&	11.062	&	0.183	&	98	\\
54365.8936	&	54608.4666	&	54487.1801	&	242.57	&	19.5	$\pm$	0.7	&	10.946	&	11.153	&	11.050	&	0.207	&	81	\\
54755.8638	&	54969.7379	&	54862.8008	&	213.87	&	19.5	$\pm$	0.9	&	11.102	&	11.437	&	11.269	&	0.335	&	63	\\
55080.1348	&	55336.7457	&	55208.4403	&	256.61	&	19.5	$\pm$	0.9	&	11.232	&	11.456	&	11.344	&	0.224	&	85	\\
55449.1376	&	55705.7368	&	55577.4372	&	256.60	&	19.5	$\pm$	0.6	&	11.365	&	11.613	&	11.489	&	0.248	&	66	\\
55825.1099	&	56065.7458	&	55945.4279	&	240.64	&	19.4	$\pm$	0.8	&	11.425	&	11.677	&	11.551	&	0.252	&	82	\\
56185.1273	&	56415.7740	&	56300.4507	&	230.65	&	19.6	$\pm$	0.6	&	11.532	&	11.712	&	11.622	&	0.180	&	73	\\
56575.1265	&	56682.9616	&	56629.0441	&	107.84	&	19.3	$\pm$	1.4	&	11.489	&	11.753	&	11.621	&	0.264	&	30	\\
57009.8632	&	57132.7543	&	57071.3088	&	122.89	&	19.5	$\pm$	1.3	&	11.431	&	11.779	&	11.605	&	0.348	&	111	\\
57285.0992	&	57499.7415	&	57392.4204	&	214.64	&	19.4	$\pm$	0.9	&	11.390	&	11.792	&	11.591	&	0.402	&	143	\\
57639.1269	&	57895.7541	&	57767.4405	&	256.63	&	19.4	$\pm$	0.7	&	11.428	&	11.668	&	11.548	&	0.240	&	137	\\
57997.1318	&	58220.7822	&	58108.9570	&	223.65	&	19.3	$\pm$	0.9	&	11.386	&	11.719	&	11.552	&	0.333	&	101	\\
58372.1080	&	58451.0433	&	58411.5757	&	78.94	&	19.7	$\pm$	2.2	&	11.380	&	11.699	&	11.539	&	0.319	&	19	\\
	&	&	&	&	&	&	&	&	&		\\										
V383\,Vir	&	&	&	&	&	&	&	&	&		\\										
52623.8560	&	52845.4759	&	52734.6660	&	221.62	&	14.0	$\pm$	0.4	&	10.156	&	10.231	&	10.193	&	0.075	&	76	\\
52996.8488	&	53189.4938	&	53093.1713	&	192.64	&	14.5	$\pm$	0.4	&	10.118	&	10.201	&	10.160	&	0.083	&	40	\\
53387.8449	&	53575.5043	&	53481.6746	&	187.66	&	14.7	$\pm$	0.3	&	10.116	&	10.277	&	10.197	&	0.161	&	26	\\
54079.1257	&	54319.4656	&	54199.2957	&	240.34	&	14.4	$\pm$	0.5	&	10.132	&	10.275	&	10.203	&	0.143	&	68	\\
54461.8549	&	54674.4738	&	54568.1644	&	212.62	&	14.3	$\pm$	0.5	&	10.199	&	10.303	&	10.251	&	0.104	&	54	\\
54822.8532	&	55040.4717	&	54931.6625	&	217.62	&	14.8	$\pm$	0.5	&	10.103	&	10.246	&	10.174	&	0.143	&	40	\\
55572.1075	&	55682.8917	&	55627.4996	&	110.78	&	13.9	$\pm$	0.6	&	10.071	&	10.224	&	10.147	&	0.153	&	22	\\
55951.0409	&	56091.8590	&	56021.4500	&	140.82	&	14.4	$\pm$	0.5	&	10.061	&	10.203	&	10.132	&	0.142	&	36	\\
57357.0975	&	57601.7468	&	57479.4221	&	244.65	&	14.5	$\pm$	0.4	&	10.053	&	10.101	&	10.077	&	0.048	&	78	\\
	&	&	&	&	&	&	&	&	&		\\										
BD+04\,3503	&	&	&	&	&	&	&	&	&		\\										
52443.6524	&	52548.5211	&	52496.0868	&	104.87	&	8.4	$\pm$	0.3	&	9.462	&	9.624	&	9.543	&	0.162	&	15	\\
52684.8874	&	52933.4962	&	52809.1918	&	248.61	&	8.5	$\pm$	0.1	&	9.466	&	9.564	&	9.515	&	0.098	&	65	\\
53063.8970	&	53219.5369	&	53141.7170	&	155.64	&	8.5	$\pm$	0.3	&	9.440	&	9.527	&	9.484	&	0.087	&	24	\\
53424.8957	&	53650.4952	&	53537.6954	&	225.60	&	8.5	$\pm$	0.2	&	9.480	&	9.553	&	9.516	&	0.073	&	49	\\
53790.8911	&	54051.6848	&	53921.2880	&	260.79	&	8.5	$\pm$	0.2	&	9.453	&	9.542	&	9.498	&	0.089	&	66	\\
54146.1528	&	54389.4928	&	54267.8228	&	243.34	&	8.4	$\pm$	0.1	&	9.406	&	9.629	&	9.517	&	0.223	&	117	\\
54521.8919	&	54748.5111	&	54635.2015	&	226.62	&	8.5	$\pm$	0.2	&	9.429	&	9.651	&	9.540	&	0.222	&	77	\\
54854.1616	&	55033.8503	&	54944.0060	&	179.69	&	8.4	$\pm$	0.2	&	9.436	&	9.615	&	9.525	&	0.179	&	54	\\
55039.5863	&	55123.7295	&	55081.6579	&	84.14	&	8.5	$\pm$	0.3	&	9.413	&	9.592	&	9.502	&	0.179	&	32	\\
55299.0881	&	55383.9764	&	55341.5323	&	84.89	&	8.5	$\pm$	0.3	&	9.444	&	9.633	&	9.539	&	0.189	&	49	\\
55394.8460	&	55509.6889	&	55452.2675	&	114.84	&	8.4	$\pm$	0.3	&	9.410	&	9.556	&	9.483	&	0.146	&	39	\\
55587.1747	&	55742.9733	&	55665.0740	&	155.80	&	8.4	$\pm$	0.2	&	9.390	&	9.531	&	9.460	&	0.141	&	36	\\
55751.8823	&	55861.7194	&	55806.8008	&	109.84	&	8.4	$\pm$	0.3	&	9.369	&	9.620	&	9.495	&	0.251	&	37	\\
55952.1726	&	56094.9465	&	56023.5596	&	142.77	&	8.4	$\pm$	0.2	&	9.317	&	9.565	&	9.441	&	0.248	&	40	\\
56122.9147	&	56235.7049	&	56179.3098	&	112.79	&	8.3	$\pm$	0.3	&	9.314	&	9.547	&	9.431	&	0.233	&	33	\\
56329.1577	&	56587.7029	&	56458.4303	&	258.55	&	8.4	$\pm$	0.1	&	9.382	&	9.578	&	9.480	&	0.196	&	40	\\
56735.0999	&	56924.7395	&	56829.9197	&	189.64	&	8.5	$\pm$	0.2	&	9.344	&	9.463	&	9.403	&	0.119	&	55	\\
		\hline																			
	\end{tabular}%
	\begin{tabular}{cccccccccc}																				
		\hline																			
HJD 	&	HJD 	&	HJD 	& $\Delta\,t$	& P	&	max	 &	 min	&	mean	&	A	 &	N	\\				
start        &  end  	&	mean		    & (day) & (day) &  (mag) &  (mag)   &  (mag)	&  (mag) &     \\																
\hline\noalign{\smallskip}																					
BD+04\,3503	&	&	&	&	&		&	&	&	&	\\										
57061.1213	&	57195.7524	&	57128.4369	&	134.63	&	8.4	$\pm$	0.2	&	9.345	&	9.627	&	9.486	&	0.282	&	60	\\
57197.9403	&	57297.7091	&	57247.8247	&	99.77	&	8.5	$\pm$	0.4	&	9.381	&	9.572	&	9.476	&	0.191	&	32	\\
57452.1425	&	57550.9560	&	57501.5493	&	98.81	&	8.4	$\pm$	0.3	&	9.392	&	9.630	&	9.511	&	0.238	&	34	\\
57560.9541	&	57669.7154	&	57615.3348	&	108.76	&	8.4	$\pm$	0.4	&	9.371	&	9.635	&	9.503	&	0.264	&	46	\\
57781.1742	&	57928.9587	&	57855.0665	&	147.78	&	8.5	$\pm$	0.2	&	9.416	&	9.630	&	9.523	&	0.214	&	61	\\
57939.8826	&	58020.7502	&	57980.3164	&	80.87	&	8.4	$\pm$	0.5	&	9.400	&	9.543	&	9.472	&	0.143	&	22	\\
58180.0800	&	58371.7721	&	58275.9261	&	191.69	&	8.4	$\pm$	0.2	&	9.509	&	9.601	&	9.555	&	0.092	&	25	\\
	&	&	&	&	&	&	&	&	&		\\										
BD+02\,3610	&	&	&	&	&	&	&	&	&		\\										
52782.7903	&	52946.5080	&	52864.6492	&	163.72	&	12.9	$\pm$	0.3	&	11.582	&	11.769	&	11.676	&	0.187	&	39	\\
53072.8968	&	53191.8071	&	53132.3520	&	118.91	&	12.8	$\pm$	0.5	&	11.511	&	11.761	&	11.636	&	0.250	&	21	\\
53444.8895	&	53676.5003	&	53560.6949	&	231.61	&	12.7	$\pm$	0.2	&	11.463	&	11.798	&	11.630	&	0.335	&	61	\\
53802.8904	&	54063.6839	&	53933.2872	&	260.79	&	12.8	$\pm$	0.2	&	11.421	&	11.849	&	11.635	&	0.428	&	46	\\
54163.1574	&	54408.5049	&	54285.8312	&	245.35	&	12.8	$\pm$	0.3	&	11.242	&	11.744	&	11.493	&	0.502	&	108	\\
54525.1493	&	54768.5136	&	54646.8315	&	243.36	&	12.8	$\pm$	0.4	&	11.259	&	11.611	&	11.435	&	0.352	&	82	\\
54881.1589	&	54995.0737	&	54938.1163	&	113.91	&	13.2	$\pm$	0.7	&	11.286	&	11.508	&	11.397	&	0.222	&	27	\\
55006.7315	&	55157.6880	&	55082.2098	&	150.96	&	12.9	$\pm$	0.6	&	11.222	&	11.394	&	11.308	&	0.172	&	43	\\
55240.1646	&	55516.7072	&	55378.4359	&	276.54	&	12.9	$\pm$	0.2	&	11.095	&	11.287	&	11.191	&	0.192	&	60	\\
55622.1460	&	55879.7065	&	55750.9263	&	257.56	&	12.8	$\pm$	0.2	&	10.961	&	11.344	&	11.152	&	0.383	&	54	\\
56001.1171	&	56242.6934	&	56121.9053	&	241.58	&	12.9	$\pm$	0.3	&	10.935	&	11.209	&	11.072	&	0.274	&	58	\\
56360.1333	&	56593.7143	&	56476.9238	&	233.58	&	12.8	$\pm$	0.3	&	10.949	&	11.154	&	11.052	&	0.205	&	24	\\
56811.0196	&	56918.7944	&	56864.9070	&	107.77	&	13.0	$\pm$	0.8	&	10.845	&	11.117	&	10.981	&	0.272	&	22	\\
57050.1689	&	57138.9662	&	57094.5676	&	88.80	&	12.8	$\pm$	0.7	&	10.801	&	11.136	&	10.969	&	0.335	&	22	\\
57153.9284	&	57305.7317	&	57229.8301	&	151.80	&	13.1	$\pm$	0.5	&	10.816	&	10.990	&	10.903	&	0.174	&	24	\\
57420.1672	&	57716.6892	&	57568.4282	&	296.52	&	12.9	$\pm$	0.3	&	10.681	&	10.985	&	10.833	&	0.304	&	48	\\
57797.1520	&	58037.7156	&	57917.4338	&	240.56	&	12.8	$\pm$	0.3	&	10.700	&	10.937	&	10.819	&	0.237	&	59	\\
	&	&	&	&	&	&	&	&	&		\\										
GH\,Psc	&	&	&	&	&	&	&	&	&		\\										
52805.9188	&	53025.5382	&	52915.7285	&	219.62	&	35	$\pm$	3	&	9.991	&	10.167	&	10.079	&	0.176	&	54	\\
53526.9347	&	53750.5566	&	53638.7457	&	223.62	&	33	$\pm$	2	&	10.027	&	10.156	&	10.091	&	0.129	&	50	\\
54270.1107	&	54487.7640	&	54378.9374	&	217.65	&	34	$\pm$	2	&	9.968	&	10.084	&	10.026	&	0.116	&	68	\\
54628.9285	&	54864.7175	&	54746.8230	&	235.79	&	35	$\pm$	3	&	9.977	&	10.090	&	10.033	&	0.113	&	59	\\
55352.1168	&	55589.7614	&	55470.9391	&	237.64	&	35	$\pm$	2	&	9.825	&	9.994	&	9.910	&	0.169	&	42	\\
55724.1049	&	55973.7182	&	55848.9116	&	249.61	&	36	$\pm$	3	&	9.890	&	10.002	&	9.946	&	0.112	&	40	\\
56088.1099	&	56329.7400	&	56208.9250	&	241.63	&	36	$\pm$	4	&	9.821	&	9.994	&	9.907	&	0.173	&	36	\\
56817.0922	&	56944.0005	&	56880.5464	&	126.91	&	33	$\pm$	2	&	9.810	&	9.906	&	9.858	&	0.096	&	51	\\
57216.1141	&	57437.7178	&	57326.9160	&	221.60	&	35	$\pm$	2	&	9.802	&	9.955	&	9.879	&	0.153	&	78	\\
57543.1121	&	57643.8466	&	57593.4794	&	100.73	&	34	$\pm$	5	&	9.834	&	9.994	&	9.914	&	0.160	&	40	\\
57651.0251	&	57806.7136	&	57728.8694	&	155.69	&	34	$\pm$	4	&	9.793	&	9.971	&	9.882	&	0.178	&	63	\\
57898.1166	&	58139.5280	&	58018.8223	&	241.41	&	34	$\pm$	3	&	9.823	&	10.023	&	9.923	&	0.200	&	84	\\
58262.1160	&	58451.7263	&	58356.9212	&	189.61	&	34	$\pm$	2	&	9.871	&	10.112	&	9.991	&	0.241	&	26	\\
	&	&	&	&	&	&	&	&	&		\\										
TYC\,723-863-1	&	&	&	&	&	&	&	&	&		\\										
52621.7101	&	52760.4668	&	52691.0885	&	138.76	&	45	$\pm$	7	&	10.127	&	10.408	&	10.268	&	0.281	&	91	\\
52861.9243	&	53124.4644	&	52993.1944	&	262.54	&	46	$\pm$	4	&	10.119	&	10.414	&	10.267	&	0.295	&	88	\\
53355.6615	&	53468.4951	&	53412.0783	&	112.83	&	44	$\pm$	7	&	10.193	&	10.429	&	10.311	&	0.236	&	27	\\
53622.9095	&	53837.4799	&	53730.1947	&	214.57	&	45	$\pm$	3	&	10.226	&	10.373	&	10.299	&	0.147	&	36	\\
53990.1016	&	54174.5673	&	54082.3345	&	184.47	&	46	$\pm$	5	&	10.251	&	10.451	&	10.351	&	0.200	&	52	\\
54350.9041	&	54469.7513	&	54410.3277	&	118.85	&	43	$\pm$	7	&	10.246	&	10.408	&	10.327	&	0.162	&	21	\\
54473.6784	&	54574.4688	&	54524.0736	&	100.79	&	44	$\pm$	9	&	10.238	&	10.454	&	10.346	&	0.216	&	31	\\
54722.8901	&	54864.8412	&	54793.8657	&	141.95	&	43	$\pm$	5	&	10.205	&	10.397	&	10.301	&	0.192	&	33	\\
55068.1145	&	55299.7702	&	55183.9424	&	231.66	&	43	$\pm$	4	&	10.227	&	10.426	&	10.326	&	0.199	&	55	\\
55425.1286	&	55681.7392	&	55553.4339	&	256.61	&	45	$\pm$	3	&	10.269	&	10.390	&	10.329	&	0.121	&	51	\\
55793.1263	&	56045.7432	&	55919.4348	&	252.62	&	44	$\pm$	4	&	10.307	&	10.460	&	10.383	&	0.153	&	43	\\
56156.1197	&	56406.7659	&	56281.4428	&	250.65	&	43	$\pm$	4	&	10.396	&	10.591	&	10.494	&	0.195	&	39	\\
56526.1118	&	56683.9819	&	56605.0469	&	157.87	&	43	$\pm$	4	&	10.428	&	10.610	&	10.519	&	0.182	&	22	\\
57008.9183	&	57103.7544	&	57056.3363	&	94.84	&	44	$\pm$	6	&	10.436	&	10.629	&	10.532	&	0.193	&	29	\\
57238.1231	&	57478.7392	&	57358.4312	&	240.62	&	43	$\pm$	4	&	10.519	&	10.649	&	10.584	&	0.130	&	70	\\
57653.0341	&	57807.8102	&	57730.4221	&	154.78	&	45	$\pm$	4	&	10.497	&	10.680	&	10.589	&	0.183	&	70	\\
57969.1217	&	58078.0550	&	58023.5884	&	108.93	&	43	$\pm$	8	&	10.589	&	10.710	&	10.650	&	0.121	&	27	\\
58352.1009	&	58450.9988	&	58401.5499	&	98.90	&	43	$\pm$	10	&	10.556	&	10.684	&	10.620	&	0.128	&	16	\\
	&	&	&	&	&	&	&	&	&		\\										
HD\,354410	&	&	&	&	&	&	&	&	&		\\										
52786.8037	&	52942.5143	&	52864.6590	&	155.71	&	27.6	$\pm$	2.6	&	11.095	&	11.270	&	11.182	&	0.175	&	40	\\
53104.9163	&	53190.7298	&	53147.8231	&	85.81	&	27.2	$\pm$	4.6	&	11.086	&	11.340	&	11.213	&	0.254	&	19	\\
53457.9060	&	53678.5017	&	53568.2039	&	220.60	&	27.5	$\pm$	1.6	&	11.080	&	11.305	&	11.192	&	0.225	&	57	\\
53820.9129	&	54093.6917	&	53957.3023	&	272.78	&	27.4	$\pm$	1.3	&	11.026	&	11.307	&	11.167	&	0.281	&	77	\\
54168.1427	&	54277.9904	&	54223.0666	&	109.85	&	27.6	$\pm$	3.3	&	11.097	&	11.252	&	11.174	&	0.155	&	53	\\
54303.7262	&	54409.5079	&	54356.6171	&	105.78	&	28.2	$\pm$	3.2	&	11.114	&	11.307	&	11.210	&	0.193	&	74	\\
54557.9038	&	54772.5136	&	54665.2087	&	214.61	&	27.4	$\pm$	1.7	&	11.061	&	11.395	&	11.228	&	0.334	&	104	\\
54918.1385	&	55175.6952	&	55046.9169	&	257.56	&	27.3	$\pm$	1.8	&	10.991	&	11.346	&	11.168	&	0.355	&	49	\\
55257.1488	&	55537.7094	&	55397.4291	&	280.56	&	27.7	$\pm$	1.6	&	11.032	&	11.245	&	11.139	&	0.213	&	72	\\
55634.1418	&	55902.6993	&	55768.4206	&	268.56	&	27.1	$\pm$	1.0	&	11.054	&	11.209	&	11.132	&	0.155	&	57	\\
56002.1209	&	56277.6948	&	56139.9079	&	275.57	&	26.8	$\pm$	1.1	&	11.044	&	11.229	&	11.136	&	0.185	&	59	\\
56355.1490	&	56621.7399	&	56488.4445	&	266.59	&	26.8	$\pm$	1.0	&	11.011	&	11.222	&	11.117	&	0.211	&	37	\\
56811.0726	&	56917.8574	&	56864.4650	&	106.78	&	28.0	$\pm$	2.5	&	10.937	&	11.231	&	11.084	&	0.294	&	20	\\
57078.1607	&	57366.7209	&	57222.4408	&	288.56	&	27.0	$\pm$	0.8	&	11.044	&	11.248	&	11.146	&	0.204	&	61	\\
57445.1627	&	57710.7036	&	57577.9332	&	265.54	&	26.5	$\pm$	1.3	&	11.079	&	11.278	&	11.178	&	0.199	&	84	\\
57809.1644	&	58049.7140	&	57929.4392	&	240.55	&	27.0	$\pm$	1.6	&	10.981	&	11.323	&	11.152	&	0.342	&	62	\\
	&	&	&	&	&	&	&	&	&		\\										
TYC\,1094-792-1	&	&	&	&	&	&	&	&	&		\\										
52734.9017	&	52970.5151	&	52852.7084	&	235.61	&	10.1	$\pm$	0.2	&	11.258	&	11.303	&	11.280	&	0.045	&	49	\\
53465.8989	&	53669.5626	&	53567.7308	&	203.66	&	10.1	$\pm$	0.2	&	11.263	&	11.305	&	11.284	&	0.042	&	40	\\
53832.9117	&	54093.6894	&	53963.3006	&	260.78	&	10.0	$\pm$	0.1	&	11.262	&	11.359	&	11.310	&	0.097	&	39	\\
54185.1482	&	54408.5302	&	54296.8392	&	223.38	&	10.2	$\pm$	0.2	&	11.260	&	11.401	&	11.330	&	0.141	&	75	\\
56017.1221	&	56275.7274	&	56146.4248	&	258.61	&	10.2	$\pm$	0.2	&	11.296	&	11.400	&	11.348	&	0.104	&	50	\\
56444.0241	&	56571.8938	&	56507.9590	&	127.87	&	10.0	$\pm$	0.4	&	11.282	&	11.358	&	11.320	&	0.076	&	39	\\
56736.1416	&	56971.7450	&	56853.9433	&	235.60	&	10.1	$\pm$	0.2	&	11.273	&	11.352	&	11.312	&	0.079	&	43	\\
57101.1487	&	57357.6901	&	57229.4194	&	256.54	&	10.2	$\pm$	0.2	&	11.302	&	11.361	&	11.331	&	0.059	&	42	\\
57880.0301	&	58037.8425	&	57958.9363	&	157.81	&	10.1	$\pm$	0.2	&	11.296	&	11.360	&	11.328	&	0.064	&	30	\\
	&	&	&	&	&	&	&	&	&		\\										
UY\,Equ	&	&	&	&	&	&	&	&	&		\\										
52812.9033	&	52970.5311	&	52891.7172	&	157.63	&	14.1	$\pm$	0.6	&	11.576	&	11.852	&	11.714	&	0.276	&	43	\\
53554.7984	&	53703.5227	&	53629.1606	&	148.72	&	14.8	$\pm$	0.6	&	11.577	&	11.698	&	11.637	&	0.121	&	30	\\
53850.9113	&	53947.0851	&	53898.9982	&	96.17	&	14.0	$\pm$	0.8	&	11.456	&	11.751	&	11.604	&	0.295	&	26	\\
53986.8655	&	54102.6934	&	54044.7795	&	115.83	&	14.1	$\pm$	0.7	&	11.483	&	11.824	&	11.653	&	0.341	&	25	\\
54197.1305	&	54427.5313	&	54312.3309	&	230.40	&	14.1	$\pm$	0.4	&	11.441	&	11.703	&	11.572	&	0.262	&	68	\\
54696.7944	&	54801.5217	&	54749.1581	&	104.73	&	13.9	$\pm$	0.8	&	11.539	&	11.668	&	11.604	&	0.129	&	24	\\
54949.9167	&	55190.7242	&	55070.3204	&	240.81	&	14.2	$\pm$	0.5	&	11.446	&	11.718	&	11.582	&	0.272	&	65	\\
		\hline
	\end{tabular}
	
	}
\end{table*}

\addtocounter{table}{-1}
\begin{table*}
	\centering
	\caption{Continued.}
	\resizebox{17cm}{!}{%
		\begin{tabular}{cccccccccc}
		\hline
HJD 	&	HJD 	&	HJD 	& $\Delta\,t$	& P	&	max	 &	 min	&	mean	&	A	 &	N	\\				
start        &  end  	&	mean		    & (day) & (day) &  (mag) &  (mag)   &  (mag)	&  (mag) &     \\																
\hline\noalign{\smallskip}																					
UY\,Equ	&	&	&	&	&		&	&	&	&	\\										
55285.1447	&	55402.0050	&	55343.5749	&	116.86	&	14.0	$\pm$	0.6	&	11.378	&	11.684	&	11.531	&	0.306	&	23	\\
55406.9854	&	55563.7050	&	55485.3452	&	156.72	&	14.2	$\pm$	0.5	&	11.510	&	11.667	&	11.589	&	0.157	&	36	\\
55657.1283	&	55912.7219	&	55784.9251	&	255.59	&	14.1	$\pm$	0.5	&	11.400	&	11.656	&	11.528	&	0.256	&	49	\\
56018.1407	&	56275.7333	&	56146.9370	&	257.59	&	14.2	$\pm$	0.3	&	11.417	&	11.568	&	11.492	&	0.151	&	47	\\
56381.1346	&	56641.7071	&	56511.4209	&	260.57	&	14.2	$\pm$	0.3	&	11.354	&	11.518	&	11.436	&	0.164	&	40	\\
56812.0131	&	56993.6882	&	56902.8507	&	181.68	&	13.9	$\pm$	0.4	&	11.327	&	11.487	&	11.407	&	0.160	&	60	\\
57101.1485	&	57353.6942	&	57227.4214	&	252.55	&	14.1	$\pm$	0.4	&	11.302	&	11.407	&	11.354	&	0.105	&	43	\\
57464.1512	&	57586.9492	&	57525.5502	&	122.80	&	14.3	$\pm$	1.1	&	11.251	&	11.330	&	11.290	&	0.079	&	24	\\
57591.9558	&	57715.7601	&	57653.8580	&	123.80	&	14.1	$\pm$	0.6	&	11.237	&	11.355	&	11.296	&	0.118	&	45	\\
57879.0927	&	58036.8399	&	57957.9663	&	157.75	&	14.3	$\pm$	0.6	&	11.205	&	11.418	&	11.312	&	0.213	&	36	\\
	&	&	&	&	&		&	&	&	&	\\										
FP\,Psc	&	&	&	&	&		&	&	&	&	\\										
52805.9122	&	52924.7198	&	52865.3160	&	118.81	&	13.6	$\pm$	0.9	&	11.202	&	11.324	&	11.263	&	0.122	&	28	\\
54268.9217	&	54467.5377	&	54368.2297	&	198.62	&	13.5	$\pm$	0.5	&	11.137	&	11.245	&	11.191	&	0.108	&	105	\\
54992.9334	&	55238.7291	&	55115.8313	&	245.80	&	13.2	$\pm$	0.3	&	11.135	&	11.249	&	11.192	&	0.114	&	46	\\
55334.1201	&	55457.9180	&	55396.0191	&	123.80	&	13.7	$\pm$	0.7	&	11.170	&	11.371	&	11.271	&	0.201	&	26	\\
55721.1098	&	55962.7415	&	55841.9257	&	241.63	&	13.6	$\pm$	0.4	&	11.155	&	11.297	&	11.226	&	0.142	&	47	\\
56795.1118	&	57054.7536	&	56924.9327	&	259.64	&	13.4	$\pm$	0.3	&	11.188	&	11.269	&	11.228	&	0.081	&	75	\\
57169.1114	&	57439.7144	&	57304.4129	&	270.60	&	13.2	$\pm$	0.4	&	11.144	&	11.200	&	11.172	&	0.056	&	76	\\
57533.1165	&	57788.7498	&	57660.9332	&	255.63	&	13.3	$\pm$	0.4	&	11.165	&	11.250	&	11.208	&	0.085	&	68	\\
57914.0927	&	58135.7146	&	58024.9037	&	221.62	&	13.6	$\pm$	0.4	&	11.238	&	11.319	&	11.279	&	0.081	&	41	\\
	&	&	&	&	&		&	&	&	&	\\										
TYC\,1541-191-1	&	&	&	&	&		&	&	&	&	\\										
52701.8917	&	52911.5072	&	52806.6995	&	209.62	&	11.7	$\pm$	0.3	&	11.142	&	11.397	&	11.269	&	0.255	&	60	\\
53070.8966	&	53189.6945	&	53130.2956	&	118.80	&	11.6	$\pm$	0.5	&	11.160	&	11.313	&	11.237	&	0.153	&	32	\\
53431.8987	&	53649.4923	&	53540.6955	&	217.59	&	11.7	$\pm$	0.3	&	11.178	&	11.453	&	11.316	&	0.275	&	58	\\
53798.8989	&	54051.6875	&	53925.2932	&	252.79	&	11.5	$\pm$	0.3	&	11.162	&	11.340	&	11.251	&	0.178	&	58	\\
54139.1229	&	54384.4883	&	54261.8056	&	245.37	&	11.6	$\pm$	0.3	&	11.128	&	11.437	&	11.283	&	0.309	&	106	\\
54524.1304	&	54743.4964	&	54633.8134	&	219.37	&	11.6	$\pm$	0.3	&	11.225	&	11.346	&	11.286	&	0.121	&	97	\\
54853.1692	&	55126.7551	&	54989.9622	&	273.59	&	11.6	$\pm$	0.3	&	11.122	&	11.337	&	11.229	&	0.215	&	81	\\
55221.1609	&	55509.6987	&	55365.4298	&	288.54	&	11.7	$\pm$	0.2	&	11.125	&	11.380	&	11.253	&	0.255	&	86	\\
55586.1656	&	55860.7450	&	55723.4553	&	274.58	&	11.7	$\pm$	0.2	&	11.149	&	11.294	&	11.221	&	0.145	&	66	\\
55951.1618	&	56243.6924	&	56097.4271	&	292.53	&	11.6	$\pm$	0.2	&	11.166	&	11.357	&	11.262	&	0.191	&	71	\\
56320.1500	&	56597.7231	&	56458.9366	&	277.57	&	11.6	$\pm$	0.2	&	11.138	&	11.415	&	11.276	&	0.277	&	50	\\
56731.0462	&	56940.7133	&	56835.8798	&	209.67	&	11.6	$\pm$	0.3	&	11.057	&	11.331	&	11.194	&	0.274	&	70	\\
57096.0361	&	57283.7574	&	57189.8968	&	187.72	&	11.7	$\pm$	0.3	&	11.140	&	11.245	&	11.192	&	0.105	&	45	\\
57478.1151	&	57585.7944	&	57531.9548	&	107.68	&	11.5	$\pm$	0.7	&	11.154	&	11.283	&	11.219	&	0.129	&	27	\\
57595.9300	&	57703.6909	&	57649.8105	&	107.76	&	11.7	$\pm$	0.6	&	11.134	&	11.328	&	11.231	&	0.194	&	38	\\
57790.1326	&	58021.7369	&	57905.9348	&	231.60	&	11.6	$\pm$	0.3	&	11.079	&	11.348	&	11.213	&	0.269	&	59	\\
	&	&	&	&	&		&	&	&	&	\\										
V343\,Del	&	&	&	&	&		&	&	&	&	\\										
52734.9157	&	52956.5249	&	52845.7203	&	221.61	&	10.5	$\pm$	0.2	&	11.308	&	11.521	&	11.415	&	0.213	&	72	\\
53481.9100	&	53630.5173	&	53556.2137	&	148.61	&	10.6	$\pm$	0.4	&	11.321	&	11.447	&	11.384	&	0.126	&	48	\\
53848.9124	&	54101.7168	&	53975.3146	&	252.80	&	10.4	$\pm$	0.1	&	11.322	&	11.518	&	11.420	&	0.196	&	73	\\
54191.1360	&	54376.5014	&	54283.8187	&	185.37	&	10.3	$\pm$	0.2	&	11.358	&	11.471	&	11.414	&	0.113	&	74	\\
54590.9162	&	54785.5172	&	54688.2167	&	194.60	&	10.4	$\pm$	0.3	&	11.282	&	11.411	&	11.346	&	0.129	&	55	\\
54942.1224	&	55103.8006	&	55022.9615	&	161.68	&	11.1	$\pm$	0.2	&	11.347	&	11.462	&	11.405	&	0.115	&	49	\\
55283.1300	&	55429.8992	&	55356.5146	&	146.77	&	10.4	$\pm$	0.3	&	11.295	&	11.465	&	11.380	&	0.170	&	40	\\
55432.9555	&	55566.7074	&	55499.8315	&	133.75	&	10.5	$\pm$	0.5	&	11.421	&	11.529	&	11.475	&	0.108	&	31	\\
55647.1297	&	55934.7009	&	55790.9153	&	287.57	&	10.4	$\pm$	0.2	&	11.373	&	11.506	&	11.440	&	0.133	&	63	\\
56182.8956	&	56277.7237	&	56230.3097	&	94.83	&	10.2	$\pm$	0.6	&	11.297	&	11.582	&	11.439	&	0.285	&	38	\\
56391.1256	&	56652.7378	&	56521.9317	&	261.61	&	10.3	$\pm$	0.1	&	11.276	&	11.470	&	11.373	&	0.194	&	37	\\
56736.1241	&	56993.6943	&	56864.9092	&	257.57	&	10.3	$\pm$	0.3	&	11.342	&	11.477	&	11.409	&	0.135	&	64	\\
57098.1498	&	57385.7052	&	57241.9275	&	287.56	&	10.4	$\pm$	0.2	&	11.364	&	11.497	&	11.431	&	0.133	&	76	\\
57481.1227	&	57630.9719	&	57556.0473	&	149.85	&	10.3	$\pm$	0.2	&	11.403	&	11.475	&	11.439	&	0.072	&	35	\\
58257.0335	&	58416.8097	&	58336.9216	&	159.78	&	10.4	$\pm$	0.3	&	11.342	&	11.526	&	11.434	&	0.184	&	19	\\
	&	&	&	&	&		&	&	&	&	\\										
V439\,Peg	&	&	&	&	&		&	&	&	&	\\										
52755.9110	&	52942.5388	&	52849.2249	&	186.63	&	24.4	$\pm$	1.3	&	10.569	&	10.669	&	10.619	&	0.100	&	42	\\
53487.9188	&	53641.5124	&	53564.7156	&	153.59	&	24.9	$\pm$	1.9	&	10.581	&	10.707	&	10.644	&	0.126	&	31	\\
53853.9200	&	54102.7015	&	53978.3108	&	248.78	&	24.0	$\pm$	0.9	&	10.592	&	10.722	&	10.657	&	0.130	&	47	\\
54196.1349	&	54376.5014	&	54286.3182	&	180.37	&	24.2	$\pm$	1.5	&	10.636	&	10.815	&	10.726	&	0.179	&	59	\\
54590.9162	&	54780.5525	&	54685.7344	&	189.64	&	24.6	$\pm$	1.6	&	10.638	&	10.850	&	10.744	&	0.212	&	88	\\
54962.1149	&	55202.7004	&	55082.4077	&	240.59	&	23.8	$\pm$	0.8	&	10.668	&	10.824	&	10.746	&	0.156	&	72	\\
55305.1252	&	55448.8158	&	55376.9705	&	143.69	&	24.0	$\pm$	1.9	&	10.675	&	10.839	&	10.757	&	0.164	&	40	\\
55701.0627	&	55828.9585	&	55765.0106	&	127.90	&	25.1	$\pm$	2.4	&	10.730	&	10.881	&	10.805	&	0.151	&	32	\\
56030.1217	&	56275.7648	&	56152.9433	&	245.64	&	24.6	$\pm$	1.2	&	10.712	&	10.976	&	10.844	&	0.264	&	49	\\
56390.1337	&	56641.7129	&	56515.9233	&	251.58	&	24.7	$\pm$	1.1	&	10.756	&	11.078	&	10.917	&	0.322	&	40	\\
56737.1496	&	56920.8622	&	56829.0059	&	183.71	&	24.5	$\pm$	1.5	&	10.799	&	11.103	&	10.951	&	0.304	&	67	\\
57115.1390	&	57395.7062	&	57255.4226	&	280.57	&	24.2	$\pm$	1.0	&	10.875	&	11.200	&	11.037	&	0.325	&	87	\\
57475.1370	&	57758.7495	&	57616.9433	&	283.61	&	24.5	$\pm$	1.0	&	11.020	&	11.259	&	11.140	&	0.239	&	82	\\
57847.1342	&	58097.6999	&	57972.4171	&	250.57	&	24.4	$\pm$	0.7	&	11.152	&	11.213	&	11.182	&	0.061	&	51	\\
58279.9735	&	58434.7789	&	58357.3762	&	154.81	&	24.5	$\pm$	0.9	&	11.103	&	11.207	&	11.155	&	0.104	&	19	\\
	&	&	&	&	&		&	&	&	&	\\										
TYC\,1683-144-1	&	&	&	&	&		&	&	&	&	\\										
52754.9243	&	52970.5297	&	52862.7270	&	215.61	&	45	$\pm$	4	&	11.335	&	11.517	&	11.426	&	0.182	&	53	\\
53492.9222	&	53677.5650	&	53585.2436	&	184.64	&	44	$\pm$	2	&	11.309	&	11.418	&	11.363	&	0.109	&	34	\\
53858.9205	&	54101.7196	&	53980.3201	&	242.80	&	44	$\pm$	2	&	11.278	&	11.418	&	11.348	&	0.140	&	39	\\
54202.1361	&	54431.5198	&	54316.8280	&	229.38	&	44	$\pm$	4	&	11.253	&	11.389	&	11.321	&	0.136	&	66	\\
54588.9153	&	54799.5226	&	54694.2190	&	210.61	&	42	$\pm$	4	&	11.281	&	11.398	&	11.340	&	0.117	&	54	\\
54952.9212	&	55209.7027	&	55081.3120	&	256.78	&	45	$\pm$	2	&	11.242	&	11.422	&	11.332	&	0.180	&	61	\\
55310.1261	&	55517.8448	&	55413.9855	&	207.72	&	46	$\pm$	4	&	11.110	&	11.363	&	11.237	&	0.253	&	57	\\
55667.1203	&	55933.7198	&	55800.4201	&	266.60	&	43	$\pm$	4	&	11.180	&	11.399	&	11.290	&	0.219	&	54	\\
56046.1086	&	56271.7605	&	56158.9346	&	225.65	&	47	$\pm$	5	&	11.282	&	11.415	&	11.349	&	0.133	&	44	\\
56399.1256	&	56652.7506	&	56525.9381	&	253.63	&	44	$\pm$	3	&	11.377	&	11.536	&	11.457	&	0.159	&	47	\\
56751.1355	&	57009.7120	&	56880.4238	&	258.58	&	44	$\pm$	4	&	11.344	&	11.595	&	11.470	&	0.251	&	65	\\
57121.1351	&	57368.7044	&	57244.9198	&	247.57	&	45	$\pm$	2	&	11.348	&	11.633	&	11.490	&	0.285	&	56	\\
57488.1251	&	57767.6997	&	57627.9124	&	279.57	&	45	$\pm$	3	&	11.391	&	11.705	&	11.548	&	0.314	&	93	\\
57848.1386	&	58034.9142	&	57941.5264	&	186.78	&	44	$\pm$	4	&	11.400	&	11.737	&	11.569	&	0.337	&	44	\\
58248.0784	&	58449.7614	&	58348.9199	&	201.68	&	44	$\pm$	5	&	11.287	&	11.741	&	11.514	&	0.454	&	28	\\
	&	&	&	&	&		&	&	&	&	\\										
	&	&	&	&	&		&	&	&	&	\\										
	&	&	&	&	&		&	&	&	&	\\										
	&	&	&	&	&		&	&	&	&	\\										
	&	&	&	&	&		&	&	&	&	\\										
		\hline																			
	\end{tabular}%
	\begin{tabular}{cccccccccc}																				
		\hline																			
HJD 	&	HJD 	&	HJD 	& $\Delta\,t$	& P	&	max	 &	 min	&	mean	&	A	 &	N	\\				
start        &  end  	&	mean		    & (day) & (day) &  (mag) &  (mag)   &  (mag)	&  (mag) &     \\																
\hline\noalign{\smallskip}																					
BE\,Ari	&	&	&	&	&		&	&	&	&	\\										
52831.9007	&	53000.5544	&	52916.2276	&	168.65	&	21.4	$\pm$	1.3	&	9.926	&	10.073	&	10.000	&	0.147	&	37	\\
53266.7132	&	53354.6418	&	53310.6775	&	87.93	&	21.0	$\pm$	4.7	&	10.023	&	10.177	&	10.100	&	0.154	&	79	\\
53554.9116	&	53699.5273	&	53627.2195	&	144.62	&	20.9	$\pm$	1.3	&	9.941	&	10.111	&	10.026	&	0.170	&	35	\\
53913.9295	&	54166.7206	&	54040.3251	&	252.79	&	21.3	$\pm$	1.0	&	9.909	&	10.117	&	10.013	&	0.208	&	33	\\
54271.1014	&	54489.8264	&	54380.4639	&	218.72	&	21.6	$\pm$	1.1	&	9.888	&	10.056	&	9.972	&	0.168	&	77	\\
54663.9351	&	54884.7191	&	54774.3271	&	220.78	&	22.0	$\pm$	1.2	&	9.886	&	10.085	&	9.985	&	0.199	&	79	\\
55007.1005	&	55131.6675	&	55069.3840	&	124.57	&	21.0	$\pm$	2.0	&	9.962	&	10.070	&	10.016	&	0.108	&	48	\\
55724.1090	&	55985.7271	&	55854.9181	&	261.62	&	21.5	$\pm$	1.0	&	9.859	&	10.035	&	9.947	&	0.176	&	45	\\
56126.1039	&	56356.7273	&	56241.4156	&	230.62	&	22.0	$\pm$	0.8	&	9.817	&	9.989	&	9.903	&	0.172	&	35	\\
56488.1031	&	56677.7954	&	56582.9493	&	189.69	&	21.8	$\pm$	1.0	&	9.810	&	10.066	&	9.938	&	0.256	&	29	\\
56833.0949	&	57054.7311	&	56943.9130	&	221.64	&	21.6	$\pm$	0.9	&	9.909	&	10.105	&	10.007	&	0.196	&	67	\\
57191.1148	&	57422.7285	&	57306.9217	&	231.61	&	21.5	$\pm$	0.7	&	9.944	&	10.329	&	10.136	&	0.385	&	58	\\
57553.1079	&	57804.7285	&	57678.9182	&	251.62	&	21.5	$\pm$	0.8	&	10.020	&	10.219	&	10.119	&	0.199	&	71	\\
57928.0916	&	58121.8363	&	58024.9640	&	193.74	&	21.4	$\pm$	1.2	&	10.128	&	10.212	&	10.170	&	0.084	&	30	\\
58289.0944	&	58447.7326	&	58368.4135	&	158.64	&	20.1	$\pm$	1.4	&	10.138	&	10.210	&	10.174	&	0.072	&	18	\\
	&	&	&	&	&		&	&	&	&	\\										
V592\,Peg	&	&	&	&	&		&	&	&	&	\\										
52787.9299	&	52976.5345	&	52882.2322	&	188.60	&	19.2	$\pm$	1.1	&	10.710	&	10.985	&	10.848	&	0.275	&	46	\\
53272.6082	&	53353.5646	&	53313.0864	&	80.96	&	20.7	$\pm$	4.8	&	10.827	&	10.983	&	10.905	&	0.156	&	34	\\
53511.9307	&	53675.5056	&	53593.7182	&	163.57	&	19.1	$\pm$	1.2	&	10.729	&	10.970	&	10.850	&	0.241	&	38	\\
53880.9290	&	54134.7053	&	54007.8172	&	253.78	&	19.5	$\pm$	0.6	&	10.705	&	10.923	&	10.814	&	0.218	&	47	\\
54220.1181	&	54490.7304	&	54355.4243	&	270.61	&	19.4	$\pm$	0.7	&	10.748	&	10.945	&	10.847	&	0.197	&	51	\\
54603.9285	&	54797.5385	&	54700.7335	&	193.61	&	19.2	$\pm$	0.7	&	10.800	&	10.932	&	10.866	&	0.132	&	42	\\
54953.1182	&	55232.7076	&	55092.9129	&	279.59	&	19.3	$\pm$	0.7	&	10.762	&	10.941	&	10.852	&	0.179	&	111	\\
55330.1191	&	55593.7228	&	55461.9210	&	263.60	&	19.3	$\pm$	0.6	&	10.718	&	10.896	&	10.807	&	0.178	&	58	\\
55696.1255	&	55963.7089	&	55829.9172	&	267.58	&	18.7	$\pm$	0.6	&	10.730	&	10.896	&	10.813	&	0.166	&	58	\\
56066.1185	&	56329.7081	&	56197.9133	&	263.59	&	19.6	$\pm$	0.8	&	10.770	&	10.874	&	10.822	&	0.104	&	46	\\
56414.1215	&	56688.7092	&	56551.4154	&	274.59	&	18.4	$\pm$	0.5	&	10.750	&	10.936	&	10.843	&	0.186	&	37	\\
56798.0969	&	57050.7364	&	56924.4167	&	252.64	&	19.0	$\pm$	0.8	&	10.760	&	10.913	&	10.836	&	0.153	&	80	\\
57145.1190	&	57390.7003	&	57267.9097	&	245.58	&	18.9	$\pm$	0.7	&	10.760	&	10.966	&	10.863	&	0.206	&	45	\\
57505.1261	&	57772.6998	&	57638.9130	&	267.57	&	18.9	$\pm$	0.7	&	10.763	&	11.005	&	10.884	&	0.242	&	103	\\
57896.1224	&	58117.7036	&	58006.9130	&	221.58	&	19.0	$\pm$	0.7	&	10.822	&	10.948	&	10.885	&	0.126	&	38	\\
58252.1053	&	58449.8110	&	58350.9582	&	197.71	&	19.0	$\pm$	1.0	&	10.755	&	11.035	&	10.895	&	0.280	&	35	\\
	&	&	&	&	&		&	&	&	&	\\										
TYC\,4667-90-1	&	&	&	&	&		&	&	&	&	\\										
52031.9292	&	52190.6035	&	52111.2664	&	158.67	&	8.9	$\pm$	0.2	&	11.275	&	11.531	&	11.403	&	0.256	&	38	\\
52765.9179	&	53012.5664	&	52889.2422	&	246.65	&	8.8	$\pm$	0.1	&	11.319	&	11.433	&	11.376	&	0.114	&	91	\\
53522.9048	&	53736.5667	&	53629.7358	&	213.66	&	8.9	$\pm$	0.2	&	11.371	&	11.582	&	11.476	&	0.211	&	83	\\
53862.9236	&	54137.7083	&	54000.3160	&	274.78	&	8.8	$\pm$	0.1	&	11.292	&	11.641	&	11.467	&	0.349	&	42	\\
54239.1160	&	54482.7609	&	54360.9385	&	243.64	&	8.8	$\pm$	0.2	&	11.327	&	11.577	&	11.452	&	0.250	&	59	\\
54623.9319	&	54863.7084	&	54743.8202	&	239.78	&	8.8	$\pm$	0.1	&	11.334	&	11.544	&	11.439	&	0.210	&	49	\\
54966.1240	&	55233.7120	&	55099.9180	&	267.59	&	8.9	$\pm$	0.2	&	11.359	&	11.630	&	11.495	&	0.271	&	77	\\
55344.1096	&	55595.7140	&	55469.9118	&	251.60	&	8.8	$\pm$	0.1	&	11.398	&	11.679	&	11.538	&	0.281	&	58	\\
55722.0833	&	55968.7161	&	55845.3997	&	246.63	&	8.9	$\pm$	0.1	&	11.360	&	11.535	&	11.448	&	0.175	&	52	\\
56083.0915	&	56328.7117	&	56205.9016	&	245.62	&	8.8	$\pm$	0.1	&	11.310	&	11.698	&	11.504	&	0.388	&	49	\\
56488.0529	&	56682.7164	&	56585.3847	&	194.66	&	8.8	$\pm$	0.1	&	11.381	&	11.574	&	11.478	&	0.193	&	29	\\
56796.1160	&	57048.7155	&	56922.4158	&	252.60	&	8.9	$\pm$	0.1	&	11.415	&	11.650	&	11.532	&	0.235	&	61	\\
57181.1062	&	57420.7214	&	57300.9138	&	239.62	&	8.7	$\pm$	0.2	&	11.514	&	11.595	&	11.555	&	0.081	&	65	\\
57529.8962	&	57774.7437	&	57652.3200	&	244.85	&	8.8	$\pm$	0.2	&	11.493	&	11.620	&	11.557	&	0.127	&	95	\\
57879.9202	&	58126.6982	&	58003.3092	&	246.78	&	8.8	$\pm$	0.1	&	11.457	&	11.654	&	11.555	&	0.197	&	74	\\
58243.9249	&	58436.7386	&	58340.3318	&	192.81	&	8.8	$\pm$	0.2	&	11.429	&	11.665	&	11.547	&	0.236	&	36	\\
	&	&	&	&	&		&	&	&	&	\\										
BC\,Sex	&	&	&	&	&		&	&	&	&	\\										
52622.8090	&	52842.4570	&	52732.6330	&	219.65	&	15.2	$\pm$	0.5	&	11.725	&	11.826	&	11.776	&	0.101	&	58	\\
52958.8601	&	53186.4881	&	53072.6741	&	227.63	&	15.5	$\pm$	0.4	&	11.744	&	11.873	&	11.808	&	0.129	&	50	\\
53356.8180	&	53558.4726	&	53457.6453	&	201.65	&	16.1	$\pm$	0.7	&	11.714	&	11.834	&	11.774	&	0.120	&	33	\\
53703.8386	&	53837.5961	&	53770.7174	&	133.76	&	16.3	$\pm$	1.1	&	11.692	&	11.853	&	11.772	&	0.161	&	29	\\
54058.0990	&	54290.4657	&	54174.2824	&	232.37	&	15.4	$\pm$	0.4	&	11.738	&	11.836	&	11.787	&	0.098	&	70	\\
54430.8514	&	54661.4638	&	54546.1576	&	230.61	&	14.8	$\pm$	0.5	&	11.740	&	11.852	&	11.796	&	0.112	&	65	\\
54794.8573	&	55023.4622	&	54909.1598	&	228.60	&	15.5	$\pm$	0.5	&	11.765	&	11.906	&	11.835	&	0.141	&	58	\\
55129.1368	&	55383.7457	&	55256.4413	&	254.61	&	15.3	$\pm$	0.3	&	11.763	&	11.904	&	11.833	&	0.141	&	59	\\
55487.1453	&	55740.7607	&	55613.9530	&	253.62	&	15.9	$\pm$	0.5	&	11.801	&	11.927	&	11.864	&	0.126	&	30	\\
55855.1484	&	56091.7872	&	55973.4678	&	236.64	&	15.6	$\pm$	0.5	&	11.797	&	11.942	&	11.869	&	0.145	&	44	\\
56233.1393	&	56443.8411	&	56338.4902	&	210.70	&	15.7	$\pm$	0.6	&	11.808	&	11.933	&	11.871	&	0.125	&	31	\\
56615.0657	&	56817.7909	&	56716.4283	&	202.73	&	15.1	$\pm$	0.4	&	11.791	&	11.899	&	11.845	&	0.108	&	40	\\
56943.1292	&	57131.8679	&	57037.4986	&	188.74	&	15.4	$\pm$	0.8	&	11.782	&	11.886	&	11.834	&	0.104	&	52	\\
57309.1376	&	57417.8940	&	57363.5158	&	108.76	&	15.5	$\pm$	1.0	&	11.796	&	11.873	&	11.835	&	0.077	&	41	\\
57420.7789	&	57576.4712	&	57498.6251	&	155.69	&	15.6	$\pm$	0.6	&	11.719	&	11.879	&	11.799	&	0.160	&	55	\\
57672.1433	&	57939.4703	&	57805.8068	&	267.33	&	15.6	$\pm$	0.5	&	11.758	&	11.918	&	11.838	&	0.160	&	127	\\
58074.8597	&	58197.9610	&	58136.4104	&	123.10	&	15.3	$\pm$	0.9	&	11.747	&	11.871	&	11.809	&	0.124	&	29	\\
		\hline
	&		&		&				&		&		&		&		&		\\
	&		&		&				&		&		&		&		&		\\
   &		&				&		&      Identifier & $P_{rot}$ (day)	               &	       &	       &	       \\
\cline{5-6}
	&		&		&				&		&		&		&		&		\\
   &	&	&	&      V660\,Vir		&	73$\pm$3			&	       &	       &	       \\
   &	&	&	&      DG\,Ari  		&	34$\pm$1			&	       &	       &	       \\
   &	&	&	&      V1263\,Tau		&	20.4$\pm$0.5		&	       &	       &	       \\
   &	&	&	&      FK\,CMi  		&	19.4$\pm$0.1		&	       &	       &	       \\
   &	&	&	&      V383\,Vir		&	14.4$\pm$0.3		&	       &	       &	       \\
   &	&	&	&      BD+04\,3503		&	8.4$\pm$0.1		&	       &	       &	       \\
   &	&	&	&      BD+02\,3610		&	12.9$\pm$0.1		&	       &	       &	       \\
   &	&	&	&      GH\,Psc  		&	34.4$\pm$0.9		&	       &	       &	       \\
   &	&	&	&      TYC\,723-863-1 	&	44$\pm$1			&	       &	       &	       \\
   &	&	&	&      HD\,354410		&	27.3$\pm$0.5		&	       &	       &	       \\
   &	&	&	&      TYC\,1094-792-1	&	10.1$\pm$0.1		&	       &	       &	       \\
   &	&	&	&      UY\,Equ  		&	14.1$\pm$0.2		&	       &	       &	       \\
   &	&	&	&      FP\,Psc  		&	13.4$\pm$0.2		&	       &	       &	       \\
   &	&	&	&      TYC\,1541-191-1	&	11.6$\pm$0.1		&	       &	       &	       \\
   &	&	&	&      V343\,Del		&	10.4$\pm$0.2		&	       &	       &	       \\
   &	&	&	&      V439\,Peg		&	24.4$\pm$0.4		&	       &	       &	       \\
   &	&	&	&      TYC\,1683-144-1	&	44$\pm$1			&	       &	       &	       \\
   &	&	&	&      BE\,Ari  		&	21.4$\pm$0.5		&	       &	       &	       \\
   &	&	&	&      V592\,Peg		&	19.2$\pm$0.5		&	       &	       &	       \\
   &	&	&	&      TYC\,4667-90-1	&	8.8$\pm$0.0		&	       &	       &	       \\
   &	&	&	&      BC\,Sex  		&	15.5$\pm$0.4		&	       &	       &	       \\
\cline{5-6}
	\end{tabular}
	
	}
\end{table*}
\end{appendix}

\bibliographystyle{pasa-mnras}
\bibliography{rscvns}

\end{document}